\documentclass[a4paper]{article}
\usepackage[margin=25mm]{geometry}
\usepackage{graphicx}
\usepackage{amsmath} 
\usepackage{amssymb}
\usepackage{makecell}
\usepackage{xcolor}
\usepackage{dcolumn,verbatim}
\usepackage{multirow}  
\usepackage[figuresright]{rotating} 
\usepackage{pythonhighlight} 
\usepackage{bigfoot} 
\usepackage[none]{hyphenat} 
\usepackage{chngcntr} 
\usepackage{float} 
\usepackage[hidelinks]{hyperref}

\graphicspath{ {./fig/} }
\counterwithout{equation}{section} 

\title{Application of Quantum Computers in Foreign Exchange Reserves Management}
\author{Martin Vesel\'y
\thanks{Contact address: martin.vesely@cnb.cz. This work was supported by Czech National Bank Research Project No. P1/2021. The author would like to thank Simona Malovan\'a for invaluable advice and project supervision. The views expressed in this paper are those of the author and not necessarily those of the Czech National Bank}
\\Czech National Bank - Risk Management Department
}

\begin{document}
\interfootnotelinepenalty=10000

\maketitle

\begin{abstract}
The main purpose of this article is to evaluate possible applications of quantum computers in foreign exchange reserves management. The capabilities of quantum computers are demonstrated by means of risk measurement using the quantum Monte Carlo method and portfolio optimization using a linear equations system solver (the Harrow-Hassidim-Lloyd algorithm) and quadratic unconstrained binary optimization (the quantum approximate optimization algorithm). All demonstrations are carried out on the cloud-based IBM Quantum\textsuperscript{TM} platform. Despite the fact that real-world applications are impossible under the current state of development of quantum computers, it is proven that in principle it will be possible to apply such computers in FX reserves management in the future. In addition, the article serves as an introduction to quantum computing for the staff of central banks and financial market supervisory authorities.
\end{abstract}

\section{Introduction and Motivation}
Computational techniques and instruments have always been an integral part of finance and trading. Even Fibonacci's famous {\it Liber Abaci}, besides introducing Arabian (Indian) numerals to medieval Europe, served mainly as a mathematical textbook for merchants and bankers. At the dawn of the modern computer era in the 1950s, corporations quickly recognized the opportunities these machines offered and started to use them not only to speed up their daily operations, but also to build sophisticated models of markets, solve complex optimization problems and so on. This led to the development of completely new branches of science and sectors of the economy. In the last decade, quantum computers have become available to the mass market, and we are now on the cusp of a revolution similar to that initiated by classical computers in the middle of the 20\textsuperscript{th} century. For this reason, we investigate possible applications of quantum computers in central banks, in particular their use for foreign exchange reserves management. This is the main aim of this paper.  A secondary goal of the paper is to serve as an introduction to quantum computing for the staff of central banks and other financial market regulatory authorities.

Historically, there have been two main incentives to build a quantum computer. First, there are physical limits on the miniaturization of electronic parts preventing any further increase in the speed of classical computers at some point in time. Second, and more importantly, some tasks are difficult to solve on classical computers by definition (e.g.,~the traveling salesperson problem and factorization of large integers). In such cases, quantum algorithms offer significant speed-up (in other words, have lower complexity) in comparison with their classical counterparts.

The idea of a quantum computer was first proposed in the seminal paper \cite{feynman-intro} as a tool for simulating quantum systems, because this task is difficult to carry out on classical computers -- in Feynman's own words: {\it Nature isn't classical, dammit, and if you want to make a simulation of nature, you'd better make it quantum mechanical, and by golly it's a wonderful problem, because it doesn't look so easy}. The first algorithm outside theoretical physics was presented in \cite{deutsch-alg}. It is able to recognize if a binary function is either constant or balanced\footnote{A balanced binary function returns 0 for half of its inputs and 1 for the other half. A constant function returns either 0 or 1 for all inputs.} with only one evaluation of the function, regardless of the number of possible inputs. On a classical computer, by contrast, the function values for a half of its inputs have to be evaluated. Although this algorithm is a pure academic exercise, it shows that there are other tasks besides quantum system simulations where quantum computers can outperform classical ones. 

The first practical quantum algorithm allowing integers to be factorized in polynomial time was introduced in \cite{shor-alg}, while the best performing classical algorithms for factorization have exponential complexity.\footnote{Because of this exponential speed-up, Shor's algorithm is capable of breaking RSA ciphering. However, current quantum computers are still not at the necessary degree of development, and many engineering problems have yet to be solved. On top of that, quantum mechanics offers an unbreakable ciphering mechanism, discussed in \cite{quantum-crypto}, which would replace RSA. A practical implementation of a network for quantum ciphering based on Bennett's proposals is presented in \cite{china-crypto}. The network is being used experimentally by the People's Bank of China, for example.} 
Another important practical algorithm was suggested in \cite{grover-alg} to achieve quadratic  speed-up of searching in unordered databases. Grover's algorithm was later modified in \cite{durr-hoyer-alg} for finding the extremes of a function. Besides optimization problems, systems of linear equations often appear in scientific and business tasks. A quantum solver for systems of linear equations (the ``HHL algorithm'') was proposed in \cite{HHL-alg}. For certain types of linear equations, it offers exponential speed-up in comparison with classical solvers. This list of algorithms is far from exhaustive and serves mainly as a historical insight into the development of quantum computing and the possibilities it offers. The interested reader can consult an extensive overview of quantum algorithms in \cite{algos-for-beginners}. We will also discuss other algorithms in the second section of this paper.

The above-mentioned algorithms and modifications thereof can be used in several tasks in finance, for example, portfolio optimization \cite{ptf-optim}, risk management \cite{quantum-risk} and \cite{quantum-credit-risk}, interest rate derivatives pricing \cite{derivatives-pricing}, and machine learning \cite{machine-learning}. Summaries of quantum computer applications in finance are provided in \cite{prospects-in-finance}, \cite{state-of-art-finance}, and \cite{prospects-in-finance-2}. On top of that, Sveriges Riksbank has published an introduction to quantum computing for economists containing a long list of algorithms useful in finance and business generally \cite{riksbank}. Importantly, that paper also discusses the concept of tamper-proof quantum money.

In this paper, we will focus on the application of quantum computers in foreign exchange reserves management. In particular, we will discuss portfolio optimization and risk measurement. In the case of portfolio optimization, we will focus first on finding the optimal balance between bonds and equities and second on selecting equities so as to maximize portfolio returns while minimizing risk. Concerning risk measures, we will present an evaluation of VaR and CVaR and some other statistical properties. All demonstrations will be carried out on the IBM Quantum\textsuperscript{TM} computer. IBM offers a cloud-based platform with access to a real quantum computer for free, provided that it is used for educational and research purposes. IBM's quantum computer is programmed with a Python-based language called Qiskit. An introduction to the language is provided in \cite{qiskit-manual} and details can be found in the official Qiskit documentation.

In this research, we found that quantum computers are able to solve some tasks connected with FX reserves management, in particular risk measurement using the quantum Monte Carlo method and Markowitz-like portfolio optimization using a quantum linear equations solver (the HHL algorithm) and a quantum quadratic unconstrained binary optimization method (the quantum approximate optimization algorithm -- QAOA). Unfortunately, the current early stage of quantum hardware development restricted us to toy models only (for example, we used the QAOA for a task with only five binary variables). The technical imperfections of the quantum computers currently available also meant that only a few of our demonstrations were successful. On the one hand, we successfully carried out portfolio optimization with the QAOA employing five binary variables for all the cases tested. On the other hand, the application of the quantum Monte Carlo method to risk measurement generated only partially correct results, and the use of the HHL algorithm in portfolio optimization failed completely. However, running the algorithms discussed above on a quantum computer simulator (i.e.,~software simulating a quantum computer on a classical one) proved that our implementations are correct, hence we came to the conclusion that in principle nothing prevents us from deploying quantum computers in FX reserves management in the future. We only have to wait for technical progress in quantum hardware capabilities.

The rest of the paper is organized as follows. In the second section, we discuss the theoretical background of quantum computing, ranging from the very basics to the algorithms we will employ in this paper. The third section contains details of the FX reserves management tasks described above and the results of solving those tasks on IBM's quantum computer. The fourth section concludes. The appendixes contain a short description of the IBM Quantum\textsuperscript{TM}  environment and the Qiskit language, the Qiskit source codes for all the demonstrations presented in this paper, and other expanding materials.

\section{Theoretical Background}

In this section, we discuss the basics of quantum computing theory. First, we show what quantum bits (qubits) are and how we can use them to perform useful calculations.
We also introduce the terms superposition and quantum entanglement. Not only are these terms crucial for understanding quantum computing, they are also the reason why quantum computers outperform classical ones in some tasks.
After that, we introduce the algorithms we will employ in our demonstration of the application of quantum computers in FX reserves management. The first is a quantum method allowing us to calculate risk measures such as VaR and CVaR based on the historical empirical distribution of portfolio returns. The second is a quantum solver of systems of linear equations (the HHL algorithm), which will be employed in Markowitz-like portfolio optimization. The last algorithm to be discussed is the quantum approximate optimization algorithm (QAOA), which enables us to solve quadratic unconstrained binary optimization problems (e.g.,~portfolio optimization and the famous traveling salesperson problem).

\subsection{Quantum Computing Basics}

This section covers the necessary minimum for understanding the theory behind quantum computing. We explain what quantum bits are, how quantum operations (or quantum gates) work, and how quantum gates are composed to form quantum circuits representing quantum algorithms. For readers interested in gaining a deeper understanding, we recommend the classical monograph on quantum computing \cite{nielsen-chuang-book} and the introduction to computer science (including quantum computing) \cite{feynman-book}. A non-technical introduction to quantum mechanics can be found in \cite{qm-concepts-book}. The mathematical background of quantum mechanics is again discussed in \cite{nielsen-chuang-book}.

\subsubsection{Quantum Bits, Superposition, Entanglement, and Measurement}
\label{subsubsectionQubits}
The basic building block of a quantum computer is the quantum bit ({\bf qubit}). Similarly to a bit on a {\bf classical computer} (i.e.,~a non-quantum computer such as we use in daily life), which is considered to be the smallest unit of information, the qubit is the smallest piece of information in the quantum sense.

A qubit is a quantum system which can be in two distinguishable states. Such states are 0 and 1, as on a classical computer. However, a qubit can be in both states at once, written mathematically as 
\begin{equation}
|q\rangle = \alpha|0\rangle + \beta|1\rangle \label{qbitDef},
\end{equation}
where $\alpha$ and $\beta$ are complex numbers. Symbols $|0\rangle$ and $|1\rangle$ represent states 0 and 1, and $|q\rangle$ is the whole qubit. The symbol $|q\rangle$ comes from the {\bf Dirac bra-ket notation}. In fact, $|q\rangle$ ({\bf ``ket''}) is a column vector, and it holds that
\begin{equation}
|0\rangle = \begin{pmatrix}  1  \\ 0  \end{pmatrix} \\ \,\,\,\,\,\,
|1\rangle = \begin{pmatrix}  0  \\ 1  \end{pmatrix} \label{Dirac01}.
\end{equation}
The Dirac notation is mainly used as an economical way of writing down vectors. To have a complete picture of the Dirac notation, we also introduce the symbol $\langle q|$ ({\bf ``bra''}), which is the transpose conjugate of $|q\rangle$, i.e.,~a row vector composed of complex conjugates of the elements of $|q\rangle$.

With \eqref{Dirac01}, we can write the qubit defined in \eqref{qbitDef} as a vector:
\begin{equation}
|q\rangle = \begin{pmatrix}  \alpha  \\ \beta  \end{pmatrix}.
\end{equation}
Vector $|q\rangle$ is clearly a linear combination of vectors $|0\rangle$ and $|1\rangle$ -- we say that qubit $|q\rangle$ is a {\bf superposition} of states
 $|0\rangle$ and $|1\rangle$.\footnote{Superposition is often explained with a thought experiment called {\bf Schr\"{o}dinger's cat}. Suppose we have a box containing a cat, a bottle of poison, a hammer, and a radioactive isotope. The probability that the isotope emits a radioactive particle is one half. Once this happens, the hammer breaks the bottle and the cat dies. Since the box is closed, we do not know whether the cat is dead or alive. It is therefore half dead and half alive. In other words, it is in a superposition of the $|\text{dead}\rangle$ and $|\text{alive}\rangle$ states.} States $|0\rangle$ and $|1\rangle$ are called {\bf basis states}, because they form the basis for the vector space $\mathbb{C}^2$. Parameters $\alpha$ and $\beta$ are called {\bf probability amplitudes}. It also holds that the square of the absolute value of the probability amplitude\footnote{We have to square the absolute value and not only $\alpha$ and $\beta$, because $\alpha, \beta \in \mathbb{C}.$}
 is the probability that the qubit is in a particular state after measurement (we will discuss measurement in detail later). It therefore holds that
\begin{equation}
P(|0\rangle) = |\alpha|^2 \\ \,\,\,\,\,\,
P(|1\rangle) = |\beta|^2.
\end{equation}
Since $|\alpha|^2$ and $|\beta|^2$ are the probabilities of mutually exclusive outcomes, it holds that $|\alpha|^2 + |\beta|^2 = 1$. This also means that the length of the vector describing qubit $|q\rangle$ is unity. Therefore, $|q\rangle$ is a proper quantum state if and only if it has unit Euclidean length.

\begin{figure}[H]
\caption{Examples of Qubit State Visualization on the Bloch Sphere}
\begin{center}
	\begin{tabular}{cc}
		\includegraphics[scale = 0.8]{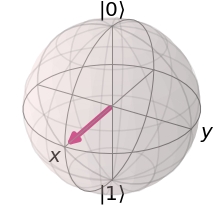} &
		\includegraphics[scale = 0.8]{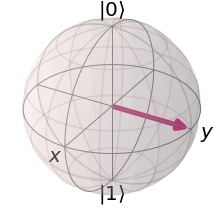}
	\end{tabular}
\end{center}
\vspace{0.1cm}
{\footnotesize \textbf{\textit{Note:}}  State $\frac{1}{\sqrt{2}}(|0\rangle + |1\rangle)$ is on the left and state $\frac{1}{\sqrt{2}}(|0\rangle + i|1\rangle)$ is on the right.}\\
{\footnotesize \textbf{\textit{Source:}} Author's own creation on IBM Quantum\textsuperscript{TM}}
\label{fig_bloch_sphere}
\end{figure}

As $\alpha$ and $\beta$ are complex numbers satisfying $|\alpha|^2+|\beta|^2 = 1$, we can use the exponential form of a complex number (i.e.,~$\alpha = |\alpha|\mathrm{e}^{i\varphi_\alpha}$) and the trigonometric identity $\sin^2 \frac{\theta}{2} + \cos^2 \frac{\theta}{2} = 1$ to rewrite \eqref{qbitDef} as
\begin{equation}
|q\rangle = \mathrm{e}^{i\varphi_\alpha}\Big(\cos \frac{\theta}{2} |0\rangle + \mathrm{e}^{i(\varphi_\beta - \varphi_\alpha)}\sin \frac{\theta}{2} |1\rangle\Big) \label{qubitPhase}.
\end{equation}
The term $\varphi_\alpha$ is called the {\bf global phase} and can be ignored, as two qubits differing in the global phase only are physically indistinguishable and are considered to represent the same state of the qubit. The term $\varphi_\beta - \varphi_\alpha$ is the {\bf relative phase} and is usually denoted by $\varphi$. The phase is a purely quantum feature of the qubit and we will see how it is used later. For the time being, consider it simply to be a parameter of the qubit. Neglecting the global phase, we arrive at the most general form of the qubit:
\begin{equation}
|q\rangle = \cos \frac{\theta}{2} |0\rangle + \mathrm{e}^{i\varphi}\sin \frac{\theta}{2} |1\rangle \label{generalQubit}. 
\end{equation}
As can be seen from \eqref{generalQubit}, the qubit is described by two angles $\theta$ and $\varphi$.
Therefore, any qubit can be visualized on a unit sphere called the {\bf Bloch sphere}.
As an example, visualizations of qubits in states $\frac{1}{\sqrt{2}}(|0\rangle + |1\rangle)$, i.e.,~having $\theta = \frac{\pi}{2}$ and $\varphi = 0$, and $\frac{1}{\sqrt{2}}(|0\rangle + i|1\rangle)$ ($\theta = \frac{\pi}{2}$ and $\varphi = \frac{\pi}{2}$) on the Bloch sphere are shown in Figure~\ref{fig_bloch_sphere}.

As can be seen in Figure~\ref{fig_bloch_sphere}, both states are superpositions of $|0\rangle$ and $|1\rangle$, because the vectors representing them lie outside the $z$ axis, where states $|0\rangle$ and $|1\rangle$ are positioned. The states differ in phase $\varphi$.  While state $\frac{1}{\sqrt{2}}(|0\rangle + |1\rangle)$ lies on the $x$ axis, the other state is located on the $y$ axis, i.e.,~it is rotated by $\pi/2$ radians. Hence, this example shows a  graphical representation of the quantum phase as a rotation of a vector on the Bloch sphere. As we will see later, rotation is one of the most important operations allowing calculations to be performed on a quantum computer.

Now that we have introduced the single-qubit states, we can move to {\bf multi-qubit states}. As two bits can form four different combinations (00, 01, 10, 11), the general two-qubit state is
\begin{equation}
|q_1q_0\rangle = \alpha |00\rangle + \beta|01\rangle + \gamma|10\rangle + \delta|11\rangle \label{twoQubitDef}, 
\end{equation}
where $|\alpha|^2 + |\beta|^2 + |\gamma|^2 + |\delta|^2 = 1$. In vector form, the basis states can be written as
\begin{equation}
|00\rangle = \begin{pmatrix}  1  \\ 0 \\ 0 \\ 0  \end{pmatrix} \\ \,\,\,\,\,\,
|01\rangle = \begin{pmatrix}  0  \\ 1 \\ 0  \\ 0 \end{pmatrix} \\ \,\,\,\,\,\,
|10\rangle = \begin{pmatrix}  0  \\ 0 \\ 1 \\ 0 \end{pmatrix} \\ \,\,\,\,\,\,
|11\rangle = \begin{pmatrix}  0  \\ 0 \\ 0  \\ 1 \end{pmatrix}.
\end{equation}
Obviously, there has to be a connection between a two-qubit state and its partial qubits, i.e.,~the single-qubit states. This connection is established through an operation called the {\bf tensor product}. We show the definition of the tensor product on an example. Consider the state $|01\rangle$, which is composed of two single qubits, the first being $|0\rangle$ and the second $|1\rangle$. We can therefore write $|01\rangle = |0\rangle \otimes |1\rangle$, where symbol $\otimes$ is the tensor product defined as follows:
\begin{equation}
|01\rangle =  |0\rangle \otimes |1\rangle =
\begin{pmatrix} 1 \\ 0 \end{pmatrix}
\otimes
\begin{pmatrix} 0 \\ 1 \end{pmatrix}
=
\begin{pmatrix}
1 \begin{pmatrix} 0 \\ 1 \end{pmatrix} \\
0 \begin{pmatrix} 0 \\ 1 \end{pmatrix} \\
\end{pmatrix}
=
\begin{pmatrix}
\begin{pmatrix} 0 \\ 1 \end{pmatrix} \\
\begin{pmatrix} 0 \\ 0 \end{pmatrix} \\
\end{pmatrix}
=
\begin{pmatrix} 0 \\ 1 \\ 0 \\ 0 \end{pmatrix} .
\end{equation}
It should be emphasized that the tensor product is not a commutative operation. For example,~$|1\rangle \otimes |0\rangle = |10\rangle$ is a clearly different state than 
$|0\rangle \otimes |1\rangle = |01\rangle$. Similarly, it is possible to define states composed of three or more qubits.\footnote{It is apparent that an $n$-qubit state is described by $2^n$ complex numbers, which means that the amount of memory needed to simulate it on a classical computer increases exponentially with the number of qubits involved. This has been the main incentive for the construction of quantum computers, as the amount of quantum memory necessary increases only linearly in the number of qubits simulated.}

Not all multi-qubit states are tensor products of other states. For example, the {\bf Bell state}
\begin{equation}
|\beta_{00}\rangle = |q_1q_0\rangle = \frac{1}{\sqrt{2}}(|00\rangle + |11\rangle) \label{bellExample}
\end{equation}
cannot be written (separated) as a tensor product of two single qubits. Such inseparable states are called {\bf entangled states}. These states occupy a special place, because, together with superposition, they are responsible for the speed-up provided by quantum computers. Another special feature of entangled states is that it is possible to infer the whole state from only a few of the qubits the state is composed of. For example, in the Bell state \eqref{bellExample}, if the first qubit is $|1\rangle$, we know that the whole state is $|11\rangle$.\footnote{Remember Schr\"{o}dinger's cat. The bottle is another quantum system in a superposition of states $|\text{broken}\rangle$ and $|\text{unbroken}\rangle$. Clearly, the cat's state depends on the state of the bottle -- we only have the combinations $|\text{broken}\rangle|\text{dead}\rangle$ and $|\text{unbroken}\rangle|\text{alive}\rangle$, so the systems are entangled. We can see a clear resemblance to the Bell state $|00\rangle + |11\rangle$. However, this analogy is not perfect, as the bottle influences the cat but not vice versa.}

Examples of other entangled, generally $n$-qubit states include
\begin{itemize}
\item GHZ states: $\frac{1}{\sqrt{2}}(|00\dots 0\rangle + |11\dots 1\rangle)$ -- for $n=2$ we get the Bell state \eqref{bellExample}
\item W states: $\frac{1}{\sqrt{n}}(|\underbrace{00\dots 01}_{n\text{ qubits}}\rangle + |00\dots 10\rangle + \dots + |01\dots 00\rangle + |10\dots 00\rangle)$
\end{itemize}
Note that GHZ and W states can be prepared using the algorithm presented in \cite{ghz-w-states}.

So far, we have considered that the qubit is in superposition. However, according to the rules of quantum mechanics, it remains so only until a {\bf measurement} is carried out. Measurement is the act of getting information (or knowledge) about the qubit. We have said that the qubit is in both states at once but that when we measure it we get a particular value, either 0 or 1. This process is also called {\bf wave function collapse}, as \eqref{qbitDef} is also referred to as a {\bf wave function} and we see that the qubit has ``collapsed'' from the superposition to one particular state.\footnote{Again recall Schr\"{o}dinger's cat. Opening the box is like making a measurement and leads to the collapse of the wave function describing the cat, i.e.,~we find out whether the cat is dead or alive.} The result of measurement is a probabilistic event -- for the single qubit described by \eqref{qbitDef} the value 0 is measured with probability $|\alpha|^2$ and 1 with probability $|\beta|^2$. Similarly, for the two-qubit state \eqref{twoQubitDef} we get 00 with probability $|\alpha|^2$, 01 with probability $|\beta|^2$, 10 with probability $|\gamma|^2$, and 11 with probability $|\delta|^2$. This raises the question of how we can perform any useful computation when the results are probabilistic. First of all, we have to get enough samples to reconstruct the probability distribution governing the quantum state saved in the qubits. This means that the state is prepared and measured several times (such repetitions are called {\bf shots} in quantum computing). Naturally, this approach partially cancels out the increase in performance gained by using a quantum computer. However, it is worth noting that suppressing the error rate in quantum computers (we will discuss this issue in detail later) reduces the number of shots needed to reconstruct the probability distribution. Moreover, there are some quantum calculations which in theory need only one evaluation, hence the high performance of quantum computers is preserved. In Section~\ref{sectionAlgorithms}, we will show how the probability distribution of the qubits can be manipulated and translated to useful figures.

\subsubsection{Quantum Gates and Quantum Circuits}

In the previous section, we considered qubits to be static objects. However, to carry out useful calculations we need a tool for changing the states of qubits. This tool is called a {\bf quantum gate}, and a collection of such gates representing a quantum algorithm is known as a {\bf quantum circuit}.\footnote{There are other models of quantum computing, for example, the single-purpose quantum annealers used in binary optimization, which do not work with gates. However, in this paper, we describe a gate-based model of a universal quantum computer as adopted by IBM and other companies such as Honeywell and Microsoft.} In the rest of this paper, we assume -- in accordance with the criteria for the physical implementation of a quantum computer introduced in \cite{divincenzo-criteria} -- that any quantum computer is able to prepare a qubit (or qubits) in state $|0\rangle$ as the initial state for subsequent calculations.

In general, the time evolution of a quantum state $|\psi\rangle$ is described by {\bf Schr\"{o}dinger's equation}
\begin{equation}
i\hbar\frac{d |\psi(t)\rangle}{d t} = \mathcal{H}|\psi(t)\rangle,
\end{equation}
where $\hbar$ is the reduced Planck constant and $\mathcal{H}$ is a Hermitian matrix\footnote{A matrix $\mathbf{A}$ is Hermitian if $\mathbf{A}^\dagger = \mathbf{A}$, where $\mathbf{A}^\dagger$ is the conjugate transpose matrix of $\mathbf{A}$, i.e.,~a~transposed matrix with all the elements switched for their complex conjugated values.}
called the {\bf Hamiltonian}. The Hamiltonian, also called the energy operator, describes how a quantum system behaves. For example, the Hamiltonian's eigenvectors represent the allowed energy levels and its eigenvalues represent the values of the energy on these levels. We will discuss Hamiltonians in detail in Section~\ref{sectionAlgorithms}.

In the case of quantum computers, we can assume that the time evolution goes in discrete time steps. As a consequence, the change from state $|\psi(t_1)\rangle$ to state $|\psi(t_2)\rangle$ is described by the equation
\begin{equation}
|\psi(t_2)\rangle = \mathbf{U}|\psi(t_1)\rangle \label{time_evolution},
\end{equation}
where $\mathbf{U}$ is a unitary matrix.\footnote{A matrix $\mathbf{A}$ is unitary if $\mathbf{A}\mathbf{A}^\dagger = \mathbf{A}^\dagger\mathbf{A} = \mathbf{I}$, where $\mathbf{I}$ is a unit matrix. Clearly, a unitary matrix always has an inverse matrix. This means that any change on a quantum computer can be reversed, the exceptions being measurement and reset (i.e.,~setting a qubit to state $|0\rangle$).}

The matrix $\mathbf{U}$ in expression \eqref{time_evolution} is the mathematical representation of a quantum gate.  The simplest gate is an {\bf identity operator} leaving a qubit unchanged.\footnote{An identity operator is similar to an idle (or empty) instruction in classical programming languages.} The identity operator is represented by a unit matrix
\begin{equation}
\mathbf{I} = 
\begin{pmatrix}
1 & 0 \\
0 & 1 
\end{pmatrix}.
\end{equation}
Later, we will see why this operator is important. The first gate actually changing the state of a qubit is a quantum {\bf NOT gate}, which converts state $|0\rangle$ to state $|1\rangle$ and vice versa. The NOT gate is described by matrix
\begin{equation}
\mathbf{X} = 
\begin{pmatrix}
0 & 1 \\
1 & 0 
\end{pmatrix}.
\end{equation}
An effect of gate $\mathbf{X}$ applied to states $|0\rangle$ and $|1\rangle$ can be easily verified by the direct calculation
\begin{equation}
\mathbf{X}|0\rangle =
\begin{pmatrix}
0 & 1 \\
1 & 0 
\end{pmatrix}  \begin{pmatrix} 1 \\0 \end{pmatrix} = 
\begin{pmatrix} 0 \\1 \end{pmatrix} = |1\rangle
\\ \,\,\,\,\,\,\,\,\,
\mathbf{X}|1\rangle =
\begin{pmatrix}
0 & 1 \\
1 & 0 
\end{pmatrix}  \begin{pmatrix} 0 \\1 \end{pmatrix} = 
\begin{pmatrix} 1 \\0 \end{pmatrix} = |0\rangle.
\end{equation}
In the previous part, we mentioned that superposition plays an important role in the high computational power of quantum computers. To prepare a qubit in superposition, we can employ, for example, the {\bf Hadamard gate}, represented by the matrix
\begin{equation}
\mathbf{H} = \frac{1}{\sqrt{2}}
\begin{pmatrix}
1 & 1 \\
1 & -1
\end{pmatrix}.
\end{equation}
Applying $\mathbf{H}$ to $|0\rangle$ and $|1\rangle$, we obtain the following states:
\begin{itemize}
\item $\mathbf{H}|0\rangle = \frac{1}{\sqrt{2}}\begin{pmatrix}1 \\ 1 \end{pmatrix} =  \frac{1}{\sqrt{2}}(|0\rangle +|1\rangle)$,
denoted $|+\rangle$.
\item $\mathbf{H}|1\rangle = \frac{1}{\sqrt{2}}\begin{pmatrix}1 \\ -1 \end{pmatrix}  = \frac{1}{\sqrt{2}}(|0\rangle -|1\rangle)$,
denoted $|-\rangle$.
\end{itemize}
States $|+\rangle$ and $|-\rangle$ are clearly in a superposition of $|0\rangle$ and $|1\rangle$. The probability of obtaining 0 and 1 after measurement is 50\% in both cases, since $(1/\sqrt{2})^2 = 0.5$. The states $|+\rangle$ and $|-\rangle$ differ only in the relative phase, which is $\pi$ in the case of $|-\rangle$ (because $-1 = \mathrm{e}^{i\textcolor{red}{\pi}}$) and zero for $|+\rangle$ ($1 =  \mathrm{e}^{i\textcolor{red}{0}}$). It is possible to convert state $|+\rangle$ to state $|-\rangle$ and vice versa with the {\bf $\mathbf{Z}$ gate} given by the matrix
\begin{equation}
\mathbf{Z} = 
\begin{pmatrix}
1 & 0 \\
0 & -1
\end{pmatrix}.
\end{equation}
It can be verified by direct calculation that $\mathbf{Z}|+\rangle = |-\rangle$ and $\mathbf{Z}|-\rangle = |+\rangle$.

In the previous section, we introduced the most general form of the single qubit \eqref{generalQubit}. To prepare this state on IBM's quantum computer, a gate denoted $\mathbf{U3}$ is used
\begin{equation}
\mathbf{U3}(\theta, \varphi, \lambda) = 
\begin{pmatrix}
\cos(\theta/2) & -\mathrm{e}^{i\lambda}\sin(\theta/2) \\
\mathrm{e}^{i\varphi}\sin(\theta/2) & \mathrm{e}^{i(\varphi+\lambda)}\cos(\theta/2)
\end{pmatrix}.
\end{equation}
Clearly, operation $\mathbf{U3}(\theta,\varphi,0)|0\rangle$ prepares the qubit  in state \eqref{generalQubit}. There are also two specialized gates based on $\mathbf{U3}$, namely, $\mathbf{U1}(\lambda) = \mathbf{U3}(0,0,\lambda)$ and $\mathbf{U2}(\varphi, \lambda) = \mathbf{U3}(\pi/2,\varphi,\lambda)$.
Note that these gates are specific to IBM. However, in this paper we will only use the IBM Quantum\textsuperscript{TM} environment.

At this point, we are able to change the state of a single qubit. To work with more qubits, we apply the single-qubit gates introduced above to more qubits. The matrix representation of such gates is given by the tensor product of the gates on the individual qubits. We demonstrate this on the example of two qubits. The Hadamard gate is applied to each of them
\begin{equation}
\mathbf{H} \otimes \mathbf{H} =
\frac{1}{\sqrt{2}}
\begin{pmatrix}
\mathbf{\textcolor{red}{1}} & \mathbf{\textcolor{blue}{1}} \\
\mathbf{\textcolor{green}{1}} & \mathbf{\textcolor{orange}{-1}}
\end{pmatrix}
\otimes
\frac{1}{\sqrt{2}}
\begin{pmatrix}
1 & 1 \\
1 & -1
\end{pmatrix} =
\frac{1}{2}
\begin{pmatrix}
\mathbf{\textcolor{red}{1}}\begin{pmatrix}1 & 1 \\ 1 & -1\end{pmatrix} & \mathbf{\textcolor{blue}{1}}\begin{pmatrix}1 & 1 \\1 & -1\end{pmatrix} \\
\mathbf{\textcolor{green}{1}}\begin{pmatrix}1 & 1 \\1 & -1\end{pmatrix} & \mathbf{\textcolor{orange}{-1}}\begin{pmatrix}1 & 1 \\1 & -1\end{pmatrix}
\end{pmatrix}
=
\frac{1}{2}
\begin{pmatrix}
1 & 1 & 1 & 1 \\
1 & -1 & 1 & -1 \\
1 & 1 & -1 & -1 \\
1 & -1 & -1 & 1 \\
\end{pmatrix}.
\end{equation}
If we apply gate $\mathbf{H} \otimes \mathbf{H}$ to two-qubit state $|00\rangle$, we get a new state $\frac{1}{2}(|00\rangle + |01\rangle + |10\rangle + |11\rangle)$. The probability of measuring each combination of 0s and 1s is 25\%, since $(1/2)^2 = 0.25$. The same goes for the other two-qubit inputs to gate $\mathbf{H} \otimes \mathbf{H}$. The states differ only in their relative phases.

Following this pattern, we are able to prepare any {\bf separable multi-qubit states}, i.e.,~those that can be written as a tensor product of individual qubits. However, to exploit the high computational performance of quantum computers, we need a tool for preparing entangled states. This can be achieved with the aid of {\bf controlled gates.} An example of such a gate is the two-qubit {\bf controlled NOT gate}, or {\bf CNOT}. When the first qubit ({\bf control qubit}) of the gate is in state $|0\rangle$, the second qubit ({\bf target qubit}) is left unchanged. But when the control qubit is in state $|1\rangle$, the NOT gate is applied to the target qubit. Note that the control qubit is unchanged. Consequently, the map of the two-qubit basis input states to the outputs has the form: 
$|00\rangle \rightarrow |00\rangle$,
$|01\rangle \rightarrow |01\rangle$,
$|10\rangle \rightarrow |11\rangle$,
$|11\rangle \rightarrow |10\rangle$.\footnote{Note that the output value of the target (second) qubit is a classical XOR function applied to both the control and target qubits, while the output value of the control (first) qubit is a copy of its input.}
The CNOT gate can also be described by the matrix
\begin{equation}
\mathbf{CNOT} =
\begin{pmatrix}
1 & 0 & 0 & 0 \\
0 & 1 & 0 & 0 \\
0 & 0 & 0 & 1 \\
0 & 0 & 1 & 0 \\
\end{pmatrix} \label{cnotMatrix}.
\end{equation}
We can see that the matrix $\mathbf{X}$ for a single-qubit NOT operation is hidden in \eqref{cnotMatrix}, because $\mathbf{CNOT} = \begin{pmatrix}\mathbf{I}_{2,2} & \mathbf{O}_{2,2} \\ \mathbf{O}_{2,2}  & \mathbf{X}\end{pmatrix}$, where $\mathbf{O}_{2,2}$ is a two-by-two zero matrix. This is a general pattern for any two-qubit controlled gate. Similarly, we can write matrices for controlled $\mathbf{Z}$, controlled $\mathbf{U3}$, and other controlled gates.

A quantum gate can have more than one control qubit. An example of such a gate is the three-qubit {\bf Toffoli gate}. In fact, the Toffoli gate is a double-controlled $\mathbf{X}$ gate, hence it is often denoted $\mathbf{CCNOT}$. The third qubit is negated when both control qubits are in state $|1\rangle$.\footnote{When the third qubit is in state $|0\rangle$, after application of the Toffoli gate it contains a value equal to the classical AND function applied to both control qubits. If the third qubit is in state $|1\rangle$, after the Toffoli gate is applied it contains the result of the NAND function applied to the control qubits.}
The matrix describing the Toffoli gate has following form:
\begin{equation}
\mathbf{CCNOT} =
\begin{pmatrix}
1 & 0 & 0 & 0 & 0 & 0 & 0 & 0  \\
0 & 1 & 0 & 0 & 0 & 0 & 0 & 0  \\
0 & 0 & 1 & 0 & 0 & 0 & 0 & 0  \\
0 & 0 & 0 & 1 & 0 & 0 & 0 & 0  \\
0 & 0 & 0 & 0 & 1 & 0 & 0 & 0  \\
0 & 0 & 0 & 0 & 0 & 1 & 0 & 0  \\
0 & 0 & 0 & 0 & 0 & 0 & 0 & 1  \\
0 & 0 & 0 & 0 & 0 & 0 & 1 & 0  \\
\end{pmatrix}
=
\begin{pmatrix}
\mathbf{I}_{4,4} & \mathbf{O}_{4,4} \\
\mathbf{O}_{4,4} & \mathbf{CNOT}
\end{pmatrix}.
\end{equation}
In the following text, we will introduce other gates. An overview of all the gates used is provided in Appendix~\ref{appendixGates}.

There are a number of relations between quantum gates. Complex multi-qubit and multi-controlled gates can be decomposed into simpler ones. An overview of those relations and decomposition techniques is provided in \cite{elementary-gates}. It is also worth noting that any quantum gate can be composed of members of a  {\bf basic gates set}, i.e.,~a small group of gates which are universal and enable any calculation to be performed on a quantum computer.\footnote{Similarly, for classical computing, the logic NAND function constitutes a single-member basic functions set allowing a classical computer to perform any calculation. As the Toffoli gate can behave like a NAND function, we are able to implement any classical algorithm on a quantum computer, too. Hence, a gate-based quantum computer is universal from both the classical and quantum perspectives.} The algorithm for constructing a gate from the basic set is called the {\bf Solovay-Kitaev algorithm} and is described in detail in \cite{solovay-kitaev}.

The composition of the quantum gates is called the {\bf quantum circuit}, and the circuit represents a quantum algorithm. Here, we provide an example of a quantum circuit for the preparation of the Bell state \eqref{bellExample}.\footnote{This circuit is sometimes considered to be the quantum analog of a {\it Hello, World!} program.} The circuit is composed of a Hadamard gate located on the first qubit, followed by a CNOT gate. While there is a Hadamard gate on the first qubit, there is {\it nothing} on the second one. This {\it nothing} is in fact the identity operator $\mathbf{I}$ (here we see the importance of the identity operator). The matrix form of the circuit is therefore
\begin{equation}
\begin{aligned}
\mathbf{BELL} = \mathbf{CNOT}(\mathbf{H}\otimes\mathbf{I})
=
\frac{1}{\sqrt{2}}
\begin{pmatrix}
1 & 0 & 0 & 0 \\
0 & 1 & 0 & 0 \\
0 & 0 & 0 & 1 \\
0 & 0 & 1 & 0 \\
\end{pmatrix}
\begin{pmatrix}
1\begin{pmatrix}1 & 0 \\0 & 1\end{pmatrix} & 1\begin{pmatrix}1 & 0 \\0 & 1\end{pmatrix} \\
1\begin{pmatrix}1 & 0 \\0 & 1\end{pmatrix} & -1\begin{pmatrix}1 & 0 \\0 & 1\end{pmatrix}
\end{pmatrix} = \\
=
\frac{1}{\sqrt{2}}
\begin{pmatrix}
1 & 0 & 0 & 0 \\
0 & 1 & 0 & 0 \\
0 & 0 & 0 & 1 \\
0 & 0 & 1 & 0 \\
\end{pmatrix}
\begin{pmatrix}
1 & 0 & 1 & 0 \\
0 & 1 & 0 & 1 \\
1 & 0 & -1 & 0 \\
0 & 1 & 0 & -1 \\
\end{pmatrix}
=
\frac{1}{\sqrt{2}}
\begin{pmatrix}
1 & 0 & 1 & 0 \\
0 & 1 & 0 & 1 \\
0 & 1 & 0 & -1 \\
1 & 0 & -1 & 0 \\
\end{pmatrix}.
\end{aligned} \label{bellCircuit}
\end{equation}
Clearly, for input $|00\rangle$, the circuit returns the Bell state \eqref{bellExample}. For inputs $|01\rangle$, $|10\rangle$, and $|11\rangle$, we get the other Bell states $|\beta_{01}\rangle = \frac{1}{\sqrt{2}}(|01\rangle + |10\rangle)$, $|\beta_{10}\rangle = \frac{1}{\sqrt{2}}(|00\rangle - |11\rangle)$, and $|\beta_{11}\rangle = \frac{1}{\sqrt{2}}(|01\rangle - |10\rangle)$, respectively. 

\begin{figure}[H]
\caption{Circuits for Preparation of Bell States on the IBM Quantum\textsuperscript{TM} Computer}
\begin{center}
\includegraphics[scale = 1]{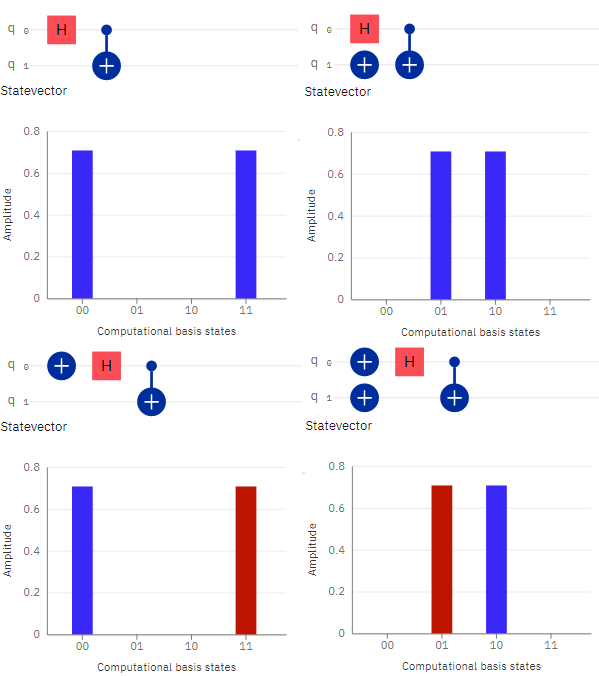}
\end{center}
\vspace{0.1cm}
{\footnotesize \textbf{\textit{Note:}} The red rectangle with an H symbolizes a Hadamard gate, the blue circle is an $\mathbf{X}$ gate, and the blue circle with a line above it is an $\mathbf{CNOT}$ gate. Only $\mathbf{H}$ and $\mathbf{CNOT}$ are employed directly in circuit \eqref{bellCircuit}. The $\mathbf{X}$ gate is used for setting the input state; for example, the input in the top right circuit is $|01\rangle$, hence the $\mathbf{X}$ gate is located on the second qubit. The other inputs are $|00\rangle$ in the top left circuit, $|10\rangle$ in the bottom left circuit, and  $|11\rangle$ in the bottom right circuit. The colors of the bars in the graphs represent the relative phase -- blue for $0$ and red-brown for $\pi$. Note that the order of the matrices in \eqref{bellCircuit}  is opposite to the order of the gates in the diagrams. Also note that the results were obtained on a quantum computer simulator (i.e.,~software simulating the behavior of a quantum computer on a classical one).}\\
{\footnotesize \textbf{\textit{Source:}} Author's own calculations on IBM Quantum\textsuperscript{TM} simulator}
\label{fig_bell_examples}
\end{figure}

Quantum circuits are often visualized with diagrams. Examples of such diagrams describing circuit \eqref{bellCircuit} and the output states for different inputs are provided in Figure~\ref{fig_bell_examples}.

\clearpage
\subsubsection{Swap Test}
A swap test is an algorithm allowing us to calculate the absolute value of the inner product of two quantum states $|\psi\rangle$ and $|\phi\rangle$, i.e.,~the value of $|\langle\psi|\phi\rangle|$. This algorithm is useful for post-processing of the results obtained when solving linear systems in general and when optimizing portfolios in particular.

The algorithm is based on a three-qubit {\bf controlled swap gate (or Fredkin gate)}. If the control qubit of the gate is in state $|1\rangle$, the states of the two target qubits are swapped. In other words, the gate's action is similar to switching two cables between two sockets.
The matrix describing the Fredkin gate is the following (note that the red part of the matrix represents the uncontrolled version of the swap gate):
\begin{equation}
\mathbf{CSWAP} =
	\begin{pmatrix}
	1 & 0 & 0 & 0 & 0 & 0 & 0 & 0  \\
	0 & 1 & 0 & 0 & 0 & 0 & 0 & 0  \\
	0 & 0 & 1 & 0 & 0 & 0 & 0 & 0  \\
	0 & 0 & 0 & 1 & 0 & 0 & 0 & 0  \\
	0 & 0 & 0 & 0 & \textcolor{red}{1} &\textcolor{red}{0} & \textcolor{red}{0} & \textcolor{red}{0}  \\
	0 & 0 & 0 & 0 & \textcolor{red}{0} &\textcolor{red}{0} & \textcolor{red}{1} & \textcolor{red}{0}  \\
	0 & 0 & 0 & 0 & \textcolor{red}{0} &\textcolor{red}{1} & \textcolor{red}{0} & \textcolor{red}{0}  \\
	0 & 0 & 0 & 0 & \textcolor{red}{0} &\textcolor{red}{0} & \textcolor{red}{0} & \textcolor{red}{1}  \\
	\end{pmatrix}.
\end{equation}

\begin{figure}[H]
\caption{General Schematic of the Swap Test}
\begin{center}
\includegraphics[scale = 0.4]{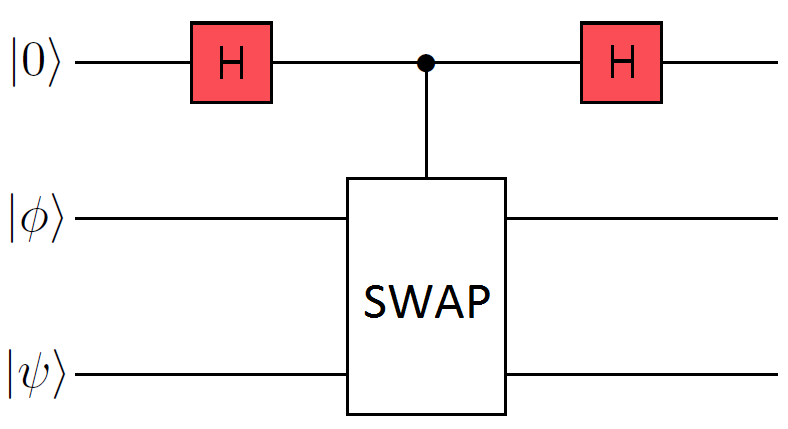}
\end{center}
\vspace{0.1cm}
{\footnotesize \textbf{\textit{Source:}} Adapted from \cite{swap-test}}
\label{fig_swap_theory}
\end{figure}

A schematic of the swap test is shown in Figure~\ref{fig_swap_theory}. For general  $n$-qubit states $|\psi\rangle$ and $|\phi\rangle$,  the ``SWAP'' block contains $n$ controlled swap gates with targets on the corresponding qubits of states $|\psi\rangle$ and $|\phi\rangle$. All these swap gates are controlled with the uppermost qubit, which is initially set to state $|0\rangle$.

The probability that the uppermost qubit is in state $|0\rangle$ is\footnote{A derivation of the equation is provided in \cite{swap-test}.}
\[
P(|0\rangle) = \frac{1}{2}(1+|\langle \psi|\phi \rangle|^2),
\]
hence for the inner product we have the formula $|\langle \psi|\phi \rangle| = \sqrt{2P(|0\rangle)-1}$.

The main disadvantage of the swap test is its inability to determine the sign of the inner product for states with real probability amplitudes.  Moreover, if the states have complex probability amplitudes, the information about the real and imaginary parts is lost. However, for the purposes of this paper, the method is sufficient, because the asset weights in portfolio optimization are always real, and if short positions are forbidden, the weights are non-negative.

\begin{figure}[hbt]
\caption{``Practical'' Swap Test}
\begin{center}
\includegraphics[scale = 0.65]{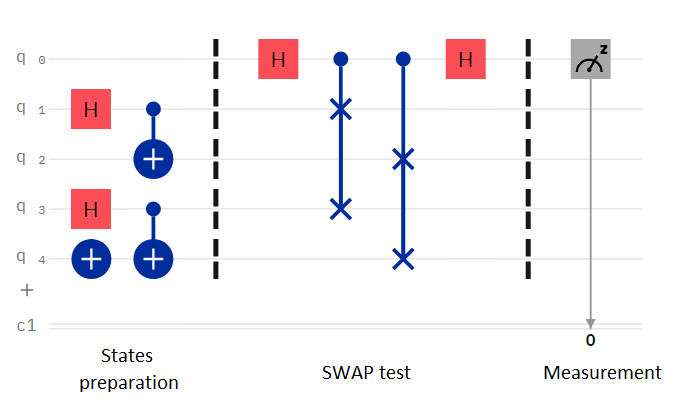}
\end{center}
\vspace{0.1cm}
{\footnotesize \textbf{\textit{Note:}} The inner product of states $|\beta_{00}\rangle$ and $|\beta_{01}\rangle$ is calculated. The blue symbols in the swap test part of the schematic represent Fredkin gates with the control on the uppermost qubit. }\\
{\footnotesize \textbf{\textit{Source:}} Author's own creation in IBM Quantum\textsuperscript{TM} environment}
\label{fig_swap_practical}
\end{figure}

In Figure~\ref{fig_swap_practical}, we provide a practical implementation of the swap test on IBM Quantum\textsuperscript{TM} for Bell states $|\beta_{00}\rangle$ and $|\beta_{01}\rangle$. In this case $\langle \psi|\phi \rangle = 0$, because the states are orthogonal, hence the probability of measuring $|0\rangle$ in the control qubit is 50\%.

\subsubsection{Quantum Fourier Transform}
In this part, we discuss the {\bf quantum Fourier transform} (QFT), a very important part of many quantum algorithms. For example, in a quantum linear equations solver, the QFT is used for extracting the eigenvalues of a matrix, which are then used in subsequent calculations. 

Note that from this point on, for the sake of simplicity, we sometimes omit the symbol $\otimes$ in tensor products. For example, the expression $|0\rangle|1\rangle$ has the same meaning as $|0\rangle \otimes |1\rangle$. For an $n$-qubit state having each qubit in either basis state $|0\rangle$ or basis state $|1\rangle$, we introduce symbol $|x\rangle$, where $x \in \{0,1,\dots 2^n-1\}$, denoting the decimal representation of the binary number stored in the $n$-qubit state. For example, for $n=3$, the expression $|5\rangle$ is equivalent to $|101\rangle=|1\rangle|0\rangle|1\rangle$.\footnote{For $n=4$, the symbol $|5\rangle$ means state $|0101\rangle = |0\rangle|1\rangle|0\rangle|1\rangle$. Similarly for any other $n$.}
We also introduce the notation $0.x_1x_2\dots x_n$, where $x_i \in \{0;1\}$, representing a {\bf binary fraction}, i.e.,~$0.x_1x_2\dots x_n$ is equivalent to $ x_1/2 + x_2/4 + \dots + x_n/2^n$ in the decimal digits system.
 
Equipped with these new symbols, we define the quantum Fourier transform for the $n$-qubit basis state $|x\rangle$ as an operation:
\begin{equation}
\text{QFT}(|x\rangle) = \frac{1}{\sqrt{2^n}}\sum_{k=0}^{2^n-1}\mathrm{e}^{2\pi ixk / 2^n}|k\rangle.
\end{equation}
As this exact definition is not easy to read and understand, an equivalent formula is used in practice:
\begin{equation}
\text{QFT}(|x\rangle) = \frac{1}{\sqrt{2^n}}
  \Big[
    \big( |0\rangle + \mathrm{e}^{2\pi i 0.x_n}|1\rangle \big)
    \big( |0\rangle + \mathrm{e}^{2\pi i 0.x_{n-1}x_n}|1\rangle \big)
    \dots
    \big( |0\rangle + \mathrm{e}^{2\pi i 0.x_1x_2\dots x_{n-1}x_n}|1\rangle \big)
  \Big] \label{QFT}.
\end{equation}
Input state $|x\rangle$ can be rewritten as $|x_1\rangle|x_2\rangle \dots |x_n\rangle$, where $|x_i\rangle \in \{|0\rangle;|1\rangle\}$, hence we see that the QFT encodes the input binary string and its sub-strings into quantum phases of the output qubits. 

As the QFT is a quantum operation, it has an inverse -- the {\bf inverse quantum Fourier transform}, abbreviated as $\text{QFT}^{-1}$ or 
$\text{QFT}^{\dagger}$. From \eqref{QFT} it follows that the inverse operation extracts the quantum phases of the input qubits and translates them into an output bit string. As mentioned in Section~\ref{subsubsectionQubits}, some quantum algorithms encode the result of the calculation into the quantum phase. To get this result, we employ $\text{QFT}^{-1}$ to translate the phase into a bit string.

\begin{figure}[H]
\caption{General Circuit Implementing the Quantum Fourier Transform}
\begin{center}
\includegraphics[scale = 0.17]{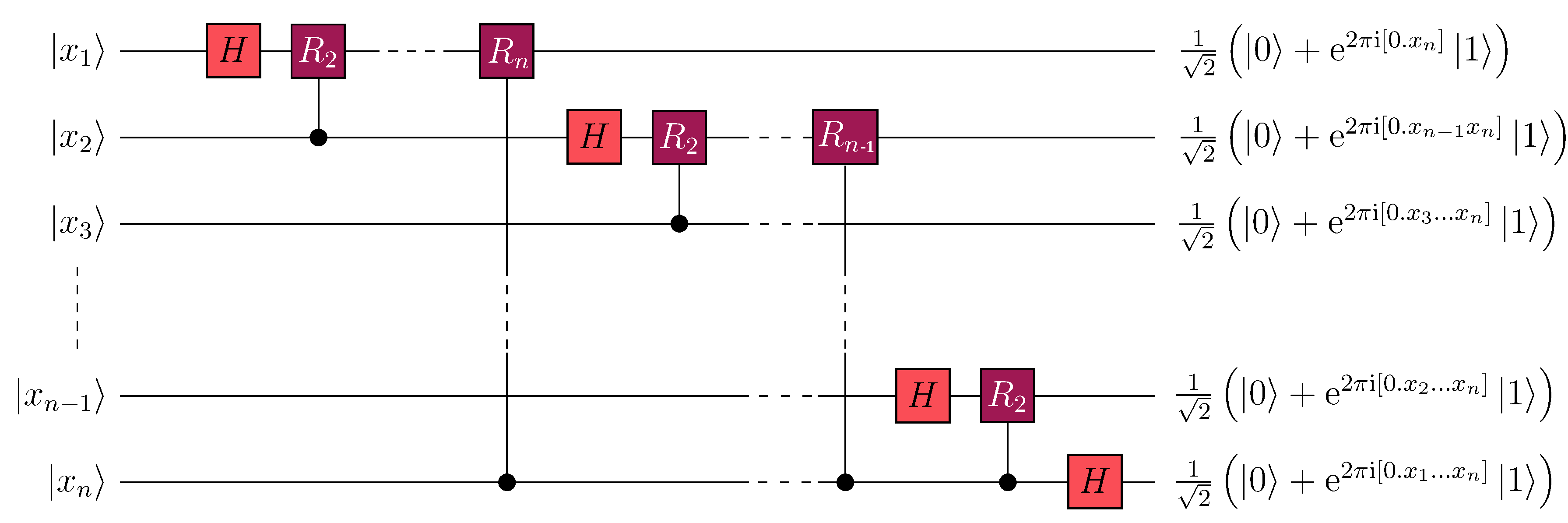}
\end{center}
\vspace{0.1cm}
{\footnotesize \textbf{\textit{Source:}} Adapted from \cite{nielsen-chuang-book}}
\label{fig_qft_gen}
\end{figure}

A general circuit implementing the QFT on a quantum computer is shown in Figure~\ref{fig_qft_gen}. The quantum gates $\mathbf{R_k}$ in Figure~\ref{fig_qft_gen} are defined as
\begin{equation}
\mathbf{R_k} =
\begin{pmatrix}
1 & 0 \\
0 & \mathrm{e^{2 \pi i/2^k}}
\end{pmatrix}.
\end{equation}
The circuit for inverse quantum Fourier transformation is the circuit in Figure~\ref{fig_qft_gen} taken from right to left, with all the rotation angles replaced by their opposite values, i.e.,~$2 \pi /2^k$ in gate $\mathbf{R_k}$ is substituted with  $-2 \pi /2^k$. The Hadamard gate is its own inverse, hence the $\mathbf{H}$ gates are left unchanged.

Note that $\mathbf{R_k}$ is equivalent to gate $\mathbf{U1}(\lambda)$, where $\lambda = \frac{2\pi}{2^k}$. Examples of three-qubit QFT and inverse QFT implementation on IBM Quantum\textsuperscript{TM} are shown in Figure~\ref{fig_qft_3qubits}. We also provide the Qiskit source code for QFT and $\text{QFT}^\dagger$ in Appendix~\ref{appendixSrc}.

\begin{figure}[H]
\caption {Three-Qubit Quantum Fourier Transform and Its Inverse}
\begin{center}
	\begin{tabular}{cc}
		\includegraphics[scale = 0.3]{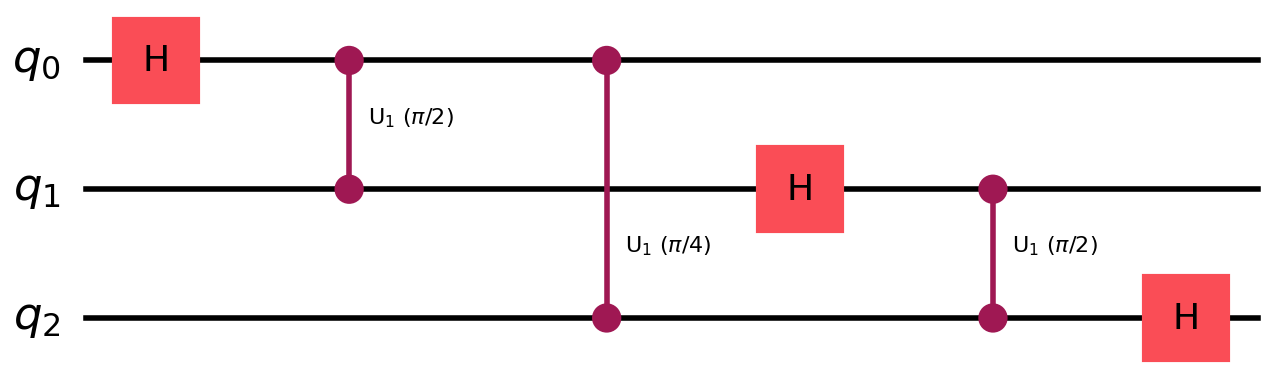}  \\
		\includegraphics[scale = 0.3]{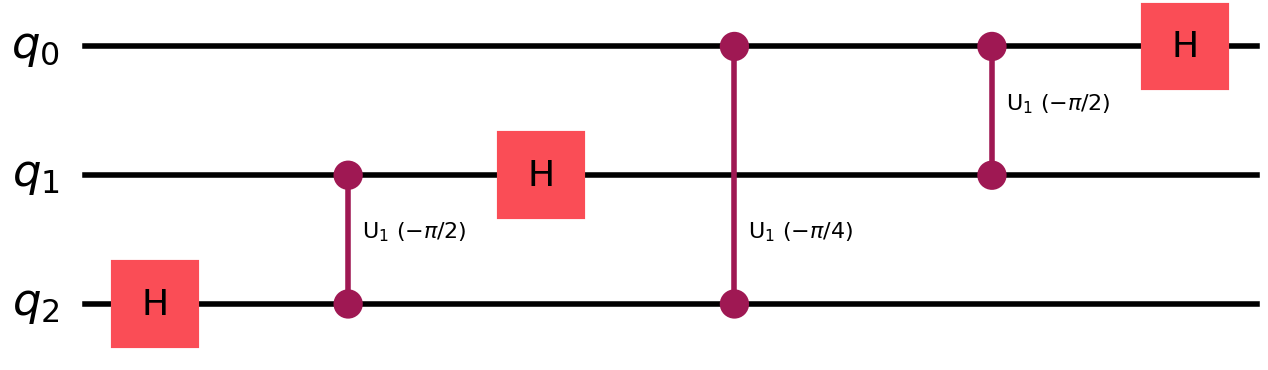}
	\end{tabular}
\end{center}
\vspace{0.1cm}
{\footnotesize \textbf{\textit{Note:}} The top diagram depicts the quantum Fourier transform and the bottom one its inverse. Dark red symbols are used for controlled $\mathbf{U1}$ gates. Note that the control and target qubits of controlled $\mathbf{U1}$ gates can be interchanged. }\\
{\footnotesize \textbf{\textit{Source:}} Author's own implementation in Qiskit}
\label{fig_qft_3qubits}
\end{figure}

\subsection{Quantum Algorithms in FX Reserves Management}
\label{sectionAlgorithms}

In this part, we will describe quantum algorithms that are useful for FX reserves management. First, we will discuss an algorithm for calculating risk metrics such as VaR and CVaR. Second, we will introduce algorithms for solving systems of linear equations and for finding optimum solutions to quadratic unconstrained binary optimization problems. We will then employ both algorithms to construct a portfolio.

\subsubsection{Calculating Risk Measures on a Quantum Computer}
\label{subsectionRisk}
Calculating risk indicators is an integral part of portfolio management. For this reason, we introduce the quantum algorithm for calculating VaR and CVaR proposed in \cite{quantum-risk} and \cite{quantum-credit-risk}. An important benefit of this algorithm is that it provides quadratic speed-up in comparison with classical calculation techniques based on Monte Carlo simulation. 
In plain language, this means that we need, for example, only 100 operations instead of the 10,000 necessary on a classical computer. In more mathematical terms, while the classical Monte Carlo method converges as $O(M^{-1/2})$, the quantum version does so as $O(M^{-1})$, where $M$ is the number of samples used.

In the rest of this section, we assume that we are dealing with discrete probability distributions with a finite number of outcomes. Without loss of generality, we also assume that the number of probability distribution outcomes is $2^n$, where $n \in \mathbb{N}$.\footnote{If the number of outcomes is not equal to a power of two, the distribution can be extended by including additional outcomes with zero probability so that the total number of outcomes is equal to a power of two.} Where a continuous distribution is involved, discretization is carried out first.  

Assume that $X$ is a random variable with a discrete probability distribution having a probability function $p$ satisfying $\sum_{i=0}^{2^n-1}p_i = 1$. Such distribution can be encoded into an $n$-qubit quantum state $|\psi\rangle = \sum_{i=0}^{2^n-1} \sqrt{p_i}|i\rangle$.
Let us denote by $f(x)$ a function $f: \{0;1\dots 2^n-1\} \rightarrow \langle 0;1\rangle$ and assume that we are able to change an $n+1$-qubit quantum state $|i\rangle |0\rangle$ into $|i\rangle(\sqrt{1-f(i)}|0\rangle + \sqrt{f(i)}|1\rangle)$.  If we apply such operation on state $|\psi\rangle|0\rangle$, we arrive at a new state
\begin{equation}
|\tilde{\psi}\rangle=
\sum_{i=0}^{2^n-1}\sqrt{p_i}|i\rangle(\sqrt{1-f(i)}|0\rangle + \sqrt{f(i)}|1\rangle) = 
\big[\sum_{i=0}^{2^n-1}\sqrt{p_i[1-f(i)]}|i\rangle\Big]|0\rangle
+\big[\sum_{i=0}^{2^n-1}\sqrt{p_i f(i)}|i\rangle\Big]|1\rangle.
\end{equation}
Clearly, the probability of measuring $|1\rangle$ in the last qubit of $|\tilde{\psi}\rangle$ is the sum of the probabilities of all outcomes having shape $|i\rangle|1\rangle$, i.e.,
\begin{equation}
P_1 = \sum_{i=0}^{2^n-1}p_i f(i), \label{eq_one_prob}
\end{equation} 
which is the expected value $E[f(X)]$ of the random variable $f(X)$.
Let us denote $N = 2^n$ and $f(x) = x/(N-1)$. In this setting, \eqref{eq_one_prob} changes to $P_1=\frac{1}{N-1}\sum_{i=0}^{N-1}i p_i$, which is $E[X/(N-1)]$, hence $E[X] = P_1(N-1)$. This means that we are able to translate the calculation of the expected value of a random variable to the measurement of a quantum state.

If we set $f(x) = x^2/(N-1)^2$, the general expression \eqref{eq_one_prob} changes to $P_1=\frac{1}{(N-1)^2}\sum_{i=0}^{N-1}i^2 p_i$, which is $E[X^2/(N-1)^2]$, hence $E[X^2] = P_1(N-1)^2$. Having $E[X^2]$ and $E[X]$, we can (classically) calculate the variance ($\text{VAR}(X)=E[X^2]-(E[X])^2$) and the standard deviation, i.e.,~$\sigma(X)=\sqrt{\text{VAR}(X)}$, of the random variable $X$. A similar pattern in setting $f(x)$ can be followed to get higher moments (for example, the skewness and kurtosis) of the random variable $X$.

A special setting of $f(x)$ also enables us to evaluate a percentile of the random variable $X$. Since Value-at-Risk (VaR) is defined as 
$\text{VaR}_\alpha(X) = \min\{x|P[X \le x] \ge 1-\alpha\}$, it is in fact the $1-\alpha$ percentile. If, for a certain $l\in \mathbb{N}$, we set $f(i) = 1$ for $i \le l$ and $f(i) = 0$ otherwise, equation~\eqref{eq_one_prob} transforms to $P_1 = \sum_{i \le l} p_i$. Setting the initial value $l = N - 1$, we start an interval bisection and iterate until the probability of measuring state $|1\rangle$ in the last qubit of $|\tilde{\psi}\rangle$ is higher than $1 - \alpha$. Once the iterative process stops, we save the value $l$, which is the desired $1-\alpha$ percentile or $\text{VaR}_\alpha(X)$. Note that when evaluating VaR, we combine quantum and classical (i.e.,~interval bisection) calculations. This approach is called a {\bf hybrid algorithm}.

Conditional VaR (CVaR) is defined as the expected value of outcomes below VaR, i.e.,
\begin{equation}
\text{CVaR}_\alpha(X) = \frac{1}{P[X\le \text{VaR}_\alpha(X)]}\sum_{i \le \text{VaR}_\alpha(X)} i p_i. 
\end{equation}
This means that we can employ an approach similar to the calculation of $E[X]$. We set $f(i) = i/\text{VaR}_\alpha(X)$ for $i \le \text{VaR}_\alpha(X)$ and $f(i) = 0$ otherwise. In this setting, \eqref{eq_one_prob} becomes
\begin{equation}
P_1=\frac{1}{\text{VaR}_\alpha(X)}\sum_{i \le \text{VaR}_\alpha(X)} i p_i. \label{eq_cvar_calc}
\end{equation}
Since $\sum_{i \le \text{VaR}_\alpha(X)}  p_i < 1$, i.e.,~the probability distribution of values below VaR is not normalized, the probabilities $p_i$ have to be replaced by $p_i/P[X\le \text{VaR}_\alpha(X)]$. Normalization, together with equation~\eqref{eq_cvar_calc}, leads to the final formula for calculating CVaR on a quantum computer:
\begin{equation}
\text{CVaR}_\alpha(X) = P_1  \frac{\text{VaR}_\alpha(X)}{P[X\le \text{VaR}_\alpha(X)]}.
\end{equation}
Note that $\text{VaR}_\alpha(X)$ is known from the previous calculations and $P[X\le \text{VaR}_\alpha(X)] = \sum_{i \le \text{VaR}_\alpha(X)} p_i$.

\begin{figure}[H]
\caption  {Circuit Implementing Three-Qubit Controlled $y$-rotations}
\begin{center}
\includegraphics[scale = 0.4]{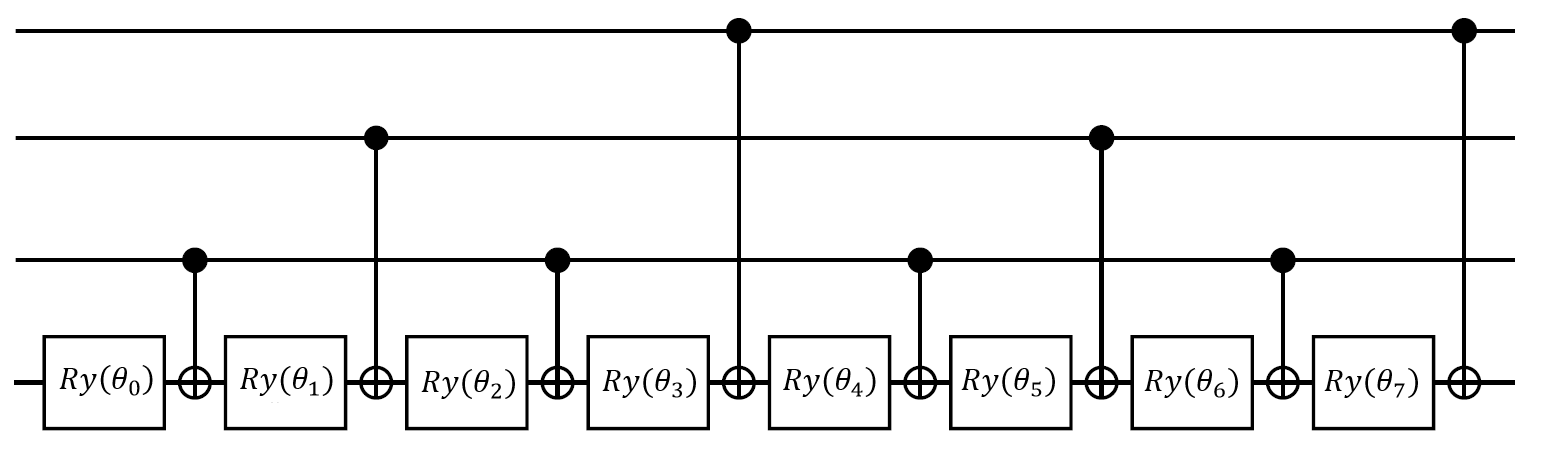}
\end{center}
\vspace{0.1cm}
{\footnotesize \textbf{\textit{Source:}} Adapted from \cite{state-preparation}}
\label{fig_risk_rotations}
\end{figure}

To carry out the calculations described above, we have to prepare state $|\tilde{\psi}\rangle$. We start by preparing the part $|i\rangle(\sqrt{1-f(i)}|0\rangle + \sqrt{f(i)}|1\rangle)$. To do so, we design a gate with $n$ controlling qubits changing the state of the $n+1$\textsuperscript{th} qubit according to an $n$-qubit state $|i\rangle$ and function $f(x)$. Clearly, the uncontrolled version of this gate (i.e.,~with $n=0$) is a single-qubit gate preparing state $\sqrt{1-p}|0\rangle + \sqrt{p}|1\rangle$ for $p \in \langle 0;1\rangle$. Such gate is called {\bf $\mathbf{y}$-rotation}\footnote{The gate rotates a single-qubit state on the Bloch sphere around the $y$-axis by angle $\alpha$.} and is described by matrix
\begin{equation}
\mathbf{Ry}(\alpha) = 
\begin{pmatrix}
\cos(\alpha /2) & -\sin(\alpha /2) \\
\sin(\alpha /2) & \cos(\alpha/2)
\end{pmatrix}.
\end{equation}
Evidently, $\mathbf{Ry}(\alpha)|0\rangle = \cos(\alpha/2)|0\rangle + \sin(\alpha/2)|1\rangle$. Setting $\alpha = 2\arccos(\sqrt{1-p})$, or equivalently $\alpha = 2\arcsin(\sqrt{p})$, we get state $\sqrt{1-p}|0\rangle + \sqrt{p}|1\rangle$. If we calculate the rotation angles $\alpha$ for all $f(i)$, we get a sequence $\alpha_i = 2\arcsin[\sqrt{f(i)}]$ for $N$ $\mathbf{Ry}(\alpha_i)$ gates. However, we have to add controls to these gates to set the proper angle $\alpha_i$ for state $|i\rangle$ on the control qubits. To do so, we employ the method described in \cite{state-preparation}. An example of a circuit implementing the preparation of $|i\rangle(\sqrt{1-f(i)}|0\rangle + \sqrt{f(i)}|1\rangle)$ for $n = 3$, i.e.,~eight possible inputs $|i\rangle$, is shown in Figure~\ref{fig_risk_rotations}.

We easily see how the $\mathbf{Ry}$ and CNOT gates are arranged. Following this pattern, we can expand the circuit for any number of control qubits.
It is important to emphasize that the rotational angles of the $\mathbf{Ry}$ gates are not $\alpha_i$ based on $f(i)$ as calculated before, but other angles $\theta_i$ obtained by means of a transform. The reason for changing the angles from $\alpha_i$ to $\theta_i$ is that there are no native $n$-qubits controlled the $\mathbf{Ry}$ gate, but such qubits can be implemented with single-qubit $\mathbf{Ry}$ gates and two-qubit CNOTs, as shown in Figure~\ref{fig_risk_rotations}. Therefore, the rotational angles have to be adapted to this substituting configuration. The transformation from angles $\alpha_i$ to angles $\theta_i$ is given by the expression
\begin{equation}
\begin{pmatrix}\theta_0 \\ \theta_1 \\ \vdots \\ \theta_{N-1}\end{pmatrix} = 
\mathbf{M}
\begin{pmatrix}\alpha_0 \\ \alpha_1 \\ \vdots \\ \alpha_{N-1}\end{pmatrix},
\end{equation}
where $\mathbf{M} \in \mathbb{R}^{N,N}$ with elements 
\begin{equation}
M_{ij} = 2^{-n}(-1)^{b(j) \bullet g(i)} \quad i,j \in \{0;1;2 \dots N-1 \},
\end{equation} 
where $b(j)$ is the binary representation of column index $j$, $g(i)$ is the Gray code representation of row index $i$ (a method for converting a binary number into Gray code is described in Appendix~\ref{appendixGray}), and operation $\bullet$ is a bit-wise product modulo 2 defined as
\begin{equation}
b(j) \bullet g(i) = [b_0(j)g_0(i) + b_1(j)g_1(i) + \dots + b_{N-1}(j)g_{N-1}(i) ] \mod 2,
\end{equation}
where $b_k(j)$ is the $k$\textsuperscript{th} bit of binary number $b(j)$ and similarly for $g(i)$.

To give an example of how the method described above works, we prepare a circuit for measuring the expected value of a distribution with eight outcomes (i.e.,~three qubits are involved). In this case, $f(i) = i/7$ and transformation matrix $\mathbf{M}_8$ is
\begin{equation}
\mathbf{M}_8 =\frac{1}{8}
\begin{pmatrix}
1 & 1 & 1 & 1 & 1 & 1 & 1 & 1 \\ 
1 & -1 & 1 & -1 & 1 & -1 & 1 & -1 \\ 
1 & -1 & -1 & 1 & 1 & -1 & -1 & 1 \\ 
1 & 1 & -1 & -1 & 1 & 1 & -1 & -1 \\ 
1 & 1 & -1 & -1 & -1 & -1 & 1 & 1 \\ 
1 & -1 & -1 & 1 & -1 & 1 & 1 & -1 \\ 
1 & -1 & 1 & -1 & -1 & 1 & -1 & 1 \\ 
1 & 1 & 1 & 1 & -1 & -1 & -1 & -1 
\end{pmatrix}.
\end{equation}
Function values $f(i)$, rotational angles $\alpha$, and transformed angles $\theta$ are shown in Table~\ref{tableRiskExample}. To prepare the whole state $|\tilde{\psi}\rangle$, we need state $|\psi\rangle$ describing the probability distribution of random variable $X$. This state can again be obtained using the method introduced in \cite{state-preparation} based on $n$-qubit controlled rotations discussed above. The interested reader can consult that paper to get an insight into how the method works. We also provide the Qiskit code implementing the method in Appendix~\ref{appendixSrc}. Alternatively, it is possible to use the implementation of this method in the {\it initialize} function obtained from the Qiskit libraries.\footnote{Note that a more sophisticated approach to preparing the initial state $|\psi\rangle$ based on a probability distribution inferred from a huge dataset with quantum machine learning was proposed in \cite{generative-modelling}.}

We now summarize the practical approach to calculating risk metrics on a quantum computer. The first step is to prepare an $n$-qubit state $|\psi\rangle$ based on the probability distribution of random variable $X$. Then, a new qubit in state $|0\rangle$ is added to state $|\psi\rangle$ to get an $n+1$-qubit state $|\tilde{\psi}\rangle =|\psi\rangle|0\rangle$. This new state is inputted into a circuit similar to that shown in Figure~\ref{fig_risk_rotations} and adapted according to the number of outcomes of random variable $X$. Finally, the $n+1$\textsuperscript{th} qubit of state $|\tilde{\psi}\rangle$ is measured and, based on the probability of measuring state $|1\rangle$, the actual value of the risk metric is calculated classically.

\begin{table}[H]
\caption{Rotational Angles in the Circuit for the Calculation of the Expected Value}
\begin{center}
{\footnotesize
\begin{tabular}{c|c|c|c|c}
\hline \hline 
\thead{$\mathbf{i}$}  & \thead{$\mathbf{f(i)}$} & \thead{\bf Respective input \\ \bf quantum state}  
& \thead{\bf Rotational \\ \bf angle $\mathbf{\alpha_i}$}  & \thead{\bf Transformed rotational \\ \bf angle $\mathbf{\theta_i}$}  \\
\hline
0 & 0.0000 & $|000\rangle$ & 0.0000 & 1.5708\\
1 & 0.1429 & $|001\rangle$ & 0.7752 & -0.2687\\
2 & 0.2857 & $|010\rangle$ & 1.1279 & 0.0000\\
3 & 0.4286 & $|011\rangle$ & 1.4274 & -0.4450\\
4 & 0.5714 & $|100\rangle$ & 1.7141 & 0.0000\\
5 & 0.7143 & $|101\rangle$ & 2.0137 & -0.1189\\
6 & 0.8571 & $|110\rangle$ & 2.3664 & 0.0000\\
7 & 1.0000 & $|111\rangle$ & 3.1416 & -0.7382\\ \hline
\end{tabular}}
\end{center}
\vspace{0.1cm}
{\footnotesize \textbf{\textit{Note:}} The table contains the rotational angles of the $\mathbf{Ry}$ gates in a circuit for calculating the expected value of a probability distribution with eight outcomes.}\\
{\footnotesize \textbf{\textit{Source:}} Author's own calculations}
\label{tableRiskExample}
\end{table}

\subsubsection{Quantum Solver of Systems of Linear Equations}
Systems of linear equations are often used in modeling and optimization. Harrow et al. designed a quantum algorithm for solving linear systems, commonly known as the {\bf HHL algorithm} \cite{HHL-alg}.\footnote{The abbreviation comes from the names of the creators of the algorithm, Harrow, Hassidim, and Lloyd.}. Later an application of the HHL algorithm in Markowitz-like portfolio optimization, which we will employ in the practical part of this paper, was proposed in \cite{ptf-optim}. The main advantage of the HHL algorithm is that it provides exponential speed-up in comparison with classical solvers. However, in some cases there is no speed-up at all, as emphasized in \cite{hhl-issues} and \cite{hhl-issues2} (we will return to this discussion at the end of this part). 

Note that we will only present an outline of the algorithm, as very technical details can hinder understanding of how it works. Moreover, in the practical part, we will use a debugged and optimized implementation of the algorithm taken from the Qiskit libraries. The interested reader can find technical details in the above-mentioned papers. Examples of how to build a quantum circuit implementing the HHL algorithm for a particular linear system are provided in \cite{hhl-example}, \cite{hhl-tutorial} (suitable for complete beginners), and \cite{algos-for-beginners}.

\begin{figure}[H]
\caption{First Step of the Phase Estimation Algorithm}
\begin{center}
\includegraphics[scale = 0.8]{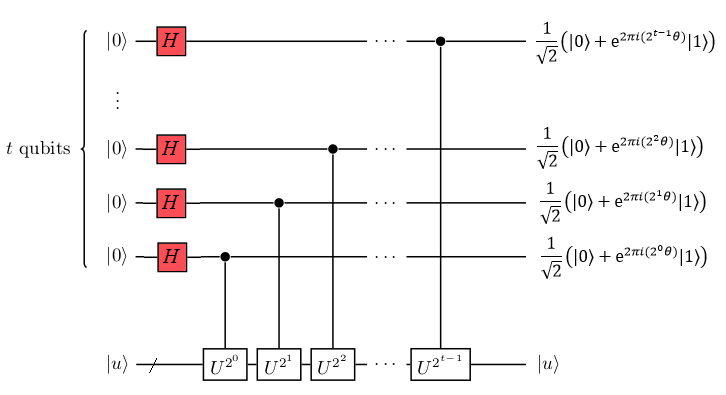}
\end{center}
\vspace{0.1cm}
{\footnotesize \textbf{\textit{Note:}}  This part of the HHL algorithm is intended for finding the phases of the unitary matrix eigenvalues. Given a unitary matrix $\mathbf{U}$ and an eigenvector $|u\rangle$, the algorithm encodes phase $\theta$ of eigenvalue $\mathrm{e}^{i\theta}$ into quantum phases of the output qubits, while eigenvector $|u\rangle$ remains unchanged. The number of qubits $t$ depends on the required precision of the binary representation of phase $\theta$. Since $\mathbf{U}$ is $\mathrm{e}^{i\mathbf{A}}$ in this case, phase $\theta$ is in fact eigenvalue $\lambda$ of matrix $\mathbf{A}$. Note that symbols $\mathbf{U}^{k}$ represent the $k$\textsuperscript{th} power of matrix $\mathbf{U}$. }\\
{\footnotesize \textbf{\textit{Source:}} Adapted from \cite{nielsen-chuang-book}}
\label{fig_phase_estimation}
\end{figure}

Suppose we want to solve a system of linear equations $\mathbf{A}|x\rangle = |b\rangle$, where $\mathbf{A} \in \mathbb{C}^{N,N}$ is a full-rank matrix, $|b\rangle \in \mathbb{C}^N$ is a non-zero vector, and $|x\rangle\in \mathbb{C}^N$. Note that we use Dirac notation to make the formulas succinct. Without loss of generality, we assume that matrix $\mathbf{A}$ is Hermitian.\footnote{If $\mathbf{A}$ is non-Hermitian, we can switch from system $\mathbf{A}|x\rangle = |b\rangle$ to 
$\begin{pmatrix}\mathbf{O} & \mathbf{A} \\ \mathbf{A^\dagger} & \mathbf{O}\end{pmatrix}
\begin{pmatrix} |0\rangle \\ |x\rangle\end{pmatrix} = \begin{pmatrix} |b\rangle \\ |0\rangle\end{pmatrix}$. Evidently, the matrix of the new system is Hermitian and part $|x\rangle$ of the new solution is the same as the original one.}
Such matrices have real eigenvalues, and their eigenvectors form the orthonormal basis of the vector space $\mathbb{C}^N$.\footnote{To be precise, the eigenvectors associated with distinct eigenvalues are orthogonal. However, it is always possible to get orthonormal vectors by normalizing such vectors by their Euclidian norm. Moreover, it holds that even in the case of degenerated eigenvalues, it is possible to find the orthogonal basis of  $\mathbb{C}^N$ composed of the matrix eigenvectors.}
Moreover, Hermitian matrices always have a {\bf spectral decomposition}, defined as\footnote{Any {\bf normal matrix} $\mathbf{A}$, i.e.,~one satisfying $\mathbf{A}\mathbf{A}^\dagger = \mathbf{A}^\dagger \mathbf{A}$, has a spectral decomposition.}
\begin{equation}
\mathbf{A} = \sum_{i=1}^N \lambda_i |u_i\rangle \langle u_i|,
\end{equation}
where $\lambda_i$ is the $i$\textsuperscript{th} eigenvalue of matrix $\mathbf{A}$ and $|u_i\rangle$ is the respective eigenvector. With the spectral decomposition, we can introduce a {\bf matrix function} $f$
\begin{equation}
f(\mathbf{A}) = \sum_{i=1}^N f(\lambda_i) |u_i\rangle \langle u_i|. \label{eq_spectral_decomposition}
\end{equation}
Setting $f(x) = x^{-1}$, we can write the inverse matrix of $\mathbf{A}$ (we assume that the matrix is full-rank, hence it is invertible) as
\begin{equation}
\mathbf{A}^{-1} = \sum_{i=1}^N \frac{1}{\lambda_i}|u_i\rangle \langle u_i|.\label{eq_inv_matrix}
\end{equation}
Since the eigenvectors of matrix $\mathbf{A}$ form the basis of space $\mathbb{C}^N$, the right-side $|b\rangle$ can be written as a linear combination of those eigenvectors $|b\rangle = \sum_{i=1}^N \beta_i |u_i\rangle$, where $\beta_i = \langle u_i | b\rangle$ is the $i$\textsuperscript{th} coordinate of $|b\rangle$ in the basis.\footnote{Note that the coordinates of $|b\rangle$ in the basis composed of the eigenvectors of $\mathbf{A}$ can be found by solving a linear system $\mathbf{U}|\beta\rangle =|b\rangle$, where $\mathbf{U}$ is a matrix, the columns of which are created from the eigenvectors of $\mathbf{A}$. Since these eigenvectors are orthonormal, $\mathbf{U}$ is unitary, hence $|\beta\rangle = \mathbf{U}^\dagger |b\rangle$. The $i$\textsuperscript{th} element of $|\beta\rangle$ is the inner product of $|b\rangle$ and the $i$\textsuperscript{th} row of $\mathbf{U}^\dagger$, which is a complex conjugated vector to the $i$\textsuperscript{th} column of $\mathbf{U}$ (i.e.,~$\langle u_i|$). This gives us the expression $\beta_i = \langle u_i | b\rangle$.}
With the help of \eqref{eq_inv_matrix}, the solution of the linear system is
\begin{equation}
|x\rangle =  A^{-1}|b\rangle = 
\Big[\sum_{i=1}^N \frac{1}{\lambda_i}|u_i\rangle \langle u_i|\Big]
\Big[ \sum_{j=1}^N \beta_j |u_j\rangle \Big] = 
\sum_{i=1}^N\sum_{j=1}^N \frac{1}{\lambda_i}\beta_j|u_i\rangle \langle u_i| u_j\rangle.
\end{equation}
Because the eigenvectors are orthonormal, $\langle u_i| u_j\rangle = 0$ for $i\ne j$ and  $\langle u_i| u_j\rangle = 1$ for $i=j$, hence the solution of the linear system is
\begin{equation}
|x\rangle = \sum_{i=1}^N \frac{1}{\lambda_i}\beta_i|u_i\rangle. \label{eq_hhl_solution}
\end{equation}
To solve $\mathbf{A}|x\rangle = |b\rangle$ on a quantum computer, we need to prepare solution \eqref{eq_hhl_solution} as a quantum state. First, we have to find eigenvalues $\lambda_i$. To do so, we employ a {\bf phase estimation} algorithm, which finds the phases of the unitary matrix eigenvalues. Since $\mathbf{A}$ is Hermitian, matrix $\mathbf{U}=\mathrm{e}^{i\mathbf{A}}$ is unitary.\footnote{It holds that $\mathbf{U}^\dagger = (\mathrm{e}^{i\mathbf{A}})^\dagger = \mathrm{e}^{-i\mathbf{A}^\dagger}$. Since $\mathbf{A}$ is Hermitian, $\mathbf{U}^\dagger = \mathrm{e}^{-i\mathbf{A}}$, therefore $\mathbf{U}\mathbf{U}^\dagger=\mathrm{e}^{i\mathbf{A}}\mathrm{e}^{-i\mathbf{A}}=\mathrm{e}^{\mathbf{O}}=\mathbf{I}$. Similarly, $\mathbf{U}^\dagger\mathbf{U} = \mathbf{I}$, hence $\mathbf{U}$ is a unitary matrix.} According to \eqref{eq_spectral_decomposition}, the eigenvalues of $\mathbf{U}$ have the form $\mathrm{e}^{i\lambda_j}$. This means that the eigenvalues of $\mathbf{A}$ are encoded into phases of the eigenvalues of matrix $\mathbf{U}$. The first step of the phase estimation algorithm is shown in Figure~\ref{fig_phase_estimation}. 

As can be seen in Figure~\ref{fig_phase_estimation}, after application of the first step of the phase estimation, eigenvalue $\lambda$ corresponding to eigenvector $|u\rangle$ of matrix $\mathbf{A}$ is encoded into phases of the output qubits. To obtain the phase (i.e.,~eigenvalue $\lambda$), we apply inverse QFT. If we put the right-side $|b\rangle$ instead of eigenvector $|u\rangle$ on the phase estimation input, we get a superposition of all the eigenvalues of $\mathbf{A}$ after inverse QFT, because $|b\rangle$ is a linear combination (superposition) of all the eigenvectors of $\mathbf{A}$.\footnote{To be precise, rather than $|b\rangle$ itself, we use its normalized version $|b\rangle/||b||$, where $||b||$ is the Euclidean norm of $|b\rangle$. To transform system $\mathbf{A}|x\rangle = |b\rangle$ to $\mathbf{\tilde{A}}|x\rangle = |b\rangle/||b||$, we have to divide all the equations in the system by $||b||$.}
The phase estimation does not change the input eigenvectors, so $|b\rangle$ is not changed either. This all means that we have quantum state $\sum_{i=1}^N \beta_i |\lambda_i\rangle |u_i\rangle$ after the phase estimation.\footnote{As can be seen in Figure~\ref{fig_phase_estimation}, the phase estimation consumes a good deal more qubits with increasing accuracy of the binary representation of the eigenvalues (i.e.,~the number of qubits $t$). The number of gates also increases enormously with increasing accuracy, due to the necessity to implement higher powers of operator $\mathbf{U}$ and a more complex inverse QFT. To tackle this bottleneck of the phase estimation process, it is possible to employ {\bf iterative phase estimation} requiring only two qubits, but at the cost of an increasing number of repetitions of the calculation. Details are provided in \cite{iterative-phase}.}
To get the inverse of the eigenvalues, we add an auxiliary qubit ({\bf ancilla qubit}) and apply controlled $\mathbf{Ry}$ gates with the target on the ancilla qubit and controlled with qubits containing the eigenvalues. The rotational angles of the $\mathbf{Ry}$ gates are set to $\theta_i = 2\arcsin(C/\lambda_i)$, where constant $C$ ensures that the resulting quantum state is normalized.\footnote{To get an idea of how to set the rotation angles of $\mathbf{Ry}$ gates in practice, consult \cite{hhl-example} and \cite{hhl-tutorial}. It is obvious that in practice we need to estimate the eigenvalues, because we do not know them in advance as they are calculated in the phase estimation process.}
After application of the $\mathbf{Ry}$ gates, we have a new quantum state
\begin{equation}
\sum_{i=1}^{N}\beta_i|\lambda_i\rangle |u_i\rangle\Big[\sqrt{1- \frac{C^2}{\lambda_i^2}}|0\rangle + \frac{C}{\lambda_i}|1\rangle\Big].
\end{equation}

\begin{figure}[H]
\caption{Schematic of the HHL Algorithm}
\begin{center}
\includegraphics[scale = 0.45]{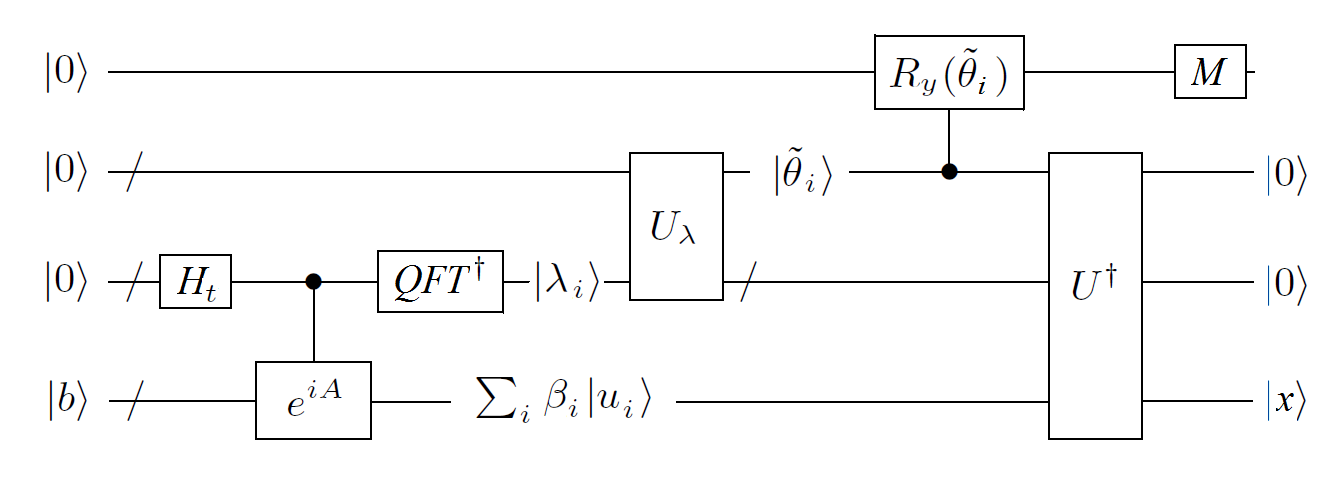}
\end{center}
\vspace{0.1cm}
{\footnotesize \textbf{\textit{Note:}}  In the first step, Hadamard gates are applied to $t$ qubits ($t$ depends on the desired precision of the representation of the eigenvalues $\lambda_i$). This is symbolized by gate $\mathbf{H}_t$. After that, the controlled version of the powers of $\mathbf{U} = \mathrm{e}^{i\mathbf{A}}$ is applied (symbolized by $\mathrm{e}^{i\mathbf{A}}$), followed by inverse QFT (denoted by $\text{QFT}^\dagger$). These steps constitute the phase estimation extracting the eigenvalues of matrix $\mathbf{A}$. Based on the eigenvalues $\lambda_i$, rotational angles $\theta_i$ are calculated (operation $\mathbf{U}_\lambda$) and controlled $\mathbf{Ry}(\theta)$ gates are applied to the ancilla qubit (the uppermost qubit). The next step (operation $\mathbf{U}^\dagger$) involves inverting the operations $\mathbf{U}_\lambda$, $\text{QFT}^\dagger$, $\mathrm{e}^{i\mathbf{A}}$, and $H_k$. Finally, the ancilla qubit is measured, and if the measurement result is $|1\rangle$ the right-side $|b\rangle$ has been changed to solution $|x\rangle$.}\\
{\footnotesize \textbf{\textit{Source:}} Adapted from \cite{hhl-example}}
\label{fig_hhl}
\end{figure}

The next step of the algorithm is to apply inverse operations to all the operations carried out so far, the $\mathbf{Ry}$ gates and the preparation of state $|b\rangle$ being exceptions.  As a result of these actions, all the qubits containing eigenvalues before are now in state $|0\rangle$ and we get quantum state
\begin{equation}
\sum_{i=1}^{N}\beta_i|0\dots 0\rangle |u_i\rangle\Big[\sqrt{1- \frac{C^2}{\lambda_i^2}}|0\rangle + \frac{C}{\lambda_i}|1\rangle\Big].
\end{equation}
The final step of the algorithm is to measure the ancilla qubit. If the result of the measurement is $|1\rangle$, neglecting qubits in state 
$|0\dots 0\rangle$, we come to state
\begin{equation}
C\sum_{i=1}^{N}\frac{1}{\lambda_i}\beta_i|u_i\rangle,
\end{equation}
which is equivalent to \eqref{eq_hhl_solution} up to the normalization constant $C$. This means that we do not get $|x\rangle$ itself, but its normalized version. If the result of the ancilla qubit measurement is $|0\rangle$, the algorithm has to be run again until $|1\rangle$ is measured, because only in this case is the solution of the linear system returned.  For better illustration, a schematic of the whole HHL algorithm is shown in Figure~\ref{fig_hhl}.

At the beginning of this section, we noted that in some cases the HHL algorithm generates no speed-up over classical linear system solvers. In particular, it cannot outperform classical solvers if we want to know all the members of solution $|x\rangle$. Obtaining the solution involves measuring all $n  = \log_2 N$ qubits representing the solution. Since $n$ qubits can be in $2^n$ basis states, complete measurement is exponentially complex and entirely cancels out the speed-up provided by the HHL algorithm.\footnote{Note that the method used to reconstruct the whole quantum state is called {\bf quantum state tomography}. The interested reader can find details in chapter~8.4.2 of \cite{nielsen-chuang-book}.}
However, quantum state $|x\rangle$ can be used as an input to another calculation. For example, if we are interested in the inner product of $|x\rangle$ and another quantum state, we employ a swap test converting the inner product calculation to the measurement of only one qubit. As a result, the exponential speed-up of the HHL algorithm is preserved. 

However, this is not the only pitfall of the HHL algorithm. The complexity of the algorithm is $O[\log(N) s \kappa / \epsilon]$, where $s$ is the {\bf sparsity} of matrix $\mathbf{A}$ (telling us how many elements in a matrix row are non-zero on average), $\kappa = |\lambda_{\max}|/|\lambda_{\min}|$ is the {\bf condition number}, and $\epsilon$ is the desired accuracy of the solution. This means that the HHL algorithm only works well for sparse matrices, i.e.,~those where the majority of their elements equal zero. At the same, the matrices should be well conditioned, i.e.,~the difference between the highest and lowest eigenvalues should be low. Ideally, the condition number should be close to 1.\footnote {If a matrix is unitary, all its eigenvalues are located on a unit circle, hence $|\lambda| = 1$ for any eigenvalue and therefore $\kappa = 1$.}
These requirements are not always satisfied, and this hinders the performance of the HHL algorithm. On top of that, if we require a high-accuracy solution, we again lose part of the performance of the HHL algorithm.

However, we now turn our attention to the application of the HHL algorithm for portfolio optimization (we will discuss how to deal with the described drawbacks later). Assume that we want to minimize the risk in a portfolio, achieve a return $G$, and spend the whole budget $B$. Let $\mathbf{C} \in \mathbb{R}^{n,n}$  be a covariance matrix of asset returns, $|R\rangle \in \mathbb{R}^n$ a vector of expected asset returns, $|P\rangle \in \mathbb{R}^n$ a vector of asset prices, and $|w\rangle \in \mathbb{R}^n$ a vector of asset weights. The task can be formulated as
\[
\min_{|w\rangle} \langle w| \mathbf{C} |w\rangle,
\]
subject to $\langle R| w \rangle = G$ and $\langle P | w \rangle = B$. Such optimization leads to minimization of the Lagrange function
\begin{equation}
L =  \langle w| \mathbf{C} |w\rangle + \lambda (\langle R| w \rangle  - G) + \mu (\langle P | w \rangle - B),
\end{equation}
where $\lambda$ and $\mu$ are Lagrange multipliers. Since function $L$ is quadratic, its derivatives are linear functions, and the optimization can be translated to the solution of a linear system\footnote{Note that $\frac{\partial L}{\partial \lambda} = \langle R| w \rangle - G$ and $\frac{\partial L}{\partial \mu} = \langle P | w \rangle - B$. Since covariance matrix $\mathbf{C}$ is symmetric (i.e.,~$c_{ij} = c_{ji}$), the derivative with respect to $w_i$ is $\frac{\partial L}{\partial w_i} = 2\sum_{j=1}^{n}c_{ij}w_j  + \lambda R_i + \mu P_i  = 2\mathbf{C}_{i \bullet} |w\rangle + \lambda R_i + \mu P_i$, where $\mathbf{C}_{i\bullet}$ is the $i$\textsuperscript{th} row of matrix $\mathbf{C}$. For all weights, the derivative can be written as $\frac{\partial L}{\partial 
|w\rangle} = 2\mathbf{C} |w\rangle  + \mu |P\rangle + \lambda |R\rangle$. Setting all derivatives equal to zero leads to the linear system \eqref{eq_lin_ptf_optim}.}
\begin{equation}
\begin{pmatrix}
0 & 0 & \langle R| \\
0 & 0 & \langle P| \\
|R\rangle & |P\rangle & \mathbf{C} 
\end{pmatrix}
\begin{pmatrix}
\lambda \\ \mu \\ |w\rangle
\end{pmatrix}
=
\begin{pmatrix}
G \\ B \\ |0\rangle
\end{pmatrix}. \label{eq_lin_ptf_optim}
\end{equation}

System \eqref{eq_lin_ptf_optim} can be solved with the HHL algorithm. However, as mentioned before, extracting the whole solution cancels out the speed-up. To exploit the high performance of the HHL algorithm, we have to use the portfolio optimization results (denoted $|x_{\text{opt}}\rangle$) in other ways, for example as suggested in \cite{ptf-optim}:
\begin{itemize}
\item {\bf To get the optimal exposure to a particular asset or sector:} 
Assume that we are interested in the weight of the $i$\textsuperscript{th} element of $|x_{\text{opt}}\rangle$. To obtain it, we calculate the inner product of $|x_{\text{opt}}\rangle$ and a vector having its $i$\textsuperscript{th} member equal to 1 and the rest as zeros. The latter vector is in fact a quantum state $|i-1\rangle$ which can be easily prepared with $X$ gates only.\footnote{Assume, for example, that our optimization involves 16 variables (i.e.,~four qubits are employed to represent the optimization results) and we are interested in the seventh one. This means that we need to calculate the inner product $\langle x_{\text{opt}}|6\rangle$ (the index runs from zero, hence the seventh position is represented by the number 6). In binary notation, 6 is represented by the string 0110 (the leading zero comes from the fact that we are dealing with four qubits), hence $|6\rangle = |0110\rangle$. Such state is prepared by applying $X$ gates to the second and third qubits of the initial state $|0000\rangle$.} Note that the first two members of state $|x_\text{opt}\rangle$ are in fact Lagrange multipliers $\lambda$ and $\mu$, therefore the first weight of an asset is the third member of $|x_\text{opt}\rangle$.
\item {\bf Compare the optimal solution with the current portfolio composition:}
To do so, we need to calculate the inner product $S = \langle x_{\text{opt}}|x_{\text{current}}\rangle$, where $|x_{\text{current}}\rangle = \sum_{i=1}^N \sqrt{x_i^{\text{current}}}|i-1\rangle$ is a quantum state representing the current composition of the portfolio. If $S = 1$, then $|x_{\text{opt}}\rangle = |x_{\text{current}}\rangle$, which means that our portfolio is optimal. In contrast, if $S = 0$, the current and optimal portfolios are orthogonal, i.e.,~totally dissimilar. Note that the Lagrange multipliers $\lambda$ and $\mu$ should also be members of state $|x_{\text{current}}\rangle$. Since we do not know the multipliers in advance, we first have to obtain them using the method described in the previous bullet point (i.e.,~we need to extract the first two members of $x_{\text{opt}}$). This may slightly attenuate the performance of the algorithm. However, the number of asset weights is much higher than two, hence we still get quantum speed-up.
\end{itemize}
Unfortunately, using the solution described above does not recover the whole speed-up. A quick check of system \eqref{eq_lin_ptf_optim} reveals that its main component is a covariance matrix which is hardly sparse. As a result, $s = N$ and the complexity of the HHL algorithm for portfolio optimization increases to $O[N\log(N) \kappa / \epsilon]$. This is still better than the best known classical algorithm introduced in \cite{hhl-best-classical}, which has complexity $O(N^{2.376})$, and the more often used algorithm designed in \cite{hhl-strassen}, which has complexity $O(N^{2.8074})$.\footnote{The most popular algorithm for solving linear systems, the Gaussian elimination algorithm, has complexity $O(N^{3})$.}

\subsubsection{Quantum Binary Optimization}
This part is dedicated to the use of quantum computers for solving a quadratic unconstrained binary optimization (QUBO) task. QUBO has several business applications, such as the traveling salesperson problem (TSP) and the knapsack problem.\footnote{Other interesting QUBO problems are discussed in \cite{ising-formulations}.} Some QUBO problems can be adapted for application in finance. For example, the TSP is used for finding currency arbitrage opportunities. Importantly for FX reserves management, QUBO is also useful in portfolio optimization.

In this part, we introduce the QUBO task quantum solver called {the \bf quantum approximate optimization algorithm (QAOA)}, designed in \cite{qaoa}. In the practical part, we employ the Qiskit implementation of the QAOA, so we only provide an outline of the algorithm. After introducing the QAOA, we will discuss binary portfolio optimization based on the work \cite{qubo-ptf-optim}, which we will use in the practical part of this paper. Note that besides the QAOA, there are other QUBO solvers, for example, the variational quantum eigensolver discussed in \cite{vqe-improved} and the solver based on Grover database search introduced in \cite{qubo-grover}. On top of that, D-Wave has designed a single-purpose quantum computer ({\bf a quantum annealer}) which implements a quantum QUBO solver on the hardware level.\footnote{More information about this device can be found in the user manual \cite{qubo-dwave}.} It is worth emphasizing that in contrast to the algorithms discussed previously, there is no proof that quantum QUBO solvers are superior to classical solvers. We will return to this issue at the end of this section.

In general, QUBO solvers aim to minimize the function
\begin{equation}
f(x_1, \dots, x_n) = \sum_{i=1}^{n}\sum_{j=1}^{n}a_{ij}x_i x_j + \sum_{i=1}^n b_i x_i + c,
\label{eq_qubo}
\end{equation}
where $x_i \in \{0;1\}$ is a binary variable and $a_{ij}$, $b_i$, and $c$ are real parameters. With the help of the Dirac notation, function \eqref{eq_qubo} is expressed in matrix form $f(x_1, \dots, x_n) = \langle x|\mathbf{A}|x\rangle + \langle b| x\rangle + c$. Note that QUBO is unconstrained optimization. Where constraints are involved, they are incorporated in the form of a penalization function (we will see this later).

On a quantum computer, the QUBO task can be translated to finding the ground state (i.e.,~the minimum energy level) of a special class of Hamiltonians ({\bf Ising Hamiltonians}) having the form\footnote{Ising Hamiltonians are connected with a special class of magnetic materials called {\bf spin glasses}. Such materials lie between paramagnetic and ferromagnetic materials. While ferromagnetics (e.g.,~iron) preserve their magnetization even after the external magnetic field is switched off, paramagnetics (e.g.~aluminum) lose their magnetization rapidly (exponentially). Spin glasses are similar to paramagnetic materials but lose their magnetization more slowly (almost linearly).}
\begin{equation}
\mathcal{H}_{\text{Ising}} = \sum_{i=1}^{n}\sum_{j=1}^{n}Q_{ij}\mathbf{Z_i} \otimes \mathbf{Z_j} + \sum_{i=1}^n c_i \mathbf{Z_i}. 
\label{eq_ising}
\end{equation}
Term $\mathbf{Z_i}$ denotes a quantum gate composed of a $\mathbf{Z}$ gate applied to the $i$\textsuperscript{th} qubit and identity operators applied to the other qubits, and term $\mathbf{Z_i} \otimes \mathbf{Z_j}$ means a quantum gate composed of two $\mathbf{Z}$ gates applied to the $i$\textsuperscript{th} and $j$\textsuperscript{th} qubits while identity operators are applied to the other qubits. $Q_{ij}$ and $c_i$ are real numbers. There is a clear resemblance between the QUBO task formulation \eqref{eq_qubo} and the Ising Hamiltonian \eqref{eq_ising} and they can be converted into each other.

The energy level associated with a quantum state $|x\rangle$ in a quantum system described by a Hamiltonian $\mathcal{H}$ is given by the expression $\langle x|\mathcal{H}|x\rangle$. The ground state $|x\rangle_\text{ground}$ minimizes this expression. Since our QUBO task has been converted into an Ising Hamiltonian, we are looking for its ground state, which in the end can be converted back into the optimal solution of the QUBO task. To find the ground state of the Ising Hamiltonian, we employ the QAOA. The algorithm is based on the {\bf adiabatic theorem}, which states that if a quantum system is described by an initial Hamiltonian $\mathcal{H}_0$, the system is in its ground state, and when the system's Hamiltonian is slowly changed, the system remains in its ground state but this new ground state is associated with a new Hamiltonian.  This means that having a Hamiltonian with a known ground state, we can find the ground state of $\mathcal{H}_\text{Ising}$. Such initial Hamiltonian is, for example,
\begin{equation}
\mathcal{H}_0 = \sum_{i=1}^n \mathbf{X_i}, \label{eq_qubo_ground}
\end{equation}
where $\mathbf{X_i}$ denotes a quantum gate composed of an $\mathbf{X}$ gate applied to the $i$\textsuperscript{th} qubit and identity operators applied to other qubits.
The ground state of Hamiltonian \eqref{eq_qubo_ground} is a uniformly distributed superposition $\frac{1}{\sqrt{2^n}}\sum_{i=0}^{2^n-1}|i\rangle$, which can be prepared by applying Hadamard gates to all $n$ qubits formerly in state $|0\rangle$ (such action is often denoted as $\mathbf{H}^{\otimes n}|0\rangle^{\otimes n}$). A simulation of a slow change from $\mathcal{H}_0$ to $\mathcal{H}_\text{Ising}$ can be carried out using a quantum gate defined as
\begin{equation}
U(\beta, \gamma) = \prod_{j=1}^{p} \mathrm{e}^{-i\beta_j \mathcal{H}_0} \mathrm{e}^{-i\gamma_j \mathcal{H}_\text{Ising}}
\label{eq_qaoa_gate}
\end{equation}
applied to state $\mathbf{H}^{\otimes n}|0\rangle^{\otimes n}$.\footnote{Note that gate \eqref{eq_qaoa_gate} is decomposed into basic gates using the method presented in \cite{vqe-improved}.} Parameter $p$ is called the algorithm depth and is set by the user. Changing $p$ can lead to better optimization results. Parameters $\beta_j$ and $\gamma_j$ have to be varied so that $\langle x| \mathcal{H}_\text{Ising}|x\rangle$ is minimized, which is done using a classical optimizer. We clearly see that the QAOA is a hybrid algorithm. The quantum part of the QAOA involves simulating the Hamiltonian encoded in gate \eqref{eq_qaoa_gate}. The classical part deals with the calculation of the energy levels of $\mathcal{H}_\text{Ising}$, which can be carried out efficiently thanks to the special structure of Ising Hamiltonians.\footnote{The method is outlined in \cite{vqe-improved}.}  After some iterations, the QAOA should find the optimal solution of the QUBO task. However, optimality of the solution is not guaranteed, because the algorithm can get stuck in a local minimum.

\begin{figure}[H]
\caption{Example of a QAOA Iteration Circuit for a Two-Variable QUBO Task}
\begin{center}
\includegraphics[scale = 1]{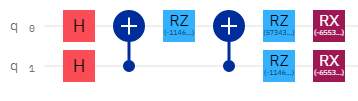}
\end{center}
\vspace{0.1cm}
{\footnotesize \textbf{\textit{Note:}} The Hadamard gates prepare a uniformly distributed quantum state. Then the Ising Hamiltonian is applied: the $\mathbf{Rz}$ gate on qubit $q_0$ together with the two CNOT gates represent a quadratic term $x_0x_1$, and the other $\mathbf{Rz}$ gates are connected with linear terms $x_0$ and $x_1$. The $\mathbf{Rx}$ gates are related to the initial Hamiltonian $\mathcal{H}_0$. }\\
{\footnotesize \textbf{\textit{Source:}} Author's own creation in the IBM Q\textsuperscript{TM} environment}
\label{fig_qaoa_example}
\end{figure}

An example of a QAOA circuit for a task with two binary variables and algorithm depth $p=1$ is shown in Figure~\ref{fig_qaoa_example}. The gates denoted as $\mathbf{Rx}(\alpha)$ and $\mathbf{Rz}(\alpha)$ are rotations around the $x$ and $z$ axes, respectively, of the Bloch sphere and are described by the matrices
\begin{equation}
	\mathbf{R_x}( \theta) = 
	\begin{pmatrix}
	\cos(\theta/2) & -i\sin(\theta/2) \\
	-i\sin(\theta/2) & \cos(\theta/2)
	\end{pmatrix}
	\,\,\,\,\,\,
	\mathbf{R_z}(\theta) = 
	\begin{pmatrix}
          \mathrm{e}^{-i\theta/2} & 0 \\
	0 & \mathrm{e}^{i\theta/2}
	\end{pmatrix}.
\end{equation}
The $\mathbf{Rz}$ gates represent exponentiation of the Ising Hamiltonian composed of $\mathbf{Z}$ gates, and the $\mathbf{Rx}$ gates are connected with the initial Hamiltonian $\mathcal{H}_0$ composed of $\mathbf{X}$ gates.\footnote{It holds that $Ra(\alpha) = \mathrm{e}^{-i\frac{\alpha}{2}\mathbf{A}}$, where $\mathbf{A}$ is either $\mathbf{X}$ or $\mathbf{Z}$.} Parameters $a_{ij}$, $b_i$, and $c$ in QUBO task \eqref{eq_qubo} and $\beta_j$ and $\gamma_j$ in \eqref{eq_qaoa_gate} are included in the rotational angles of the $\mathbf{Rx}$ and $\mathbf{Rz}$ gates.

As mentioned at the beginning of this part, there is no rigorous proof of the superiority of quantum QUBO solvers (including the QAOA). According to \cite{qubo-preskill}, there is little hope that a proof will ever be found, but for some tasks, quantum QUBO solvers may be faster than classical ones. For example, the studies  \cite{qubo-crisis-prediction} concerning the prediction of economic downturns and \cite{qubo-routing} relating to maritime routing problems, provide empirical evidence that quantum QUBO solvers perform better. Additionally, \cite{qubo-nonadiabatic} introduced the {\bf non-adiabatic quantum QUBO solver}, which should be faster for tasks with no specific internal structure. While the QAOA operates with Hamiltonian $\mathcal{H} = (1-t)\mathcal{H}_0 + t\mathcal{H}_{\text{Ising}}$, where $t$ is slowly changed from 0 to 1 (the ``adiabatic regime''), the non-adiabatic approach uses a Hamiltonian of the form $\mathcal{H} = \mathcal{H}_{\text{Ising}} - \frac{g}{t^\alpha}\mathcal{H}_{\text{init}}$, where $g$ and $\alpha$ are user-defined parameters, $\mathcal{H}_{\text{Init}}$ is the initial Hamiltonian\footnote{$\mathcal{H}_{\text{Init}}$ is generally different from $\mathcal{H}_0$ used in the QAOA and its form depends on the task being solved.} and $t$ is slowly increased from zero theoretically to infinity. This work is so far at a theoretical stage and no practical implementation on a real quantum computer has been presented yet.
On the other hand, \cite{annealing-pitfalls} revealed that in many instances QUBO problems are specially tailored (in this case to the topology of D-Wave quantum processors) to be hard for classical solvers and easier for quantum ones. On top of that, the authors of the study criticized the fact that quantum QUBO solvers are often only compared with, in principle, related {\bf classical simulated annealing} and/or derived algorithms,\footnote{Simulated annealing was introduced in \cite{annealing-metropolis} and was originally employed in the unfortunate race to develop nuclear weapons (one of its creators was Edward Teller, father of the H-bomb).} 
while several other classical heuristics are neglected. They also showed that in an assignment problem, a classical exact {\bf minimum-weight perfect matching algorithm}\footnote{See the detailed description of the algorithm in \cite{optim-physics}.} outperforms the D-Wave quantum processor for instances with more than 20 variables.  Overall, we should use quantum QUBO solvers as a complement to, not a substitute for, classical algorithms. This means we should try to employ both classical and quantum heuristics and, in the end, pick the best solution found regardless of which approach produced it.

The QAOA will be practically demonstrated on Markowitz-like portfolio optimization. Let us denote as $x_i$ a binary variable indicating whether the $i$\textsuperscript{th} asset is a member of the portfolio ($x_i = 1$) or not ($x_i=0$). In contrast to the optimization described in the previous part, we do not look for particular weights of assets, but only for information on whether to invest in some asset or not. Assume that the return on the $i$\textsuperscript{th} asset is $r_i$, the covariance of the $i$\textsuperscript{th} and $j$\textsuperscript{th} asset returns is $c_{ij}$, $M_i$ is the maximum amount of money we are willing to invest in the $i$\textsuperscript{th} asset, and $B$ is the total budget. Our aim is simultaneously to maximize the portfolio return $\sum_{i=1}^n r_i x_i$, to minimize risk $\sum_{i=1}^n \sum_{j=1}^n c_{ij} x_i x_j$, and to invest the whole budget, i.e.,~$\sum_{i=1}^n M_i x_i = B$. These three aims have importance expressed by coefficients $\lambda_1$, $\lambda_2$, and $\lambda_3$, respectively. The equivalent QUBO task is to minimize the function
\begin{equation}
f(x_1,\dots,x_n) = -\lambda_1 \sum_{i=1}^n r_i x_i + \lambda_2\sum_{i=1}^n \sum_{j=1}^n c_{ij} x_i x_j + \lambda_3\Big(B-\sum_{i=1}^n M_i x_i\Big)^2.\label{eq_qubo_ptf_optim}
\end{equation}
The minus sign before the first term (return) indicates that this part of the function has to be maximized. The third term is a penalization function expressing the budget constraint. The square in the third term ensures that the constraint is fulfilled in both directions.

As discussed in \cite{qubo-integer}, even binary optimization allows one to work with integer asset weights instead of only binary flags indicating whether to include an asset in a portfolio or not. The integer weights represent, for example, the amount of money invested in a particular asset or percentage weights. In order for integer variables to be incorporated into the QUBO task, they have to be expressed with binary variables as follows:
\begin{equation}
w_i = \sum_{j=0}^{m-1} x_{ij}2^j, \label{eq_bin_rep}
\end{equation}
where $x_{ij}$ is a binary variable representing the $j$\textsuperscript{th} bit in the binary representation of the $i$\textsuperscript{th} weight. 
It is important to emphasize that the number of bits used in representation \eqref{eq_bin_rep} constitutes a natural ceiling on weight $w_i$, because the maximum integer value represented with $m$ bits is $2^m-1$. Hence, an upper limit can be introduced on a weight without the use of a penalization function. However, note that integer optimization consumes a good deal more qubits than the binary case.

\section{Practical Demonstration}
In this part, we focus on applying the algorithms introduced in the previous section to FX reserves management problems. First, we will measure the VaR and CVaR of a chosen portfolio and compare the results with those obtained using classical methods. Second, we will apply the HHL algorithm to find the optimal ratio between equities and fixed-income instruments. Third, we will use the QAOA to find the best performing assets in an index while minimizing risk. In both cases of portfolio optimization, we will again compare the results obtained on a quantum computer with classical methods. In our demonstration, we will employ both real quantum computers and their simulators. We will also discuss the state of development of the underlying quantum technologies and their current limitations for practical deployment.

\subsection{Notes on the Current State of Development of Quantum Hardware}
\label{subsectionQuatumHardware}
Before we proceed to actual applications of quantum computers to FX reserves management tasks, we have to discuss the limitations and hurdles that quantum computers suffer from in their current state of development. We will discuss the consequences of noise, the limited connectivity of qubits, and the unavailability of quantum memories. It is important to emphasize that in contrast to the inherent limitations of some of the algorithms discussed earlier (such as the HHL algorithm), the said hardware issues are merely questions of further research and development and will be overcome in the end. Therefore, we will also present a metric used to benchmark quantum processors, allowing us to compare processors with each other and track their development over time.

{\bf Noise and quantum state decoherence:}
A quantum system remains in its state until it is disturbed by an external factor, which can be intentional, that is, measuring or resetting a qubit to state $|0\rangle$, or unwanted, such as thermal noise or cosmic radiation. Since natural noises are all around us, a quantum computer has to be shielded. In the case of IBM Quantum\textsuperscript{TM}, the quantum processors are cooled to almost 15~mK to avoid thermal noise.\footnote{Note that the temperature of interstellar space is 2.7~K (i.e.,~2 700~mK). To reach milli-Kelvin temperatures, a special device called a {\bf dilution refrigerator} is used. The refrigerator is based on spontaneous separation of the helium isotopes $^3\text{He}$ and $^4\text{He}$ into two layers at temperatures below 1~K (above this temperature, ``classical'' cryogenic techniques are used). The $^4\text{He}$ layer contains a small amount of $^3\text{He}$. Much higher partial pressure of $^3\text{He}$ allows to pump $^3\text{He}$ out of the $^4\text{He}$ layer and forces $^3\text{He}$ from $^3\text{He}$ layer to diffuse into the $^4\text{He}$ layer. This diffusion is an endothermic process, hence it drains energy from the surrounding environment and makes it colder. Further information on dilution refrigerators and other cryogenic techniques can be found in \cite{cryo}.} There is also shielding against electromagnetic fields and so on. Despite these measures, some noise is inevitable in a quantum processor, because its temperature is always above 0~K and the radioactive background cannot be shielded out perfectly. What is more, quantum computers are probabilistic machines and errors can occur without any external force because of ubiquitous quantum fluctuations. As a result, a calculation carried out by a quantum processor is prone to errors. There are two basic types of error -- bit-flip and phase-flip (referred to together as {\bf quantum state decoherence}). The former refers to state $|0\rangle$ switching to $|1\rangle$ or vice versa, while the latter is connected with a change of the phase of a state, for example, $|+\rangle$ switching spontaneously to $|-\rangle$ or vice versa.

The resilience of a quantum processor to flips is expressed by the {\bf relaxation time $T_1$}, i.e.,~the expected time a qubit remains in state $|1\rangle$ before switching (relaxing) to $|0\rangle$, and the {\bf dephasing time $T_2$}, i.e.,~the expected time for which a qubit retains its phase.\footnote{Note that spontaneous relaxation and dephasing are governed by the exponential law. The probability of an event occurring in a time interval of length $t$ is $P(t)=1-\mathrm{e}^{-t/T}$, where $T$ is either $T_1$ or $T_2$.} Obviously, increasing times $T_1$ and $T_2$ leads to higher quality of quantum processors and allows more complex calculations to be performed. Currently, the relaxation and dephasing times are aproaching 1 millisecond, which is still relatively low for solving real-world tasks. However, at the beginning of the millennium, the decoherence times of IBM-like processors were in the tens of nanoseconds.\footnote{See the conclusion of \cite{superconductingQubitsProspects} for a short discussion of the {\it development of decoherence times} over the last two decades.}

\begin{figure}[H]
\caption{Demonstration of Decoherence Effects}
\begin{center}
	\begin{tabular}{cc}
		\includegraphics[scale = 0.57]{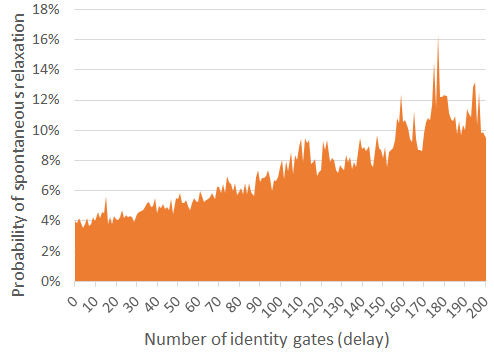}
		\includegraphics[scale = 0.57]{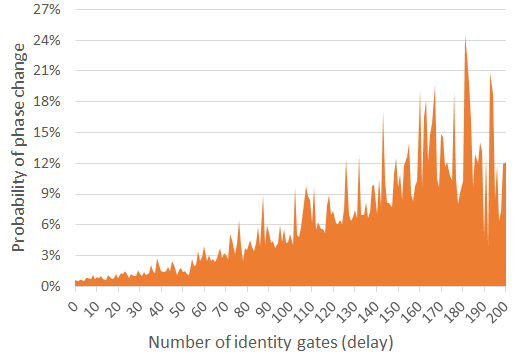}
	\end{tabular}
\end{center}
\vspace{0.1cm}
{\footnotesize \textbf{\textit{Note:}} The results for spontaneous relaxation are depicted in the left figure and those for a spontaneous phase change are shown in the right one. Both figures show that with an increasing delay (or an increasing number of identity gates) between the preparation and the measurement of the quantum state, the probability of error occurrence increases as well.}\\
{\footnotesize \textbf{\textit{Source:}} Author's own calculations on IBM's Lima processor}
\label{fig_decoherence}
\end{figure}

To demonstrate decoherence in practice, we can prepare two simple circuits. The first consists of an $\mathbf{X}$ gate applied to a qubit followed by several identity operators $\mathbf{I}$. These operators do not change the quantum state but introduce a time delay between the preparation and measurement of the state. In an ideal world, the qubit would remain in state $|1\rangle$ indefinitely, but in the real world, an increasing number of $\mathbf{I}$ gates heightens the probability of spontaneous relaxation, as demonstrated in Figure~\ref{fig_decoherence} (left). The second circuit consists of an $\mathbf{H}$ gate changing state $|0\rangle$ to $|+\rangle$, again several $\mathbf{I}$ gates, and finally an $\mathbf{H}$ gate (switching $|+\rangle$ back to $|0\rangle$). In an ideal world, we should measure state $|0\rangle$ with 100\% probability regardless of the number of $\mathbf{I}$ gates. However, the higher the number of $\mathbf{I}$ gates, the higher the probability that $|+\rangle$ changes to $|-\rangle$ and in the end we measure $|1\rangle$ instead of the expected $|0\rangle$. This is illustrated in Figure~\ref{fig_decoherence} (right). 

Note that bit flips and phase flips can be eliminated with the error correction schemes discussed in detail, for example, in \cite{nielsen-chuang-book}. According to the {\bf threshold theorem}, with a sufficient number of auxiliary qubits, it is possible to achieve any accuracy of a quantum computation. This means that once we have quantum processors with thousands of qubits, we can implement corrections to eliminate decoherence effects.\footnote{\cite{googleErrorMitigation} showed in practice on a Google quantum processor that an increasing number of qubits employed in an error correction scheme leads to exponential attenuation of the error rate. However, protecting a quantum processor against errors requires a good deal more qubits than the protected calculation.}
The number of qubits is currently in the higher tens/lower hundreds (IBM Washington has 127 qubits and Google Bristlecone has 72).

\begin{figure}[hbt]
\caption{Schematic of Qubit Connections in the Yorktown IBM Processor}
\begin{center}
\includegraphics[scale = 0.3]{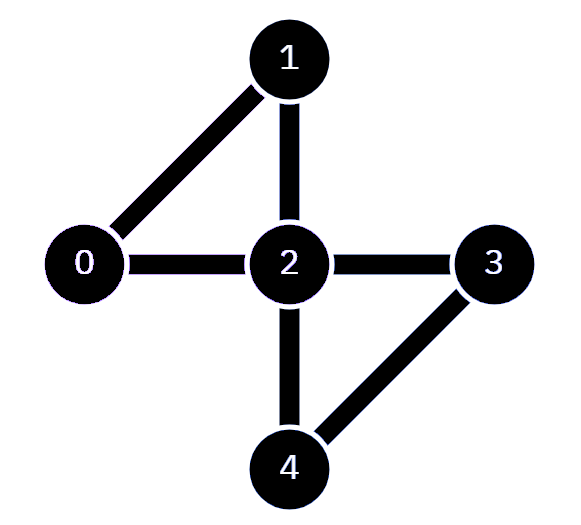}
\end{center}
\vspace{0.1cm}
{\footnotesize \textbf{\textit{Source:}} Adapted from the IBM Quantum\textsuperscript{TM} environment}
\label{fig_limited_connectivity}
\end{figure}

{\bf Limited qubit connectivity:}
To be able to implement controlled gates, qubits have to communicate with each other. However, quantum processors, especially those based on superconducting qubits,\footnote{Details on the state-of-the-art technical realization of quantum computers based on superconducting qubits are provided, for example, in \cite{superconductingQubitsStateOfTheArt}. The future prospects of superconducting qubits, including error mitigation on the hardware level, are widely discussed in \cite{superconductingQubitsProspects}. Other ideas for improving the design of superconducting quantum processors are offered in \cite{superconductingQubitsProspects2}.} are planar devices, which prevents connections being established between all qubits. An example of this limitation in the case of the Yorktown IBM Quantum\textsuperscript{TM} processor is shown in Figure~\ref{fig_limited_connectivity}. 

If we want to establish a line of communication between two qubits that are not directly connected, we have to use intermediate qubits and swap gates. For example, in the case of the Yorktown processor, if we want to control the gate acting on qubit~4 with qubit~0, we have to first swap the states of qubits~0 and 2, then put the controlled gate on qubits~2 (control) and 4 (target), and again swap the states of qubits~2 and 0. Adding new swap gates to the circuit makes it more complex and hence more prone to noise. Together with decoherence, limited connectivity is hindering the application of quantum computers to real-world problems for the time being. However, a design with all-to-all connectivity has been tested and examined, for example, in \cite{connectivityIssue}, and a similar design for a quantum processor with all-to-all connectivity is being developed, for example, by Honeywell (note, however, that these processors have worse decoherence times than those based on superconducting technology).

{\bf Quantum Random Access Memory (qRAM):} Like any other computer, a quantum computer also needs a memory for storing data and programs. However, in contrast to classical computers, data and programs are stored directly in a quantum processor, because there is currently no functional quantum equivalent to an operational memory ({\bf quantum RAM} or {\bf qRAM}). This means that the quantum processor has to be reprogrammed when the input data are changed. 
Besides its obvious ability to store data, qRAM is important for a quantum computer for another three reasons. First, qRAM is able to prepare  quantum states based on stored data. It translates a superposition of addresses $\sum_{i=1}^n a_i|i\rangle$ to a superposition of useful data $\sum_{i=1}^n a_i|i\rangle|D_i\rangle$, where $|D_i\rangle$ is a quantum state stored at address $i$. While the method for preparing a general quantum state proposed in \cite{state-preparation} shows an exponential increase in the number of gates with an increasing number of qubits, qRAM is able to prepare the general state efficiently. Without qRAM, we can only prepare several types of quantum states with polynomially increasing resources in the number of qubits involved, for example:
\begin{itemize}
\item any basis state, such as $|1\rangle$ or $|101\rangle$ (we initialize all qubits to state $|0\rangle$ and put $\mathbf{X}$ gates on those we want to have in state $|1\rangle$)
\item W states and GHZ states, using the method proposed in \cite{ghz-w-states}
\item states representing a discretized Gaussian distribution, using the algorithm designed in \cite{gaussian-states}
\end{itemize} 
Second, some algorithms, for example the HHL one, assume that the input data are stored in qRAM, otherwise the expected speed-up is not necessarily achieved. As mentioned in \cite{hhl-issues2}, in the HHL algorithm, qRAM would help with the preparation of the right-side $|b\rangle$ and even with the controlled gate $\mathrm{e}^{i\mathbf{A}}$. Without qRAM, the HHL algorithm is unusable for real-world problems.\footnote{Note, however, that even with qRAM there are some constraints on the problems that the HHL algorithm can solve with exponential speed-up. Specifically, vector $|b\rangle$ should be more or less uniform and matrix $\mathbf{A}$ should be sparse.}
Third, in some cases, algorithms equipped with qRAM are faster than their original versions -- as shown in \cite{expQram}, a Grover database searching algorithm employing qRAM can achieve exponential speed-up instead of only quadratic.
Currently, one of the biggest problems preventing qRAM from working is decoherence of stored quantum states. New designs have been devised in \cite{qram1} and \cite{qram2}, but these remain rather experimental. Fortunately, Qunnect has introduced photonic qRAM for quantum communication networks\footnote{This memory is based on light ``stopping''. In the writing phase, data are encoded into photons, which are subsequently absorbed by electrons in atoms. In the reading phase, specifically modulated laser pulses are used to stimulate the electrons to emit the absorbed photons. If we simply waited for the electrons to relax spontaneously, the emitted photons would have random properties and the stored information would be lost, hence the need for a ``reading'' laser. The most limiting factor for this type of qRAM is again the decoherence time -- we need to preserve the electrons in the excited state for as long as possible.} and the company plans to commercialize this product \cite{qram-qunnect}.

{\bf Quantum volume:} We have said that it is mainly the decoherence of quantum states which is responsible for the limited capabilities of current quantum processors. However, the relaxation time $T_1$ and the dephasing time $T_2$ are not the only important factors for assessing the quality of quantum processes. The overall quality also depends on the implementation of gates, on their speed and their ability to perform the expected operations correctly ({\bf gate fidelity}), on the connectivity between qubits, and on other factors. To take all these factors into account, a metric called {\bf quantum volume} has been introduced in \cite{quantumVolume}. In simple terms, the quantum volume is the maximum size $n$ of a quantum circuit consisting of $n$ qubits with $n$ gates on each of them which can be implemented without adverse effects on the calculation carried out by the circuit. Note that the gates are native ones (i.e., implemented on the hardware level of the quantum processor). The quantum volume is actually expressed as the number of basis states the circuit can operate with. So, if the quantum volume is, say, 128, then $n$ is 7, because 7 qubits can be in 128 basis states ($128 = 2^7$). For example, the quantum volume of the free-to-use IBM Quantum\textsuperscript{TM} processor is between 8 and 32, which means that circuits consisting of 3 to 5 qubits with a maximum of 3 to 5 gates on each can be run without major errors. Note that in demonstrating decoherence, we used circuits with 200 identity gates. However, an identity gate is a simple wait instruction, and no other action which could further contribute to decoherence is performed.

Note, however, that quantum volume is a somewhat artificial benchmark, because it is based on the generation of random quantum circuits. A quantum processor can behave differently in practical applications such as HHL or QAOA algorithms. Moreover, quantum volume is based on square circuits (i.e.,~the number of qubits is the same as the number of gates applied to each of them), a condition hardly fulfilled in practice (most circuits are rectangular). As proposed in \cite{quantumComputersBenchmarks}, to fully assess the capabilities of a quantum processor, its performance in different practical and theoretical tasks has to be evaluated and benchmarking should not be reduced to only one number. 

{\bf A closing remark on the state of development of quantum hardware:} It is worth emphasizing that it is possible to overcome some of the above-mentioned obstacles by making changes to the design of the algorithms. For example, \cite{hhl-hybrid} introduced a hybrid version of the HHL algorithm exploiting the iterative phase estimation designed in \cite{iterative-phase}. This HHL algorithm enhancement significantly reduces the number of qubits used, hence the noise and the inter-qubit connectivity requirements are reduced as well. Although this approach is not a panacea, as it does not remove other issues connected with the design of the HHL algorithm (for example, the requirement for matrix sparsity), it may help demonstrate the capabilities of quantum computers, enable them to be used to tackle some real-world problems sooner, and facilitate the path to mature quantum computers.

\subsection{Measuring Risk in a Fixed-Income Portfolio}
In this part, we present the results of the calculation of risk metrics using the algorithm introduced in Section~\ref{subsectionRisk}. In particular, we calculate the expected value, the standard deviation, historical Value-at-Risk (VaR), and historical CVaR (both VaR and CVaR are calculated for the 95\% and 99\% significance levels). The underlying data are the daily profits and losses (P/L) of a portfolio expressed in basis points. 

\begin{figure}[H]
\caption{Histogram of Daily P/L for the CNB USD Fixed-Income Investment Portfolio}
\begin{center}
\includegraphics[scale = 0.7]{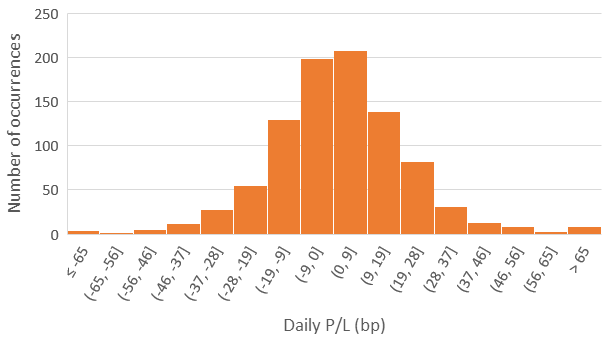}
\end{center}
\vspace{0.1cm}
{\footnotesize \textbf{\textit{Note:}}  The portfolio does not contain investments in MBS. The underlying time series begins on January~2\textsuperscript{nd}, 2018 and ends on August~4\textsuperscript{th}, 2021.}\\
{\footnotesize \textbf{\textit{Source:}} Author's own calculations based on CNB internal reporting data} 
\label{fig_histogramPL}
\end{figure}

For our demonstration, we chose the USD fixed investment portfolio (excluding mortgage-backed securities).\footnote{The Czech National Bank reserves consist of two tranches -- a liquidity one and an investment one. The former is used for monetary policy implementation and is composed of highly liquid short-term instruments denominated in EUR and USD only, while the latter is considered to be a gain generator. The investment tranche has two main parts -- fixed-income (AUD, CAD, CNY, EUR, GBP, SEK, and USD) and equity (AUD, CAD, EUR, GBP, JPY, and USD). The investment tranche also comprises a small portion of USD-denominated mortgage-backed securities (MBS) and physical gold. The structure of the reserves is depicted in a diagram in Appendix~\ref{appendixResStruct}.} 
The time series begins on January~2\textsuperscript{nd}, 2018\footnote{The tranching was introduced in August 2017, hence 2018 is the first full year it was active.} and ends on October~4\textsuperscript{th}, 2021 and comprises 927 observations in total. A histogram of the daily P/L observations is depicted in Figure~\ref{fig_histogramPL}. Note that P/L values below -65~bp and above 65~bp are considered to be outliers, as they occur only four and eight times, respectively.

Preparing a histogram is the first step in calculating risk measures on a quantum computer, because the input distribution for the calculation has to be discrete (the histogram is the discretized version of the continuous distribution). In the next step, an integer value (starting with 0) is assigned to each histogram bin, the integers are then expressed as binary numbers, and the numbers of occurrences are replaced by relative counts. In other words, the histogram is converted into an empirical probability function of a discrete probability distribution. We carried out the discretization for histograms with 8 and 16 bins, as these numbers are powers of two. In Table~\ref{tableRiskResClassical}, we show the expected values of the risk measures as calculated classically (in MS Excel), first with the assumption that the underlying distribution is continuous and second after discretization. Once the distribution is discretized, we get the value of each risk metric expressed as the number of the histogram bin.\footnote{For example, the expected value is calculated as $\sum_{i=0}^{N-1} i p_i$, where $i$ is the number of the $i$\textsuperscript{th} bin and $p_i$ is the probability that the P/L value belongs to the $i$\textsuperscript{th} bin.}
Based on this number, and knowing the lower and upper bounds of the bins, we get an estimate of the value of the metric. As expected, the numbers in Table~\ref{tableRiskResClassical} show that finer discretization returns more accurate estimates of the values of the metrics. Note that in the case of the standard deviation, the bin number means ``the size of the deviation expressed in bins'', that is, to get the actual standard deviation, the bin number is multiplied by the bin width.

The approach described above enables us to encode the discretized distribution into a quantum state $|\psi\rangle = \sum_{i=0}^{N-1} \sqrt{p_i}|i\rangle$, where $p_i$ is the relative count in the $i^\text{th}$ bin and $|i\rangle$ is a basis state representing the $i^\text{th}$ bin. This state is an input to the quantum calculation of the risk metrics. Note that $|\psi\rangle$ is a three-qubit state for eight bins ($8 = 2^3$) and a four-qubit state for 16 bins ($16 = 2^4$). One additional qubit is necessary for the calculation, hence four and five qubits are used, respectively.

\begin{table}[H]
\catcode`\-=12
\caption{Risk Metric Values Calculated Classically}
\begin{center}
{\footnotesize
\begin{tabular}{c|c|ccc|ccc}
\hline\hline 
\multirow{2}{*}{\thead{ \\ \bf Metrics}} & \multirow{2}{*}{\thead{ \\ \bf Continuous}} & \multicolumn{3}{c|}{\thead{\bf Discretized -- 8 bins}} & \multicolumn{3}{c}{\thead{\bf Discretized -- 16 bins}} 
\\  \cline{3-8}
                 &                   &   \thead{\bf Bin \\ \bf  number}     &  \thead{\bf Lower \\ \bf  bound}    &  \thead{\bf Upper \\ \bf  bound}    &  \thead{\bf  Bin \\ \bf number}    &  \thead{\bf  Lower \\ \bf  bound}    & \thead{\bf Upper \\ \bf  bound}       \\ \hline
Expected value & 1.554 & 3.584 & -21.667 & 21.667 & 7.666 & -9.286 & 9.286\\
Standard deviation & 21.067 & 0.948 & 20.542 &  & 2.089 & 19.400 & \\
VaR (95\%) & -28.848 & 2.000 & -43.333 & -21.667 & 4.000 & -37.143 & -27.857\\
VaR (99\%) & -47.383 & 1.000 & -65.000 & -43.333 & 2.000 & -55.714 & -46.429\\
CVaR (95\%) & -44.032 & 1.802 & -65.000 & -21.667 & 3.137 & -46.429 & -27.857\\
CVaR (99\%) & -76.020 & 0.667 & -167.53 & -43.333 & 1.091 & -65.000 & -46.429\\ \hline
\end{tabular}}
\end{center}
\vspace{0.1cm}
{\footnotesize \textbf{\textit{Note:}} The numbers refer to the USD fixed-income investment portfolio (excluding MBS).}\\
{\footnotesize \textbf{\textit{Source:}} Author's own calculations based on CNB internal reporting data}
\label{tableRiskResClassical}
\end{table}

First, we ran the calculation on the IBM Quantum\textsuperscript{TM} simulator to check the correctness of the algorithm design. Note that the simulator behaves as an error-free quantum computer with all-to-all qubit connectivity.  As can be seen in Table~\ref{tableRiskResQuantum}, the results are in accordance with the classical calculation presented in Table~\ref{tableRiskResClassical}. Small deviations are caused by rounding and sampling errors in the simulation. 

After checking on the simulator, we switched to actual quantum hardware. We employed four IBM Quantum\textsuperscript{TM} processors available for free, namely, Yorktown, Lima, Belem, and Bogota, all of them five-qubit devices.\footnote{Note that other processors -- Santiago, Manila, and Quito -- are also available for free, but their characteristics are similar to the named ones so they were not used in this experiment. IBM also provides the Armonk processor, but it is a one-qubit device and is therefore unsuitable for our experiments employing more qubits.}
While Yorktown is a representative of an older generation of quantum processors (family code name {\it Canary}) and was retired shortly after our experiments were carried out, others are more modern IBM processors  (family code name {\it Falcon}).\footnote{The state-of-the-art IBM processor (not available for free) is the 127-qubit Washington (family code name {\it Eagle}).}
The quantum volume is 8 in the case of Yorktown and Lima, 16 for Belem, and 32 for Bogota. The decoherence times are close to 100~$\mu$s for the Falcon family processors and around 50 $\mu$s for the Yorktown processor. The processors also differ in qubit connectivity. While Bogota is a single row of qubits, Lima and Belem have their qubits arranged in a T-shaped lattice, and a schematic of the Yorktown connections is shown in Figure~\ref{fig_limited_connectivity}.

The results of the calculations on the above-mentioned devices are shown in Table~\ref{tableRiskResQuantum}. As can be seen, the expected value was calculated almost correctly for both the three- and four-qubit states, with the exception of the Lima processor. Unfortunately, the evaluation of the standard deviation and the 95\% VaR is totally incorrect. As a result, CVaR is also evaluated wrongly. Although the 99\% VaR for the three-qubit case seems to be correct, we are skeptical about this result, as it looks like the VaR is nested in the value 1. The most probable cause of this issue is the minimum possible change in the rotation angle of the $\mathbf{Ry}$ gates. As we approach lower percentiles, the rotational angles get smaller and the difference in the angles between consecutive iterations of the algorithm is below the lowest distinguishable level. This makes it impossible to distinguish between low percentiles of the distribution. Interestingly, if the 99\% VaR was really equal to 1, the 99\% CVaR calculation is not far from the correct value for the three-qubit case on the Lima and Belem processors.

\begin{table}[H]
\caption{Risk Metric Values Calculated with the Quantum Algorithm}
\catcode`\-=12 
\begin{center}
{\footnotesize
\begin{tabular}{c|c|cccc}
\hline\hline 
\multirow{2}{*}{\thead{ \\ \bf Metrics}} & \multirow{2}{*}{\thead{\bf Simulator \\ \bf (3 qubits)}} & \multicolumn{4}{c}{\thead{\bf Real quantum processor (3 qubits)}}  
\\  \cline{3-6}
                 &                   &   \thead{\bf Lima}     &   \thead{\bf Yorktown}    &  \thead{\bf Belem}    &  \thead{\bf Bogota}      \\ \hline
Expected value & 3.564 & 3.140 & 3.620 & 3.436 & 3.560  \\
Standard deviation & 0.937 & 2.751 & 2.990 & 3.157 & 4.079  \\
VaR (95\%) & 2.000 & 1.000 & 1.000 & 1.000 & 1.000  \\
VaR (99\%) & 1.000 & 1.000 & 1.000 & 1.000 & 1.000  \\
CVaR (95\%) & 2.065 & 0.771 & 0.659 & 0.711 & 0.968  \\
CVaR (99\%) & 0.655 & 0.588 & 0.874 & 0.681 & 0.973  \\ \hline
\vspace*{0.5 cm}
\end{tabular}}

{\footnotesize
\begin{tabular}{c|c|cccc}
\hline\hline 
\multirow{2}{*}{\thead{ \\ \bf Metrics}} & \multirow{2}{*}{\thead{\bf Simulator \\ \bf (4 qubits)}} & \multicolumn{4}{c}{\thead{\bf Real quantum processor (4 qubits)}}  
\\  \cline{3-6}
                 &                   &   \thead{\bf Lima}     &   \thead{\bf Yorktown}    &  \thead{\bf Belem}    &  \thead{\bf Bogota}      \\ \hline
Expected value & 7.630 & 5.986 & 7.537 & 7.150 & 7.377  \\
Standard deviation & 2.181 & 6.679 & 8.839 & 6.916 & 8.028  \\
VaR (95\%) & 4.000 & 1.000 & 1.000 & 1.000 & 1.000  \\
VaR (99\%) & 2.000 & 1.000 & 1.000 & 1.000 & 1.000  \\
CVaR (95\%) & 3.208 & 0.675 & 1.015 & 0.788 & 1.035  \\
CVaR (99\%) & 1.290 & 0.904 & 1.055 & 0.907 & 0.874   \\ \hline
\end{tabular}}
\end{center}
\vspace{0.1cm}
{\footnotesize \textbf{\textit{Note:}} The risk metric values were calculated for the USD fixed-income investment portfolio (excluding MBS). The P/L distribution was discretized to 8 bins (top table) and 16 bins (bottom table), i.e.,~three qubits and four qubits, respectively, were used for the probability function encoding. The tables show the results for several IBM Quantum\textsuperscript{TM} processors. The number of shots of each calculation was 8,192. }\\
{\footnotesize \textbf{\textit{Source:}} Author's own calculations on IBM Quantum\textsuperscript{TM} based on CNB internal reporting data}
\label{tableRiskResQuantum}
\end{table}

Based on these results, it seems meaningless to employ quantum computers to evaluate risk metrics, but, as discussed in Section~\ref{subsectionQuatumHardware}, quantum computers are relatively new devices and some further development is necessary. However, we can extract several useful conclusions from the results:
\begin{itemize}
\item We provided a proof-of-concept that the quantum algorithm for calculating risk measures works. We showed this on the simulator for all metrics and on real quantum processors for the expected value. This is the most important conclusion, because it shows that quantum computers have the potential to be employed in FX reserves management (or portfolio management in general).
\item Qubit connectivity plays an important role in error mitigation. While the Falcon family processors have better decoherence times, the results obtained are often worse in comparison with the older Yorktown processor (for example, the expected value for the four-qubit state). As can be seen in Figure~\ref{fig_limited_connectivity}, the number of connections is six in the case of Yorktown and only four in the case of the T-shaped (Belem and Lima) and row (Bogota) lattices. A lower number of connections leads to an increase in the number of swap gates, which makes the circuits more complex and hence prone to decoherence.
\item The quantum volume of a processor has to be used carefully. Although the quantum volumes of the processors we employed differ significantly, the results are similar. One would expect Bogota, which has the highest quantum volume, to be the best processor, but this is not the case. The reason is that quantum volume assumes square circuits, whereas our circuit widths (the maximum number of gates on one qubit\footnote{Circuit depth is sometimes used as a synonym for circuit width.}) are far higher than our circuit heights (the number of qubits). For example, the circuit for the expected value calculation (the three-qubit case) on the Lima processor has a width of 69 but a height of only 4.
\end{itemize}

We provide the complete source code for the risk metric calculation in Appendix~\ref{appendixSrc}. Note that we implemented the preparation of the quantum state $|\psi\rangle$ based on the method described in \cite{state-preparation}. We omitted the part for setting the quantum phase, as this is not necessary for the risk metric calculation. One could also employ the {\it initialize} function available in the Qiskit libraries. However, for educational purposes, we wrote the quantum state preparation routine ourselves.

\subsection{Finding the Optimal Balance Between Bonds and Equities}
\label{subsectionEqVsBondsOptim}
In this section, we present portfolio optimization based on solving a system of linear equations with the HHL algorithm. In particular, we demonstrate the balancing of the ratio between fixed-income assets and equities for a required expected return.

\begin{figure}[hbt]
\caption{Returns on USD Fixed-Income and Equity Portfolios}
\begin{center}
\includegraphics[scale = 0.7]{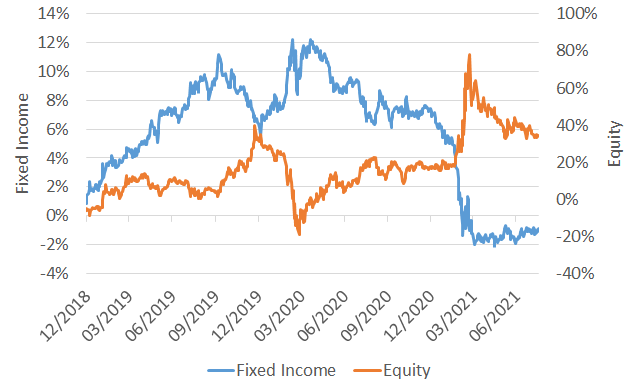}
\end{center}
\vspace{0.1cm}
{\footnotesize \textbf{\textit{Note:}}  Comparison of one-year moving returns for the USD fixed-income investment portfolio (excluding MBS) and the USD equity portfolio. The underlying time series begins on January~2\textsuperscript{nd}, 2018 and ends on August~4\textsuperscript{th}, 2021.}\\
{\footnotesize \textbf{\textit{Source:}} Author's own calculations based on CNB internal reporting data}
\label{fig_equity_vs_bonds}
\end{figure}

For the demonstration, we used our USD fixed-income investment portfolio (excluding MBS), which consists of investment-grade bonds (issued by governments, agencies, and supranational bodies) and plain-vanilla interest rate swaps, and our USD equity portfolio, which replicates the S\&P 500 index. First, we calculated the one-year moving returns for each day of the period from January~2\textsuperscript{nd}, 2019 to August~4\textsuperscript{th}, 2021 for both portfolios. For illustration, we present the returns on both portfolios in Figure~\ref{fig_equity_vs_bonds}. The graph indicates very well the negative correlation between bond and equity returns. The market risk-off regime introduced at the beginning of the Covid-19 pandemic (March 2020) and the subsequent economic recovery in spring 2021 are clearly visible as well. 

Based on the moving returns, we calculated the average yearly returns for fixed-income securities (5.86\%) and equities (16.78\%). These figures serve as the expected returns forming vector $|R\rangle$. The covariance matrix, again based on the moving returns, is
\[
\mathbf{C}=
\begin{pmatrix}
0.15\% & -0.43\% \\
-0.43\% & 2.46\%
\end{pmatrix}.
\]
Note that 0.15\% is the variance of the returns on the fixed-income portfolio and 2.46\% is that of the returns on the equity portfolio. We set the price vector $\langle P| = (1, 1)$ and the budget $B = 1$ to impose the constraint $w_\text{fix. inc.} + w_\text{equity} = 1$. The required return $G$ is a free parameter of the optimization task and is set by the user. Substituting all these figures into \eqref{eq_lin_ptf_optim}, we get a linear system describing the optimization task:\footnote{Note that the condition number of our matrix is 1.11, which is close to 1. This means that even though the matrix is not sparse, the structure of its eigenvalues does not hinder the performance of the HHL algorithm further.}
\begin{equation}
\begin{pmatrix}
0.0000 &  0.0000 & \textcolor{red}{0.0586} & \textcolor{red}{0.1678} \\
0.0000 &  0.0000 &\textcolor{blue}{1.0000} & \textcolor{blue}{1.0000} \\
\textcolor{red}{0.0586} & \textcolor{blue}{1.0000} & \textcolor{orange}{0.0015} & \textcolor{orange}{-0.0043} \\
\textcolor{red}{0.1678} & \textcolor{blue}{1.0000} & \textcolor{orange}{-0.0043} &  \textcolor{orange}{0.0246} \\
\end{pmatrix}
\begin{pmatrix}
\textcolor{red}{\lambda} \\ \textcolor{blue}{\mu} \\ \mathbf{w_\text{fix. inc.}} \\ \mathbf{w_\text{equity}}
\end{pmatrix}
=
\begin{pmatrix}
\textcolor{red}{G} \\ \textcolor{blue}{1.0000} \\ 0.0000 \\ 0.0000
\end{pmatrix}.
\label{eq_portfolio_optim_system}
\end{equation}
As can be seen in equation~\eqref{eq_portfolio_optim_system}, the linear system breaks up into two separate systems in the case of only two assets. The first is composed of the equations $0.0586w_\text{fix. inc} + 0.1678w_\text{equity} = G$ and $w_\text{fix. inc} + w_\text{equity} = 1$. This allows us to evaluate the weights for a given expected return $G$ regardless of the rest of \eqref{eq_portfolio_optim_system}, i.e.,~the risk part. For example, for $G = 7\% = 0.07$, we get $w_\text{fix. inc} = 89.53\%$ and $w_\text{equity} = 10.47\%$.  Although this task is trivial and can be solved quickly by hand, its purpose is to demonstrate the capabilities and identify the bottlenecks of quantum computers and the HHL algorithm in portfolio optimization tasks.

As in the risk measurement case present in the previous section, we first run the algorithm on a quantum computer simulator. Note that despite the loss of the speed-up delivered by a quantum computer, we decided to extract the whole solution of \eqref{eq_portfolio_optim_system} to assess the correctness of the implementation of the algorithm. Unfortunately, when we ran the HHL solver for $G = 7\%$, we obtained totally wrong results -- $w_\text{fix. inc} = 10.29\%$ and $w_\text{equity} = 18.41\%$. To debug the code, we ran the HHL solver on a simple system with matrix $\mathbf{A} = \text{diag}(1,2,3,4)$ and right-side $\langle b| = (1,1,1,1)$ having the solution $\langle x|  = (1,1/2,1/3,1/4)$. In this case, the solver returned the correct result. Interestingly, for $\mathbf{A} = \text{diag}(-1,2,3,4)$ and  $\langle b| = (-1,1,1,1)$ with the same solution $\langle x|$, the HHL solver failed again, as the returned $\langle x| = (0.1429, 1/2, 1/3, 1/4)$. After several attempts, we realized that the correctness of the solution is linked with the matrix eigenvalues. In the case of a diagonal matrix, the eigenvalues are the diagonal elements. While in the former case they are all positive, in the latter there is one negative eigenvalue. To examine this conjecture further, we prepared matrix $(\mathbf{H}\otimes\mathbf{H})\text{diag}(1,2,3,4)(\mathbf{H}\otimes\mathbf{H})$, where $\mathbf{H}$ is a Hadamard gate. The resulting matrix is derived from the diagonal one by unitary transformation $\mathbf{H}\otimes\mathbf{H}$, hence its eigenvalues are the same as those of matrix $\text{diag}(1,2,3,4)$. Setting $\langle b| = (0,1,1,0)$, we get the correct solution $\langle x|=(3/8, 5/8, 5/8, 3/8)$. However, changing the first element of the diagonal matrix to -1 led again to incorrect results. After some additional investigation, we realized that the current Qiskit implementation of the HHL algorithm is only able to work with matrices where all the eigenvalues are positive.\footnote{See the unofficial statement on this issue by IBM staff here: https://github.com/Qiskit/qiskit-terra/issues/6880, cited on August~27\textsuperscript{th}, 2021.} We calculated the eigenvalues of the matrix in system \eqref{eq_portfolio_optim_system} and found that they are both positive and negative, specifically: -16.90792562,  -0.36203117,  1.03367322,  18.84628357. As a result, we were unable to carry out the portfolio optimization task with the HHL algorithm successfully for the time being.

For illustration purposes, we tried to solve a simple system with the diagonal matrix $\text{diag}(1,2,3,4)$ on a real quantum processor. However, the results suffered from decoherence and were meaningless. Note that the circuit width was 159 gates. In the case of the portfolio optimization task \eqref{eq_portfolio_optim_system}, the width even reached 13,762 gates. Hence, regardless of the issue with the eigenvalues, the task would still be unsolvable on the current quantum hardware, because the circuit is too deep and complex.

\begin{figure}[H]
\caption{HHL Algorithm Circuit for a 2x2 System of Linear Equations}
\begin{center}
\includegraphics[scale = 0.6]{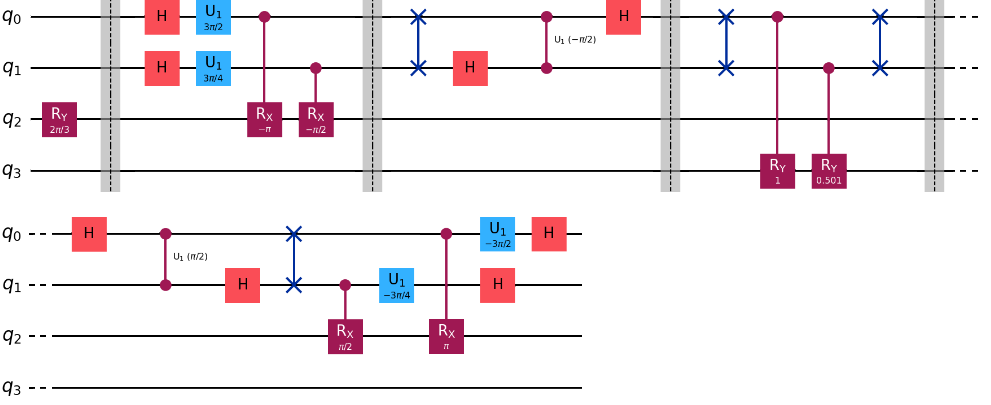}
\end{center}
\vspace{0.1cm}
{\footnotesize \textbf{\textit{Note:}}  The circuit implements the HHL algorithm for system \eqref{eq_2by2_system} with $\theta = \pi/3$. The first part of the circuit implements the preparation of the right-side $|b\rangle$ with the $\mathbf{Ry}(2\theta)$ gate. The second part represents controlled rotations of the phase estimation algorithm. The algorithm continues with inverse QFT in the third part, and the fourth part implements the inverse of the eigenvalues. The last part provides the uncomputation of the working qubits $q_0$ and $q_1$. Note that $q_3$ is the ancilla qubit and the solution is stored in qubit $q_2$.}\\
{\footnotesize \textbf{\textit{Source:}} Author's own creation in Qiskit based on \cite{hhl-example} and \cite{machine-learning}}
\label{fig_hhl_2_by_2}
\end{figure}

To assess the current capabilities of free-to-use quantum processors in solving linear systems, we tried to run a circuit for the linear system
\begin{equation}
\begin{pmatrix}
1.5 & 0.5\\
0.5 & 1.5
\end{pmatrix}
\begin{pmatrix}
x_0 \\ x_1
\end{pmatrix}
=
\begin{pmatrix}
\cos(\theta) \\ \sin(\theta)
\end{pmatrix}
\label{eq_2by2_system}
\end{equation}
discussed in \cite{hhl-example} and \cite{machine-learning}. The implementation of the circuit is shown in Figure~\ref{fig_hhl_2_by_2}. Note that the right side of \eqref{eq_2by2_system} is the normalized quantum state and can be easily prepared with an $\mathbf{Ry}$ gate applied to state $|0\rangle$. We run the circuit for several angles $\theta$: $\pi/3$, $\pi/4$, $\pi/6$, and $\pi/7$. Regardless of the IBM quantum processor used (Lima, Quito, and Bogota were employed), the results were fully obscured because of decoherence. The circuit width after adaptation to the connectivity of the real processor and its basic gate set reached 102 gates for all the processors tested.

Based on the results presented above, we conclude that the quantum hardware we tested is not able to perform the HHL algorithm successfully. The first issue is decoherence of the quantum states, caused by deep circuits and short decoherence times. The second issue is the missing implementation of the HHL algorithm for matrices with negative eigenvalues. The latter issue will be overcome soon, once IBM developers expand the capabilities of the Qiskit implementation of the HHL algorithm. However, the former issue could take more time to resolve, as it is connected with further research in the field of quantum processor construction. It is worth noting that we only employed quantum processors that are available to use free of charge. However, the decoherence times of all the IBM devices are close to 100~$\mu$s.\footnote{As of August 27\textsuperscript{th}, 2021.}  For educational purposes and hopefully future use, we provide the source code for portfolio optimization using the HHL algorithm and the circuit for solving system \eqref{eq_2by2_system} in Appendix~\ref{appendixSrc}.

\subsection{Picking the Best Performing Assets}
\label{subsectionQAOADemonstration}
This part is devoted to picking a subset of assets from a larger set, for example, a benchmark index, subject to some constraints and objectives. To do so, we employ quadratic unconstrained binary optimization and the QAOA algorithm. In our case, we want to choose three shares from a set containing five shares in total, and at the same time to maximize the return and minimize the market risk. The number of shares in the larger set is limited by the number of qubits on the quantum processors available to us (i.e.,~five). The shares we deal with in the demonstration are five semiconductor industry companies contained in the S\&P 500 index, namely, AMD, Intel, Qualcomm, Analog Devices, and Texas Instruments.\footnote{There is no particular reason for selecting these companies. {\bf In fact, the Czech National Bank's equity investments follow a passive index replication strategy and we do not choose particular shares to invest in.} Therefore, this example serves only to demonstrate the capabilities of the QAOA on real market data and is not connected with the actual investment decision-making at the Czech National Bank.}

To perform the task described above, we modified objective function \eqref{eq_qubo_ptf_optim} by setting the budget equal to the number of shares we want to invest in, i.e.,~$B=m$, and the maximum amount of money invested in the $i$\textsuperscript{th} share was set to 1. Moreover, we need to rewrite \eqref{eq_qubo_ptf_optim} to the general form of a quadratic binary optimization objective function, i.e.,~$f(x) = \langle x|\mathbf{A}|x\rangle + \langle b|x\rangle + c$. After algebraic manipulation, we end up with the objective function
\begin{equation}
f(x) = \langle x|(\lambda_2\mathbf{C}+\lambda_3 \mathbf{1}_{n,n})|x\rangle - (\lambda_1 \langle r| + 2m\lambda_3 \langle 1_{n}|) |x\rangle + \lambda_3 m^2 \rightarrow \min,\label{eq_bin_optim_particular}
\end{equation}
where $\langle r|$ is a vector of share returns, $\mathbf{C}$ is a covariance matrix, $m$ is the number of shares we want to invest in, $\mathbf{1}_{n,n}$ is an $n$-by-$n$ matrix having all elements equal to 1, and $\langle 1_{n}|$ is a row vector with all $n$ members equal to 1. The returns vector and the covariance matrix are based on one-year moving returns calculated similarly to those in Section~\ref{subsectionEqVsBondsOptim}. The results of this calculation are shown in Table~\ref{tableShareReturnCovariance}. Parameters $\lambda$ describing the importance and scale of each part of the objective function were set as follows: $\lambda_1 = \lambda_3 = 1$ and $\lambda_2 = 4$ (the risk term is scaled by $\lambda_2 = 4$ to be similar in magnitude to the return term).

To compare the results obtained on the quantum computer, we first calculated the minimum value of function~\eqref{eq_bin_optim_particular} classically using the brute force method, i.e.,~all the solutions were listed and the objective function was evaluated. Since we are dealing with five equities, there are $2^5 = 32$ possible solutions, each denoted with a bit string representing the equities included in the portfolio (the $i$\textsuperscript{th} bit is equal to one if the $i$\textsuperscript{th} share is included). However, only solutions having exactly three bits equal to one are feasible solutions, as we require $m=3$. The optimal solution is $\langle x_\text{opt}| = (1,1,0,1,0)$, i.e.,~we should invest in AMD, Intel, and Analog Devices. All the possible solutions are presented in Table~\ref{tableQAOAadditional} in the Appendix.

\begin{table}[H]
\caption{Returns Vector and the Covariance Matrix used for Binary Portfolio Optimization}
\begin{center}
{\footnotesize
\begin{tabular}{l|rrrrr}
\hline\hline 
{\it Company} &\thead{\bf AMD} & \thead{\bf Intel} & \thead{\bf Qualcomm} &
\thead{\bf Analog \\ \bf Devices} & \thead{\bf Texas \\ \bf Instruments} \\
\hline
{\it Average returns} & 95.2\% & 5.3\% &39.8\% & 20.4\% & 21.3\%\\
\hline
{\it Price variances and covariances} & & & & & \\
{\bf AMD}&\textcolor{red}{20.2\%}&&&&\\
{\bf Intel}&0.3\%&\textcolor{red}{2.5\%}&&&\\
{\bf Qualcomm}&3.3\%&-0.3\%&\textcolor{red}{9.9\%}&&\\
{\bf Analog Devices}&-1.5\%&0.4\%&3.6\%&\textcolor{red}{3.5\%}&\\
{\bf Texas Instruments}&-1.9\%&0.3\%&5.3\%&3.7\%&\textcolor{red}{5.0\%} \\ \hline
\end{tabular}}
\end{center}
\vspace{0.1cm}
{\footnotesize \textbf{\textit{Note:}}  The table contains the one-year average returns, price variances (highlighted in red), and covariances of the shares used in the demonstration of quantum binary optimization. The underlying time series begins on January~2\textsuperscript{nd}, 2018 and ends on August~4\textsuperscript{th}, 2021}\\
{\footnotesize \textbf{\textit{Source:}} Author's own calculations based on CNB internal reporting data}
\label{tableShareReturnCovariance}
\end{table}

We carried out the calculation on all the available five-qubit real quantum processors with success. The only difference among the processors was the number of iterations needed to find the solution, as can be seen in Table~\ref{tableBinOptimIters}. The number of iterations depends on the quantum volume of the processor and is lowest for processors with quantum volume 32. It is worth noting that the number of iterations is higher than the number of possible solutions of the task (i.e.,~32). This means that the QAOA seems to be less effective than the brute force method. However, we have to take into account the fact that the current quantum computers still suffer from noise and, as a result, more iterations are needed. Despite this issue, we found that the QAOA can be performed successfully on the current real quantum hardware. 

\begin{table}[h]
\caption{Number of Iterations of the QAOA on Different IBM Quantum\textsuperscript{TM} Processors}
\begin{center}
{\footnotesize
\begin{tabular}{l|cc}
\hline \hline 
\thead{\bf Processor} & \thead{\bf Number of \\ \bf iterations} & \thead{\bf Quantum \\ \bf volume} \\
\hline
{\bf Lima} & 171 & 8\\
{\bf Belem} & 184 & 16\\
{\bf Quito} & 184 & 16\\
{\bf Bogota} & 132 & 32\\
{\bf Manila} & 116 & 32\\
{\bf Santiago} & 116 & 32 \\ \hline
\end{tabular}}
\end{center}
\vspace{0.1cm}
{\footnotesize \textbf{\textit{Note:}}  The table depicts the number of iterations needed to find the solution to the task with the QAOA. The number of shots was 1,024.}\\
{\footnotesize \textbf{\textit{Source:}} Author's own calculations on  IBM Quantum\textsuperscript{TM} platform}
\label{tableBinOptimIters}
\end{table}

Interestingly, a quantum circuit for a five-qubit QAOA (see Appendix~\ref{appendixQAOA}) is comparable in terms of the number of gates to circuits performing the other algorithms discussed in this paper. However, in contrast to those, we recorded 100\% success with the QAOA. This shows the importance of application-oriented benchmarking of quantum computers. In other words, we should assess the quality of the current quantum processors based on a particular application and not only on a single number like the quantum volume, even though it is also an important indicator.

\section{Conclusion}
In this article, we investigated possible applications of quantum computers in foreign exchange reserves management. We discussed risk measurement and portfolio construction. For the former task, we employed the quantum Monte Carlo method, while for the latter we used the HHL algorithm and quadratic unconstrained binary optimization with the QAOA. We also discussed technical issues connected with the currently available quantum computers. All tests and demonstrations were carried out on IBM Quantum\textsuperscript{TM} processors. Our calculations were based on real data taken from the Czech National Bank's daily internal reporting on its foreign exchange reserves. Besides analyzing the capabilities of the current quantum computers, we also provided a detailed introduction to the mathematical background and basic building blocks of quantum computing, explained how several quantum algorithms work, and supplemented our discussion with examples. Moreover, we shared all the Qiskit source codes we used to help the reader with studying the basics of quantum computing on her/his own.

We found that, in principle, all the algorithms discussed are suitable for the purposes for which we employed them. First, we ran the algorithms on a quantum computer simulator and all the results obtained were in accordance with those obtained using classical methods.

After the simulations, we performed the tasks on several five-qubit real quantum processors. In this case, the results were mixed. Quadratic binary optimization for five variables using the QAOA returned the correct results on all the available processors. This means that even on the current noisy quantum computers, the QAOA ran successfully. Based on this result and several other studies mentioned in the theoretical part, it seems that binary optimization will be among the first real-world problems that quantum computers are used for. However, it is worth noting that the number of iterations needed to find the solution depended on the quantum volume of the quantum processor. In other words, noise still plays some role in algorithm efficiency.

On the other hand, portfolio optimization leveraging the HHL algorithm failed completely. This is because the current implementation of the HHL algorithm in Qiskit cannot work with matrices that have negative eigenvalues. Therefore, we tried to solve a simple linear system with a 2x2 matrix with all the eigenvalues positive. However, despite the fact only four qubits were used, wrong results were returned for all the right sides we used. The most probable reason for this is the complexity of the quantum circuit representing the HHL algorithm, especially the phase estimation part, which consists of many controlled quantum gates.

The quantum risk measurement algorithm returned the correct results for the expected value but failed for the other risk parameters evaluated. Note that we tried to calculate risk measures with two different levels of input probability distribution quantization -- 8 bins (three qubits) and 16 bins (four qubits). The performance of the algorithm was better in the case of three qubits because of a less complex circuit, which meant it was less impacted by noise. There is also an issue with limited accuracy of the rotational angles of the $\mathbf{Ry}$ gates, which prevented us from making a finer distinction between the percentiles in the Value-at-Risk calculation.

Interestingly, despite the fact we used ``free'' quantum processors, we had access to state-of-the-art technology, because even IBM processors with higher numbers of qubits show similar decoherence times and gate error rates to smaller five-qubit devices.  For example, the quantum volume of the 127-qubit Washington processor is 32, i.e.,~the same as for the five-qubit Manila processor we used in our simulations. If we had access to a bigger processor, we would be able to implement, for example, binary optimization with more variables. However, noise would still negatively impact the results. Overall, this means that currently the biggest limiting factor of quantum processors is noise, not the number of qubits. 
Although the presented results and discussion could suggest that quantum computers are of low practical value, we should keep in mind that they represent a very new technology, and intensive research work continues to enhance the parameters of quantum processors.  Recent advances in the quality of quantum processors allowed us to perform at least some calculations correctly. Had we carried out similar simulations a few years ago, we would have failed in all cases.

Let us briefly discuss future applications of quantum computers in the finance industry and their impact on central banks and other financial market regulatory authorities. The possible fields of application of quantum computers range from complex derivatives pricing to credit risk analysis (including loan underwriting). In these cases, the quantum Monte Carlo method can be deployed. Because of their promised high computational performance, quantum computers would be interesting to high-frequency traders. As such investors bother about every microsecond, they would again employ the quantum Monte Carlo or quick portfolio optimization methods (i.e.,~the HHL algorithm or the QAOA). Clearly, banks and other financial companies have incentives to exploit quantum computing.  All the topics discussed in this article are therefore important not only for the FX reserves management arms of central banks, but also for their supervision and regulation departments and other supervisory authorities, because the latter have to examine financial firms' pricing, credit, and market models closely to prevent misconduct, both intentional and unintentional, and ensure the stability of financial systems.

It is true that many technological steps to advance quantum computers have to take place before the deployments discussed above can materialize. For example, decoherence times and the number of qubits have to be increased, gate fidelity has to be improved, and, above all, a functional qRAM has to be built. Despite the rapid development, there is still a long way to a fully mature quantum computer. However, central banks and financial market regulators in general should be prepared for the massive deployment of quantum technologies, as it is not a question of if but when this happens.

\nocite{*}
\bibliographystyle{ieeetr}
\bibliography{ms}

\clearpage
\appendix
\section{Czech National Bank FX Reserves Structure}

\setcounter{table}{0}
\setcounter{figure}{0}
\renewcommand{\thetable}{A\arabic{table}}
\renewcommand{\thefigure}{A\arabic{figure}}

\label{appendixResStruct}

\begin{figure}[h]
\caption{ Structure of Czech National Bank Foreign Exchange Reserves as of August 2021}
\begin{center}
\includegraphics[scale = 0.6]{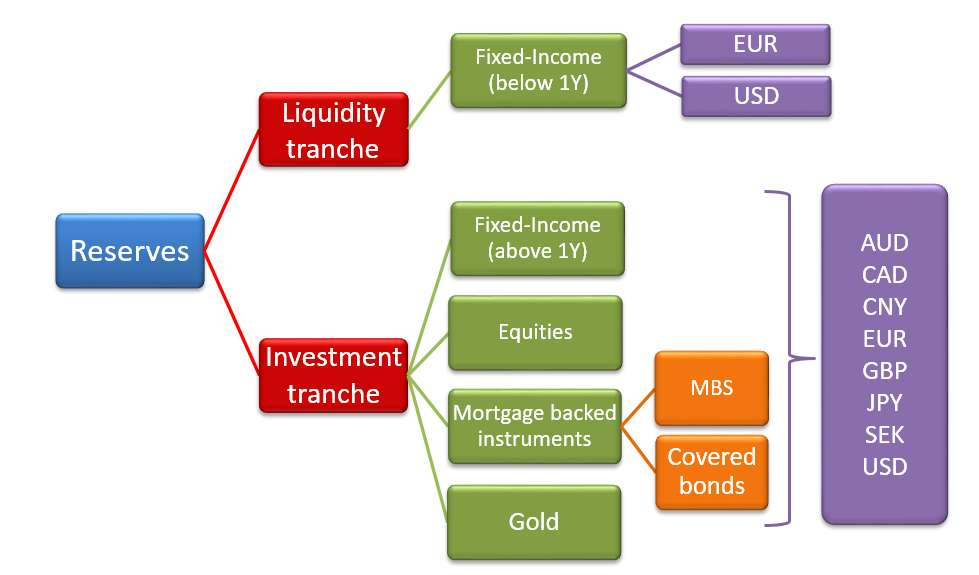}
\end{center}
\vspace{0.1cm}
{\footnotesize \textbf{\textit{Source:}}  Author's own creation}
\end{figure}

\clearpage
\section{Quantum Gates Overview}

\setcounter{table}{0}
\setcounter{figure}{0}
\renewcommand{\thetable}{B\arabic{table}}
\renewcommand{\thefigure}{B\arabic{figure}}

\label{appendixGates}
In this appendix, we present a list of quantum gates. This list is not exhaustive, but it contains the usually used gates and their widely accepted symbols. Note that in the case of controlled gates, the part of the matrix corresponding to the non-controlled version of the gate is highlighted in red.
\begin{itemize}
\item {\bf Pauli gates}
	\[
	\mathbf{I} = 
	\begin{pmatrix}
	1 & 0 \\
	0 & 1
	\end{pmatrix}  
	\,\,\,\,\,\,
	\mathbf{X} = 
	\begin{pmatrix}
	0 & 1 \\
	1 & 0
	\end{pmatrix}  
	\,\,\,\,\,\,
	\mathbf{Y} = 
	\begin{pmatrix}
	0 & -i \\
	i & 0
	\end{pmatrix}  
	\,\,\,\,\,\,
	\mathbf{Z} = 
	\begin{pmatrix}
	1 & 0 \\
	0 & -1
	\end{pmatrix}  
	\]
	Note that for any Pauli gate $\mathbf{A}$, it holds that $\mathbf{A^\dagger} = \mathbf{A}$, hence $\mathbf{A}^2 = \mathbf{I}$.
\item{\bf Square root of Pauli $\mathbf{X}$ gate}
	\[
	\sqrt{\mathbf{X}} = 
	\frac{1}{2}
	\begin{pmatrix}
	1+i & 1-i \\
	1-i & 1+i
	\end{pmatrix} 
	\]
	Note that the square root of $\mathbf{X}$ can be obtained with a spectral decomposition. Matrix $\mathbf{X}$ has eigenvalues 1 and -1 and 		
        respective eigenvectors $|+\rangle$ and $|-\rangle$. Setting $f(x) = \sqrt{x}$, we get 
	$\sqrt{\mathbf{X}} = \sqrt{1}|+\rangle\langle +| + \sqrt{-1}|-\rangle\langle -| = |+\rangle\langle +| + i|-\rangle\langle -|$ .
\item{\bf $\mathbf{H}$, $\mathbf{S}$, and $\mathbf{T}$ gates}
	\[
	\mathbf{H} = \frac{1}{\sqrt{2}}
	\begin{pmatrix}
	1 & 1 \\
	1 & -1
	\end{pmatrix}
	\,\,\,\,\,\,
	\mathbf{S} = 
	\begin{pmatrix}
	1 & 0 \\
	0 & i
	\end{pmatrix}
	\,\,\,\,\,\,
	\mathbf{T} = 
	\begin{pmatrix}
	1 & 0 \\
	0 & \mathrm{e}^{i\frac{\pi}{4}}
	\end{pmatrix}
	\]
           \\
	\[
	\mathbf{S^\dagger} = 
	\begin{pmatrix}
	1 & 0 \\
	0 & -i
	\end{pmatrix}
	\,\,\,\,\,\,
	\mathbf{T^\dagger} = 
	\begin{pmatrix}
	1 & 0 \\
	0 & \mathrm{e}^{-i\frac{\pi}{4}}
	\end{pmatrix}
	\]
	Note that $\mathbf{H^\dagger} = \mathbf{H}$, hence $\mathbf{H}^2 = \mathbf{I}$.
\item{\bf Rotations around axes of Bloch sphere }
	\[
	\mathbf{R_x}( \alpha) = 
	\begin{pmatrix}
	\cos(\alpha/2) & -i\sin(\alpha/2) \\
	-i\sin(\alpha/2) & \cos(\alpha/2)
	\end{pmatrix}
	\,\,\,\,\,\,
	\mathbf{R_y}( \alpha) = 
	\begin{pmatrix}
	\cos(\alpha/2) & -\sin(\alpha/2) \\
	\sin(\alpha/2) & \cos(\alpha/2)
	\end{pmatrix}
	\]
	\[
	\mathbf{R_z}(\alpha) = 
	\begin{pmatrix}
          \mathrm{e}^{-i\alpha/2} & 0 \\
	0 & \mathrm{e}^{i\alpha/2}
	\end{pmatrix}
	\]
	Note that:
	\begin{itemize} 
		\item $\mathbf{R_a^\dagger}(\alpha) = \mathbf{R_a}(-\alpha)$ for any rotation around axis $a \in \{x,y,z\}$
		\item $\mathbf{R_a}(\alpha)\mathbf{R_a}(\beta) = \mathbf{R_a}(\alpha + \beta)$ for any rotation around axis $a \in \{x,y,z\}$
		\item The rotations are connected with Pauli matrices via formula $R_a(\alpha) = \mathrm{e}^{-i\frac{\alpha}{2}\mathbf{A}}$,
			where $\mathbf{A} \in \{\mathbf{X};\mathbf{Y};\mathbf{Z}\}$
	\end{itemize}
\item{\bf $\mathbf{U}$ gates on IBM Quantum\textsuperscript{TM}}
	\[
	\mathbf{U1}( \lambda) = 
	\begin{pmatrix}
	1 & 0 \\
	0 & \mathrm{e}^{i\lambda}
	\end{pmatrix}
	\,\,\,\,\,\,
	\mathbf{U2}(\varphi, \lambda) = \frac{1}{\sqrt{2}}
	\begin{pmatrix}
          1 & -\mathrm{e}^{i\lambda} \\
	\mathrm{e}^{i\varphi} & \mathrm{e}^{i(\varphi+\lambda)}
	\end{pmatrix}
	\]
	\[
	\mathbf{U3}(\theta, \varphi, \lambda) = 
	\begin{pmatrix}
	\cos(\theta/2) & -\mathrm{e}^{i\lambda}\sin(\theta/2) \\
	\mathrm{e}^{i\varphi}\sin(\theta/2) & \mathrm{e}^{i(\varphi+\lambda)}\cos(\theta/2)
	\end{pmatrix}
	\]
\item{\bf $\mathbf{CNOT}$ and controlled $\mathbf{Z}$ gates}	
	\[
	\mathbf{CNOT} =
	\begin{pmatrix}
	1 & 0 & 0 & 0 \\
	0 & 1 & 0 & 0 \\
	0 & 0 & \textcolor{red}{0} & \textcolor{red}{1} \\
	0 & 0 & \textcolor{red}{1} & \textcolor{red}{0} \\
	\end{pmatrix}
	\,\,\,\,\,\,
	\mathbf{CZ} =
	\begin{pmatrix}
	1 & 0 & 0 & 0 \\
	0 & 1 & 0 & 0 \\
	0 & 0 & \textcolor{red}{1} & \textcolor{red}{0} \\
	0 & 0 & \textcolor{red}{0} & \textcolor{red}{-1} \\
	\end{pmatrix}
	\]
	Note that $\mathbf{CNOT}^\dagger = \mathbf{CNOT}$ and $\mathbf{CZ^\dagger} = \mathbf{CZ}$.
\item{\bf Swap gate}
	\[
	\mathbf{SWAP} =
	\begin{pmatrix}
	1 & 0 & 0 & 0 \\
	0 & 0 & 1 & 0 \\
	0 & 1 & 0 & 0 \\
	0 & 0 & 0 & 1 \\
	\end{pmatrix}
	\]
	Note that $\mathbf{SWAP}^\dagger = \mathbf{SWAP}$.
\item{\bf Toffoli ($\mathbf{CCNOT}$) and Fredkin (controlled swap) gates}
	\[
	\mathbf{CCNOT} =
	\begin{pmatrix}
	1 & 0 & 0 & 0 & 0 & 0 & 0 & 0  \\
	0 & 1 & 0 & 0 & 0 & 0 & 0 & 0  \\
	0 & 0 & 1 & 0 & 0 & 0 & 0 & 0  \\
	0 & 0 & 0 & 1 & 0 & 0 & 0 & 0  \\
	0 & 0 & 0 & 0 & 1 & 0 & 0 & 0  \\
	0 & 0 & 0 & 0 & 0 & 1 & 0 & 0  \\
	0 & 0 & 0 & 0 & 0 & 0 & \textcolor{red}{0} & \textcolor{red}{1}  \\
	0 & 0 & 0 & 0 & 0 & 0 & \textcolor{red}{1} & \textcolor{red}{0}  \\
	\end{pmatrix}
	\,\,\,\,\,\,
	\mathbf{CSWAP} =
	\begin{pmatrix}
	1 & 0 & 0 & 0 & 0 & 0 & 0 & 0  \\
	0 & 1 & 0 & 0 & 0 & 0 & 0 & 0  \\
	0 & 0 & 1 & 0 & 0 & 0 & 0 & 0  \\
	0 & 0 & 0 & 1 & 0 & 0 & 0 & 0  \\
	0 & 0 & 0 & 0 & \textcolor{red}{1} &\textcolor{red}{0} & \textcolor{red}{0} & \textcolor{red}{0}  \\
	0 & 0 & 0 & 0 & \textcolor{red}{0} &\textcolor{red}{0} & \textcolor{red}{1} & \textcolor{red}{0}  \\
	0 & 0 & 0 & 0 & \textcolor{red}{0} &\textcolor{red}{1} & \textcolor{red}{0} & \textcolor{red}{0}  \\
	0 & 0 & 0 & 0 & \textcolor{red}{0} &\textcolor{red}{0} & \textcolor{red}{0} & \textcolor{red}{1}  \\
	\end{pmatrix}
	\]
	Note that $\mathbf{CCNOT}^\dagger = \mathbf{CCNOT}$ and $\mathbf{CSWAP^\dagger} = \mathbf{CSWAP}$.
\end{itemize}

Here we provide examples of universal gate sets, i.e.,~groups of gates allowing any unitary operation to be implemented on a quantum computer:
\begin{itemize}
\item $\mathbf{H}$, $\mathbf{T}$, and $\mathbf{CNOT}$ (sometimes enriched by gates $\mathbf{X}$ and $\mathbf{S}$ for easier implementation)
\item $\mathbf{R_x}(\theta)$, $\mathbf{R_y}(\theta)$, $\mathbf{R_z}(\theta)$, and $\mathbf{CNOT}$
\item $\mathbf{H}$ and $\mathbf{CCNOT}$
\item $\mathbf{H}$ and $\mathbf{CSWAP}$
\item $\mathbf{U3}(\theta, \varphi, \lambda)$ and $\mathbf{CNOT}$
\item $\mathbf{I}$, $\mathbf{Rz}(\alpha)$, $\mathbf{X}$, $\sqrt{\mathbf{X}}$, and $\mathbf{CNOT}$ (an IBM Quantum\textsuperscript{TM} processor gate set implemented on the hardware level)
\end{itemize}

\clearpage
\section{Gray Code}

\setcounter{table}{0}
\setcounter{figure}{0}
\renewcommand{\thetable}{C\arabic{table}}
\renewcommand{\thefigure}{C\arabic{figure}}

\label{appendixGray}
In this appendix, we describe the binary number encoding called {\bf Gray code}. In comparison with standard binary number representation, the code has the special property that the immediate successor of a binary number differs from that number only in one bit. This property for four-bit numbers can be easily seen in Table~\ref{tableGray}.

\begin{table}[h]
\caption{Example of Gray Code for Four-Bit Numbers}
\begin{center}
{\footnotesize
\begin{tabular}{c|c|c||c|c|c}
\hline\hline 
\thead{\bf Decimal\\\bf number} & \thead{\bf Binary\\ \bf representation} & \thead{\bf Gray\\ \bf code} &
\thead{\bf Decimal\\\bf number} & \thead{\bf Binary\\ \bf representation} & \thead{\bf Gray\\ \bf code} \\
\hline
0 &  0 0 0 0 & 0 0 0 0  &
8 &  1 0 0 0 & 1 1 0 0 
\\
1 &  0 0 0 1 & 0 0 0 1  &
9 &  1 0 0 1 & 1 1 0 1 
\\
2 &  0 0 1 0 & 0 0 1 1  &
10 &  1 0 1 0 & 1 1 1 1 
\\
3 &  0 0 1 1 & 0 0 1 0  &
11 &  1 0 1 1 & 1 1 1 0 
\\
4 &  0 1 0 0 & 0 1 1 0  &
12 &  1 1 0 0 & 1 0 1 0 
\\
5 &  0 1 0 1 & 0 1 1 1  &
13 &  1 1 0 1 & 1 0 1 1 
\\
6 &  0 1 1 0 & 0 1 0 1  &
14 &  1 1 1 0 & 1 0 0 1 
\\
7 &  0 1 1 1 & 0 1 0 0  &
15 &  1 1 1 1  & 1 0 0 0  \\ \hline
\end{tabular}}
\end{center}
\vspace{0.1cm}
{\footnotesize \textbf{\textit{Source:}} Author's own calculations based on the method described in the main text}
\label{tableGray}
\end{table}

Before we describe how to convert a binary number to Gray code and vice versa, we define an XOR function\footnote{Sometimes denoted  exclusive-OR or non-equivalence.} denoted by symbol $\oplus$. For two input bits $a$ and $b$, the XOR returns 1 if $a \ne b$ and 0 if $a = b$, i.e.,~$0 \oplus 0 = 0$, $0 \oplus 1 = 1$, $1 \oplus 0 = 1$, and $1 \oplus 1 = 0$.

Let $b$ be an $n$-bit binary number and $b_i$ its $i$\textsuperscript{th} bit. Assume that $b_n$ is the leftmost bit. Then the representation of $b$ in Gray code is given by the equations
\[
\begin{split}
g_{n} = & \,\,b_{n}\\
g_{n-1} = & \,\,b_{n} \oplus b_{n-1} \\
\dots \\
g_{2} = & \,\,b_{3} \oplus b_{2} \\
g_{1} = & \,\, b_{2} \oplus b_{1} \\
\end{split}.
\]
In practice, the conversion can be carried out using the following procedure based on the equations above:\footnote{We employ this procedure in our risk metrics calculation. A Python function converting a binary number into Gray code is available in Appendix~\ref{appendixSrc}.}
\begin{itemize} 
\item take number $b$ and the same number shifted by one bit to the right
\item remove the rightmost bit of the shifted number
\item apply the XOR function to the equivalent bit positions in both numbers (bit-wise XOR)
\end{itemize}
We illustrate the procedure on the conversion of \textcolor{red}{101}\textcolor{blue}{1} (i.e.,~the decimal number 11) to Gray code. Shifting the number to the right, we get 0\textcolor{red}{101}(\textcolor{blue}{1}). The rightmost bit of the shifted number (i.e.,~\textcolor{blue}{1}) is forgotten, hence we are left with 0\textcolor{red}{101}. The next step is to apply the bit-wise XOR function to numbers 1011 and 0101, which delivers 1110, because the bits on the first (leftmost), second, and third places differ from each other, while the last bits (the rightmost ones) do not.

The conversion from Gray code back to standard binary representation is done as follows:
\[
\begin{split}
b_{n} = & \,\,g_{n}\\
b_{n-1} = & \,\,g_{n} \oplus g_{n-1} \\
\dots \\
b_{2} = & \,\,g_{n} \oplus g_{n-1} \oplus \dots \oplus g_{2} \\
b_{1} = & \,\, g_{n} \oplus g_{n-1} \oplus \dots \oplus g_{2} \oplus g_{1}  \\
\end{split}
\]

\clearpage
\section{The IBM Quantum\textsuperscript{TM} Environment and the Qiskit Language}

\setcounter{table}{0}
\setcounter{figure}{0}
\renewcommand{\thetable}{D\arabic{table}}
\renewcommand{\thefigure}{D\arabic{figure}}

\label{appendixIBM}
IBM Quantum\textsuperscript{TM} is a cloud-based quantum computing platform. It allows users to use real quantum processors and quantum computer simulators (i.e.,~software running on a classical computer and simulating a quantum computer). Some of the processors are available for free for education and research, provided that the results obtained are not used for commercial purposes. Currently, the fee-free processors comprise one single-qubit machine and six five-qubit processors. A 127-qubit processor is available on a partnership basis, and IBM plans to build a 1,000-qubit processor by 2023.

There are several ways users can program the quantum processors and simulators. The first is by using a {\bf quantum composer}. This feature allows the user to build a quantum circuit using the drag-and-drop method. The composer also supports the {\bf QASM} assembly language, which is a text description of the gates used in the circuits. The composer mode also provides visualization of the resulting quantum state and the probability distribution of the circuit's output. The composer is useful for the beginners in quantum computing and for circuit debugging.

A more sophisticated tool for the development of quantum algorithms is the Python-based language {\bf Qiskit}. IBM provides {\bf a Quantum Lab}, which is web-based environment using interactive Jupyter notebooks. Users can write Python code here, debug it, and run it on the quantum processors. All the {\it classical Python} libraries are also available to be combined with the Qiskit ones. IBM provides several modules for implementing the most important quantum algorithms, along with specialized libraries for finance, machine learning, optimization, and quantum chemistry. Qiskit can be installed on a local computer and included in Python development tools, too.

\clearpage
\section{Binary Optimization with the QAOA -- Additional Materials}

\setcounter{table}{0}
\setcounter{figure}{0}
\renewcommand{\thetable}{E\arabic{table}}
\renewcommand{\thefigure}{E\arabic{figure}}

\label{appendixQAOA}
In this appendix, we provide a list of all the possible portfolios which can occur in the demonstration of binary optimization using the QAOA presented in Section~\ref{subsectionQAOADemonstration}. We also show a quantum circuit implementing one iteration of the QAOA for five qubits.

\begin{table}[h]
\caption{Possible Portfolios in Binary Optimization with Five Assets}
\begin{center}
{\footnotesize
\begin{tabular}{c|ccccc|ccc|c|c}
\hline\hline 
\thead{\bf Portfolio} & \thead{\bf AMD} & \thead{\bf Intel}  & \thead{\bf Qual.}  & \thead{\bf An. Dev.}  & \thead{\bf TI}  &
\thead{\bf Return}  & \thead{\bf Risk}  & \thead{\bf Max. \\ \bf assets}  & \thead{\bf Total \\ \bf obj. func.}  & \thead{\bf Feasible \\ \bf solution} 
 \\
\hline \vspace{0.05cm}
0&0&0&0&0&0&0.0000&0.0000&9&9.0000&No\\
1&0&0&0&0&1&-0.2132&0.0497&4&3.9857&No\\
2&0&0&0&1&0&-0.2044&0.0348&4&3.9348&No\\
3&0&0&0&1&1&-0.4176&0.1595&1&1.2205&No\\
4&0&0&1&0&0&-0.3975&0.0991&4&3.9989&No\\
5&0&0&1&0&1&-0.6107&0.2546&1&1.4077&No\\
6&0&0&1&1&0&-0.6019&0.2052&1&1.2190&No\\
7&0&0&1&1&1&-0.8151&0.4357&0&0.9278&Yes\\
8&0&1&0&0&0&-0.0530&0.0248&4&4.0462&No\\
9&0&1&0&0&1&-0.2662&0.0808&1&1.0572&No\\
10&0&1&0&1&0&-0.2574&0.0684&1&1.0164&No\\
11&0&1&0&1&1&-0.4706&0.1995&0&0.3273&Yes\\
12&0&1&1&0&0&-0.4505&0.1178&1&1.0205&No\\
13&0&1&1&0&1&-0.6637&0.2795&0&0.4545&Yes\\
14&0&1&1&1&0&-0.6549&0.2327&0&0.2760&Yes\\
15&0&1&1&1&1&-0.8681&0.4695&1&2.0099&No\\
16&1&0&0&0&0&-0.9515&0.2021&4&3.8569&No\\
17&1&0&0&0&1&-1.1646&0.2131&1&0.6880&No\\
18&1&0&0&1&0&-1.1558&0.2075&1&0.6742&No\\
19&1&0&0&1&1&-1.3690&0.2936&0&-0.1948&Yes\\
20&1&0&1&0&0&-1.3490&0.3667&1&1.1176&No\\
21&1&0&1&0&1&-1.5621&0.4835&0&0.3717&Yes\\
22&1&0&1&1&0&-1.5534&0.4434&0&0.2203&Yes\\
23&1&0&1&1&1&-1.7665&0.6352&1&1.7744&No\\
24&1&1&0&0&0&-1.0045&0.2325&1&0.9255&No\\
25&1&1&0&0&1&-1.2177&0.2499&0&-0.2182&Yes\\
{\bf 26}&{\bf 1}&{\bf 1} &{\bf 0}&{\bf 1}&{\bf 0}&{\bf -1.2089}&{\bf 0.2468}&{\bf 0}&{\bf -0.2218}&{\bf Yes}\\
27&1&1&0&1&1&-1.4220&0.3391&1&0.9344&No\\
28&1&1&1&0&0&-1.4020&0.3909&0&0.1617&Yes\\
29&1&1&1&0&1&-1.6152&0.5140&1&1.4410&No\\
30&1&1&1&1&0&-1.6064&0.4765&1&1.2997&No\\
31&1&1&1&1&1&-1.8196&0.6746&4&4.8789&No\\ \hline
\end{tabular}}
\end{center}
\vspace{0.1cm}
{\footnotesize \textbf{\textit{Note:}}  The best portfolio is highlighted in bold.}\\
{\footnotesize \textbf{\textit{Source:}} Author's own calculations}
\label{tableQAOAadditional}
\end{table}

\begin{figure}[hbt]
\caption{Quantum Circuit Implementing One Iteration of the QAOA for Five Binary Variables}
\begin{center}
\includegraphics[scale = 0.9, angle =90]{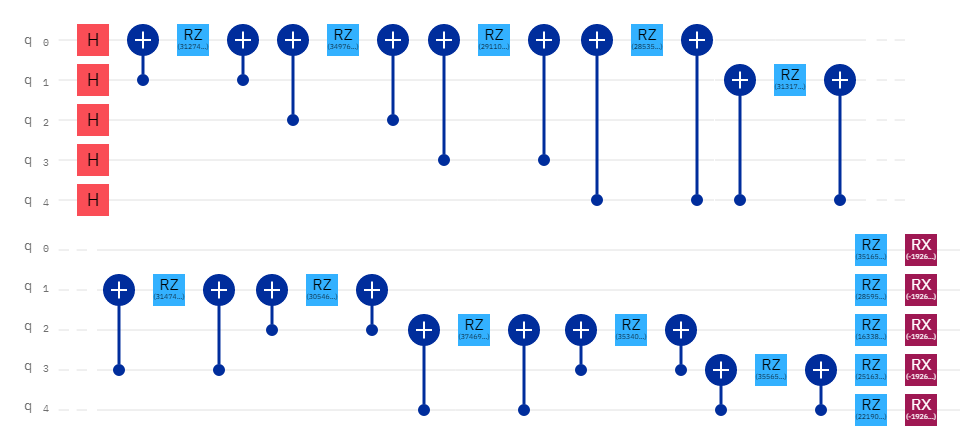}
\end{center}
\vspace{0.1cm}
{\footnotesize \textbf{\textit{Source:}} Author's own creation on IBM Quantum\textsuperscript{TM}}
\label{fig_qaoa_5_qubits}
\end{figure}

\clearpage
\section{Qiskit Source Codes}

\setcounter{table}{0}
\setcounter{figure}{0}
\renewcommand{\thetable}{F\arabic{table}}
\renewcommand{\thefigure}{F\arabic{figure}}

\label{appendixSrc}

\subsection{Quantum Fourier Transform and Its Inverse}
\vspace*{0.5 cm}
\begin{python}
# Importing standard Qiskit libraries
from qiskit import ClassicalRegister, QuantumRegister, QuantumCircuit 
from qiskit import Aer, execute, IBMQ
import math as m #Python math functions

def QFT(qubits, inverse, q, c):
    #if inverse = True than inverse QFT is returned
    n = len(qubits)
    circuit = QuantumCircuit(q, c)
    
    if inverse:
        for i in range(n-1,-1,-1):            
            for j in range(n-1,i,-1):
                #add controlled Rk gates (U1 on IBM Q) 
                #(older version of Qiskit cu1, now cp)
                circuit.cp(-m.pi/2**(j - i),q[qubits[j]],q[qubits[i]])
            circuit.h(q[qubits[i]])
    else:
        for i in range(0,n-1):
            circuit.h(q[qubits[i]])
            for j in range(i+1,n):
                circuit.cp(m.pi/2**(j - i),q[qubits[j]],q[qubits[i]])
    
        circuit.h(q[qubits[n-1]])
        
    return circuit

n = 4 #no of qubits
q = QuantumRegister(n, name = 'q')
c = ClassicalRegister(n, name = 'c')
#apply QFT on qubits no 0,1,2,3, use registers q and c
#prepare QFT, not inverse (parameter inverse set on False)
circuit = QFT([0, 1, 2, 3], False, q, c)
circuit.draw() #draw circuit
\end{python}

\subsection{Decoherence Experiments}
\vspace*{0.5 cm}
\begin{python}
#connection to IBM Q simulators and loading libraries
from qiskit import QuantumRegister, ClassicalRegister, QuantumCircuit
from qiskit import execute, Aer, IBMQ
from qiskit.compiler import transpile, assemble
Provider = IBMQ.load_account()

#connecting to quantum processor
#simulator
#Backend = Aer.backends(name='qasm_simulator')[0] 
#real quantum processor
Backend = Provider.backends(name='ibmq_lima')[0]

#initialization of a circuit composed of 1 qubit #
q = QuantumRegister(5, name = 'q') #the real processor has 5 qubits
c = ClassicalRegister(1, name = 'c') #only one qubit is measured
qbit = 2 #position of qubit used in the experiment

t = 200 #max. number of I gates
#number the algorithm repetition (to get data for output distribution)
Shots = 8192 

#spontaneous relaxation
#make circuits with different number of I operators
for i in range(0,t+1): 
    qc = QuantumCircuit(q, c, name = 'CircInit')
    qc.x(q[qbit]) #X gate to prepare |1>
    for j in range(0,i): #adding identity operators
        qc.id(q[qbit])
    qc.measure(q[qbit],c[0])
    job = execute(qc, backend = Backend, optimization_level= 0, 
               shots = Shots)
    results = job.result().get_counts(qc)
    print(['No of Ids:', i, ':', results])

#spontaneous dephasing
for i in range(0,t+1):
    qc = QuantumCircuit(q, c, name = 'CircInit')
    qc.h(q[qbit]) #Hadamard gate to prepare |+>
    for j in range(0,i):
        qc.id(q[qbit])
    qc.h(q[qbit]) #Hadamard gate to return |+> back to |0>
    qc.measure(q[qbit],c[0])    
    job = execute(qc, backend = Backend, optimization_level= 0, 
               shots = Shots)
    results = job.result().get_counts(qc)
    print(['No of Ids:', i, ':', results])
\end{python}

\subsection{Risk Measures Calculation}
\vspace*{0.5 cm}
\begin{python}
#-------import libraries------
from qiskit import ClassicalRegister, QuantumRegister, QuantumCircuit
from qiskit import  Aer, execute, IBMQ
from qiskit.compiler import transpile, assemble
import numpy as np
import math as m

#-------functions-------

#conversion of decimal integer to binary
def dec2bin(decimal, bits):
    b = bin(decimal)
    b = b[2:len(b)]
    while len(b) < bits:
        b = '0' + b
    return b

#conversion of binary number to Gray code (see appendix B)
def bin2Gray(binary):
    #bin to Gray: b XOR (b moved to right by 1 bit)
    n = int(binary,2) #binary to decimal
    l = len(binary)
    b = bin(n ^ (n>>1)) #^ -> XOR; n>>1 - move by 1 bit to the right
    b = b[2:len(b)]
    while len(b) < l:
        b = '0' + b
    return b

def bitwiseDotProduct(a, b):
    #dot product of a and b mod 2
    aa = int(a,2)
    bb = int(b,2)
    c = bin(aa & bb) #product of coresponding bits
    c = c[2:len(c)]
    #logical OR of all bits in c
    s = 0
    for x in c:
        s = (s + int(x,2)) 
    return s

#uniform rotations construction (according to Mottonen et al. (2005))
#prepare CNOTs structure for uniform rotations
def _uniformRot(ctrlQubits, trgQubit, circuit):
    if ctrlQubits == 1: #elementary circuit
        circuit.i(q[trgQubit])
        circuit.cx(q[trgQubit - 1],q[trgQubit])
        circuit.i(q[trgQubit])
        circuit.cx(q[trgQubit - 1],q[trgQubit])
    else:
        circuit = _uniformRot(ctrlQubits - 1, trgQubit, circuit)
        del circuit.data[len(circuit.data) - 1]
        circuit.cx(q[trgQubit - ctrlQubits],q[trgQubit])
        
        circuit = _uniformRot(ctrlQubits - 1, trgQubit, circuit)
        del circuit.data[len(circuit.data) - 1]        
        circuit.cx(q[trgQubit - ctrlQubits],q[trgQubit])
    return circuit

#set angles of Ry gates and return uniform rotations circuit
def uniformRot(theta, q, c): 
    qubits = m.ceil(m.log(len(theta),2))
    N = 2**qubits
 
    circuit = QuantumCircuit(q,c)
    #prepare structure of CNOTs
    circuit = _uniformRot(qubits, qubits, circuit)
    hlpCircuit = QuantumCircuit(q) #used for preparing Ry gate

    #here thetas are modified to phis with matrix M
    M = np.zeros((N,N))
    c = 1/N
    for i in range(0,N):
        for j in range(0,N):
            #matrix M definition in the paper
            b = dec2bin(j, qubits)
            g = bin2Gray(dec2bin(i, qubits))
            M[i][j] = c*((-1)**bitwiseDotProduct(b,g)) 
    
    phi = M.dot(theta) #angles modification, dot - matrix multiplication
    
    #modify circuit (replace I with Ry)
    for i in range(0, len(circuit.data)):
        if i 
           #prepare auxiliary circuit
            if len(hlpCircuit.data) > 0: del hlpCircuit.data[0] 
            hlpCircuit.ry(phi[int(i/2)],q[qubits]) #add Ry gate
            gate = hlpCircuit.data[0] #save gate definition...
            
            del circuit.data[i] #...del I gate...
            circuit.data.insert(i,gate) #...and insert Ry gate
    
    return circuit

#state preparation (according to the paper Mottonen et al. (2005))
#quantum phase is ignored as it is not necessary for our purposes
#calculate alphas angles based on probability amplitudes
def calculateAlphas (a):
    qubits = m.ceil(m.log(len(a),2))
    gates = 2**(qubits - 1)
    
    alphas = np.zeros((qubits, gates))
    
    for k in range(1, qubits + 1):
        for j in range(1, 2**(qubits - k) + 1):
            s1 = 0
            e = 2**(k - 1)
            for l in range(1, e + 1):
                s1 += (a[(2*j - 1)*e + l - 1])**2            
            
            s2 = 0
            e = 2**k
            for l in range(1, e + 1):
                s2 += (a[(j - 1)*e + l - 1])**2
            
            alphas[k-1, j-1] = 2*m.asin(m.sqrt(s1)/m.sqrt(s2))
    
    return alphas

#make circuit for state preparation, based on alphas and uniform rot.
def statePreparationCircuit(amplitudes, q, c):
    #calc. alphas based on amplitudes
    alphas = calculateAlphas(amplitudes) 
    qubits = m.ceil(m.log(len(amplitudes),2))
    
    circuit = QuantumCircuit(q, c, name = 'q')        
    circuit.ry(alphas[qubits-1][0],q[0]) #initial gate on q[0]
    
    k = 0
    for i in range(qubits-2, -1, -1):
        k += 1              
        #build rest of circuit with uniform rotations
        #inplace = True => result of uniformRot is added to 
        #variable circuit and to var. circuit
        circuit.compose(uniformRot(alphas[i,0:2**k], q, c), 
                             inplace = True)
        
    return circuit

#converts probabilities to probability amplitudes    
def probToAmplitudes (prob): 
    a = np.zeros(len(prob))
    for i in range(0, len(prob)):
        a[i] = m.sqrt(prob[i]/100)
        
    return a

#risk measures calculation itself
def riskMeasuresCalculation 
                (probabilities, alphaVar, processor, Shots, q, c):  
   #number of necessary qubits
    qubits = m.ceil(m.log(len(probabilities),2))
    N = 2**qubits #number of quantum states
    
    res = {}; #empty dictionary for results
    
    #converting probabilities to quantum amplitudes
    amplitudes = probToAmplitudes(probabilities) 
    #state preparation based on the prob. distribution
    circuitStatePrep = statePreparationCircuit(amplitudes, q, c)

    #calculation of EX
    alphas = np.zeros(N) #alphas for EX calculation
    for i in range(0, N): alphas[i] = 2*m.acos(m.sqrt(1-i/(N-1)))
    #connect EX cal. circuit to state prep. circuit and add measurement
    #circuit = circuitStatePrep + uniformRot (alphas, q, c) 
    circuit = circuitStatePrep.compose(uniformRot (alphas, q, c), 
                         inplace = False)
    circuit.measure(q[qubits],c[0])
    M = execute(circuit, 
                 processor, shots = Shots).result().get_counts(circuit)
    #EX calculation post-processing    
    EX = (M['1']/Shots)*(N-1) #M['1']/shots = P(|1>)
    res['Expected'] = round(EX ,3)

    #calculation of EX^2
    N2 = N**2 #N^2
    alphas = np.zeros(N) #alphas for EX calculation
    for i in range(0, N): alphas[i] = 2*m.acos(m.sqrt(1-i**2/(N2-1)))
    #circuit = circuitStatePrep + uniformRot (alphas, q, c) 
    circuit = circuitStatePrep.compose(uniformRot (alphas, q, c), 
                          inplace = False)
    circuit.measure(q[qubits],c[0])
    M = execute(circuit, 
                 processor, shots = Shots).result().get_counts(circuit)
    EX2 = (M['1']/Shots)*(N2-1) #M['1']/shots = P(|1>)
    #stdDev = sqrt[E(X^2) - (EX)^2]
    res['StdDev'] = round(m.sqrt(EX2 - EX**2),3) 
    
    #VaR calculation
    a = 0
    b = len(probabilities) - 1
    alpha = 1 - alphaVar #converting VaR to percentil

    #if alpha < a probability of the first value
    if alpha <= probabilities[0]/100: 
        VaR = 0
    else: #interval bisection
        while (b - a > 1):
            VaR = m.floor((a + b)/2)
            alphas = np.zeros(N)
            #f(i) = 1 for states 0..s, otherwise f(i)=0...
            for i in range(0,VaR + 1): alphas[i] = m.pi 
            #circuit = circuitStatePrep + uniformRot (alphas, q, c)
            circuit = circuitStatePrep.compose(uniformRot (alphas, q, c),
                                  inplace = False)
            circuit.measure(q[qubits],c[0])
            M = execute(circuit, 
                   processor, shots=Shots).result().get_counts(circuit)
            #...P(1) = sum_{i=0}^{s} p_i => cum. distrib. function...                 
            P1 = (M['1']/Shots) 
            #...comparing cum distribution with alpha => percentil 
            #calculation and interval bisection
            if P1 > alpha:
                b = VaR
            else:
                a = VaR        
    res['VaR'] = round(VaR,3)

    #CVaR calculation
    if VaR != 0:
        alphas = np.zeros(N) #alphas for CVaR calculation
        for i in range(0, N): 
            if i > VaR:
                alphas[i] = 0
            else:
                alphas[i] = 2*m.acos(1 - m.sqrt(i/VaR))
        #circuit = circuitStatePrep + uniformRot (alphas, q, c) 
        circuit = circuitStatePrep.compose(uniformRot (alphas, q, c), 
                                inplace = False)
        circuit.measure(q[qubits],c[0])
        M = execute(circuit, 
                 processor, shots = Shots).result().get_counts(circuit)
        #CVaR calculation post-processing        
        res['CVaR'] = round(VaR*(M['1']/Shots)/P1,3) #P1 = P[X<=VaR]
    else:
       #VaR = 0 => all values under VaR also zero => CVaR = 0
        res['CVaR'] = 0 
    
    return res

#-------main program-------
#Processor selection:
#processor = Aer.backends(name='qasm_simulator')[0] #simulator
#real quantum processors
Provider = IBMQ.load_account()
processor = Provider.backends(name='ibmq_belem')[0]

# 3 qubits
probabilities=[0.431, 0.863, 7.443, 38.188, 40.885, 9.493, 1.834, 0.863]
# 4 qubits
#probabilities=[0.431, 0.216, 0.539, 1.294, 3.02, 5.933, 14.024,
#                 21.467, 22.438, 14.995, 8.846, 3.344, 1.402, 
#                 0.863, 0.324, 0.863]

shots = 8192
alphaVar = 0.95
#alphaVar = 0.99

l = len(probabilities)
qubits = m.ceil(m.log(l,2)) #number of necessary qubits    
#prepare registers in quantum computer
q = QuantumRegister(qubits + 1, name = 'q') 
c = ClassicalRegister(1, name = 'c')

resQuantum = riskMeasuresCalculation (probabilities, alphaVar, 
                                         processor, shots, q, c)
print(resQuantum)

\end{python}

\subsection{Solving 2x2 Linear System with HHL}
\vspace*{0.5 cm}
\begin{python}
from qiskit import ClassicalRegister, QuantumRegister, QuantumCircuit
from qiskit import Aer, execute, IBMQ
from qiskit.compiler import transpile, assemble
import numpy as np
from numpy import linalg as la
import math as m

#solving system of equations Ax = b for #A = [[1.5 0.5],[0.5 1.5]]
def hhlExample(vector_b, processor, Shots):
    #only for non-negative solutions as we do measurement only in 
    #computational basis    
    #based on paper by Cao et al. (2012)
    q = QuantumRegister(5, name = 'q')
    c = ClassicalRegister(5, name = 'c')
    
    hhl = QuantumCircuit(q, c, name = 'hhl')

    #preparing initial state for phase estimation 
    hhl.h(q[0])
    hhl.h(q[1])
    #preparing state representing vector b (assuming b is normalized)
    hhl.ry(2*np.arccos(vector_b[0]), q[2])
    #hhl.initialize(vector_b, q[2])
    #powers of operator exp(i*pi*A) in the phase estimation
     #A = [[1.5 0.5],[0.5 1.5]], A = 0.5X + 1.5I
     #exp(c*A)=exp(c*(0.5X+1.5I))=exp(c1*X)*exp(c2*I)=Rx(2*c1)*exp(c2)
     #we have controlled Rx(2*c1) and controlled global phase U1(c2)
     #exp(i*pi*A)
    hhl.u1(3*np.pi/2, q[0])
    hhl.crx(-np.pi, q[0], q[2])
        #exp(i*pi*A/2)
    hhl.u1(3*np.pi/4, q[1])
    hhl.crx(-np.pi/2, q[1], q[2])
    #invQFT
    hhl.swap(q[0],q[1])
    hhl.h(q[1])
    hhl.cu1(-np.pi/2,q[0],q[1])
    hhl.h(q[0])
    #inverting eigenvalues
    hhl.swap(q[0],q[1])
    #rotations on ancilla qubit
    #increase r => decrease prob. of ancilla being in |1>, 
    #but increase accuracy of calc. |x>
    #value of r depends on minimum eigenvalue of matrix A
    #according to original paper r >= 2.65
    r = 2.65
    hhl.cry(2*np.pi/2**r,q[0],q[3])
    hhl.cry(np.pi/2**r,q[1],q[3])

    #uncomputing qubits 0 to 2
        #eigenvalues
    hhl.swap(q[0],q[1])
        #QFT
    hhl.h(q[0])
    hhl.cu1(np.pi/2,q[0],q[1])
    hhl.h(q[1])
    hhl.swap(q[0],q[1])
        #inverting matrix exp (i.e.,~rest of phase estimation)
    hhl.crx(np.pi/2, q[1], q[2])
    hhl.u1(-3*np.pi/4, q[1])
    hhl.crx(np.pi, q[0], q[2])
    hhl.u1(-3*np.pi/2, q[0])
        #inverting Hadamards (H is its own inversion, i.e.,~HH=I)
    hhl.h(q[0])
    hhl.h(q[1])                   
        #measurement
    hhl.measure(range(4),range(4))   
    
    M = execute(hhl, processor, shots = Shots).result().get_counts(hhl)
    
    #|1> post selection in ancilla qubit 
         #first qubit is unused, hence always in state |0>
         #second qubit is the ancilla - has to be in state |1>
         #third qubit contains results - we need both |0> and |1>
         #fourth and fifth qubits are working ones - has to be |0>   
    x_0 = M['01000']/Shots #probabilities
    x_1 = M['01100']/Shots
    
    #obtained state with ancilla |1> is "unnormalized" state, 
    #so we need to normalize it
     #x_0/norm and x_1/return state with sum of probs. equal 1
    normSquared = x_0 + x_1
    #we need amplitudes, instead of probabilities
    x_0 = round(m.sqrt(x_0/normSquared),4) 
    x_1 = round(m.sqrt(x_1/normSquared),4)
    #now x_0 and x_1 contains normalized solution |x>
    x = [x_0, x_1]
    
    #expceted results (normalized)
    expected_x = la.solve(np.array([[1.5, 0.5],[0.5, 1.5]]), vector_b)
    expected_x = expected_x/la.norm(expected_x)
    for i in range(0,len(expected_x)): 
                expected_x[i] = round(expected_x[i],4)
    
    return [x, expected_x, hhl]

#angles for preparing normalized right side vectors
#(sin^2+cos^2=1)
thetas = [np.pi/7, np.pi/6, np.pi/4, np.pi/3]
processor = Aer.backends(name='qasm_simulator')[0]

#provider = IBMQ.load_account()
#processor = provider.backends(name='ibmq_lima')[0]

shots = 8192

for theta in thetas:    
   #normalized right sides 
    vector_b = np.array([m.cos(theta), m.sin(theta)]) 
    x, expcted_x, circuit = hhlExample(vector_b, processor, shots)
    print('-----------------')
    print('Theta: ' + str(round(theta,4)))
    print('Vector b: ' + str(vector_b))
    print(x)
    print(expcted_x)
\end{python}

\subsection{Portfolio Optimization with HHL Algorithm}
\vspace*{0.5 cm}
\begin{python}
from qiskit import ClassicalRegister, QuantumRegister, QuantumCircuit
from qiskit import Aer, execute, IBMQ
from qiskit.compiler import transpile, assemble
from qiskit.quantum_info import Statevector

from qiskit.algorithms.linear_solvers.hhl import HHL

import numpy as np
from numpy import linalg as la
import math as m

#prepare matrix for optimization based on returns, prices 
#and covariance matrix
def ptfOptimLinearSystem(CovarianceMatrix,Prices,Returns,Gain,Budget):
    assets = len(CovarianceMatrix[:,0])

    A = np.block([[Returns.transpose()],[Prices.transpose()]])
    A = np.block([np.zeros((2,2)), A])
    A = np.block([[A],[Returns,Prices,CovarianceMatrix]])
    
    b = np.array([Gain, Budget])
    b = np.block([b,np.zeros((1,assets))])    
    
    return [A, b[0,:]]

#auxiliary function: conversion from dec to bin
def dec2bin(decimal, bits): 
    b = bin(decimal)
    b = b[2:len(b)]
    while len(b) < bits:
        b = '0' + b
    return b

#function for linear systems solving, based on HHL implemented in Qiskit
def hhlSolver (A, b, Processor, Shots, Verbose = False, 
                gatesSet = ['id', 'rz', 'sx', 'x', 'cx']):
    n = len(A[1,:])
    qubitsForResults = m.ceil(m.log(n,2))

    bNorm = la.norm(b) #norm of right-side

    hhlSolver = HHL(quantum_instance = Processor)
    #prepare circuit for given A and b    
    hhlCircuit = hhlSolver.construct_circuit(A, b) 
       
    #get state vector representing result of HHL algorithm    
    stateVector = Statevector(hhlSolver.solve(A, b).state).data    
        
    #adapt the HHL circuit to real quantum hardware, 
    #default settings: the gate set implemented on IBM 
    #processors (sx = sqrt(X))
    circuitToRun = transpile(hhlCircuit, basis_gates = gatesSet)    
    #add measurement
    circuitToRun.measure_all()
    
    M = execute(circuitToRun, 
            Processor, shots = Shots).result().get_counts(circuitToRun)

    #width - num of qubits and classical bit summed
    usedQubits = circuitToRun.width()/2 
    zeros = dec2bin(0,usedQubits - qubitsForResults - 1)        
    
    amplitudes = np.zeros(n)
    x = np.zeros(n)
    
    for i in range(0,n):
        #ancilla qubit has to be |1>, working qubits |0>, results are in 
        #last "qubitsForResults" qubits
        #if qubitsForResults = 2 and total qubits is 6, then 3 are 
        #working qubits (1 qubit is ancilla),
        #results are stored in |1 000 00>, |1 000 01>, 
        #|1 000 10> and |1 000 11>
        indx = '1' + zeros + dec2bin(i,qubitsForResults)
        #bNorm*m.sqrt(M[indx]/Shots) - amplitudes needed => sqrt
        #and adjust them by norm of vector b to
        #get "unnormalized" solution x        
        x[i] = np.round(bNorm*m.sqrt(M[indx]/Shots),4)        
        #"denormalization" of state vector amplitudes
        #abs in state vector - to eliminate a global phase
        amplitudes[i] = np.round(bNorm*abs(stateVector[int(indx,2)]),4) 
    
    expected_x = la.solve(A, b)
    matrixEigenvalues = np.sort(la.eigvals(A))
    
    if Verbose:
        print('HHL results on processor ' + str(processor) + ':')
        print(x)
        
        print('\nState vector aplitudes after "denormalization by 
                          ||b|| and elimination of global phase:')
        print(amplitudes)

        #solution of Ax = b obtained by a classical method
        print('\nExpected solution:')
        print(expected_x)

        #eigenvalues of a matrix A
        print('\nMatrix A eigenvalues:')
        print(matrixEigenvalues)
    
    return [x, amplitudes, expected_x, matrixEigenvalues, 
                         hhlCircuit, circuitToRun]

C = np.array([[0.15, -0.43],[-0.43, 2.46]]) #covariance matrix
P = np.array([[1],[1]]) #prices
R = np.array([[5.86],[16.78]]) #returns

G = 7 #expected gain
B = 1 #budget

A, b = ptfOptimLinearSystem(C, P, R, G, B)

processor = Aer.backends(name='qasm_simulator')[0]
shots = 8192

hhlSolver(A, b, processor, shots, True)
\end{python}

\subsection{Binary Optimization with QAOA}
\vspace*{0.5 cm}
\begin{python}
# Importing standard Qiskit libraries and configuring account
from qiskit import QuantumCircuit, QuantumRegister, ClassicalRegister
from qiskit import execute, Aer, IBMQ
from qiskit_optimization import QuadraticProgram #quadratic optimization
#optimization algorithms
from qiskit_optimization.algorithms import MinimumEigenOptimizer 
from qiskit.algorithms import QAOA, NumPyMinimumEigensolver 
import numpy as np #Python numeric library

def quboQaoaSolver(A, b, c, processor):
    variables = len(b) #number of required binary variables
    
    #specify a qubo task
    task = QuadraticProgram(name = 'QUBO on QC')
    
    for i in range(0,variables):
        task.binary_var(name = 'x' + str(i)) #add variables
        
    #set objective function
    task.minimize(linear = b, quadratic = A, constant = c) 
    print('Task formulation:')
    print(task.export_as_lp_string()) #show the QUBO task definition
    
    print('\nIsing Hamiltonian representing QUBO task:')
    #show Ising Hamiltonian related to the QUBO task
    operator, offset = task.to_ising() 
    print(operator)
    
    #run on real quantum processor
    qaoa_solver=MinimumEigenOptimizer(QAOA(quantum_instance=processor)) 
     #classical solver for comparison
    exact_solver = MinimumEigenOptimizer(NumPyMinimumEigensolver())
        
    print('\nResults on QAOA:')
    qaoa_results = qaoa_solver.solve(task)
    print(qaoa_results)
    
    print('\nResults on classical solver:')
    exact_result = exact_solver.solve(task)
    print(exact_result)

#prepare A, b and c based on returns, covariance matrix and lambdas
def portfolioBinOptTaskDef(returns, covariance_matrix, assets_number, 
             lambda_returns, lambda_risk, lambda_asset_number):

    n = len(returns)
    
    A = lambda_risk*covariance_matrix + 
                          lambda_asset_number*np.ones((n,n))
    b = -(lambda_returns*returns + 
                 lambda_asset_number*2*assets_number*np.ones((1,n)))
    #after calculation before a matrix 1-by-n is returned, 
    #however, the solver expects a vector                 
    b = b[0,:] 
    c = lambda_asset_number*assets_number**2
    
    return [A, b, c]

returns = np.array([0.951464,0.05303,0.397515,0.204385,0.21317])
covariance_matrix = np.array([
    [0.202087,0.002804,0.032734,-0.014687,-0.019333],
    [0.002804,0.024813,-0.00308,0.00442,0.003153],
    [0.032734,-0.00308,0.099099,0.03567,0.05288],
    [-0.014687,0.00442,0.03567,0.034792,0.037495],
    [-0.019333,0.003153,0.05288,0.037495,0.049725]])

lambda_returns = 1
lambda_risk = 4
lambda_asset_number = 1
assets_number = 3

A, b, c = portfolioBinOptTaskDef(returns, covariance_matrix, 
                        assets_number, lambda_returns, lambda_risk, 
                        lambda_asset_number)

#connecting a simulator
#processor = Aer.backends(name = 'qasm_simulator')[0] 

#real quantum processor
provider = IBMQ.load_account()
processor = provider.backends(name='ibmq_lima')[0]

quboQaoaSolver(A, b, c, processor)
\end{python}

\end{document}